# Survey of Nodeless Regular Almost-Everywhere Holomorphic Solutions for Exactly Solvable Gauss-Reference Liouville Potentials on the Line

## I. Subsets of Nodeless Jacobi-Seed Solutions Co-Existent with Discrete Energy Spectrum

**G. Natanson**

ai-solutions Inc.

2232 Blue Valley Dr.

Silver Spring MD 20904

U.S.A.

greg_natanson@yahoo.com

The paper presents a complete list of nodeless regular almost-everywhere holomorphic (AEH) solutions for a subset of rational canonical Sturm-Liouville equations (RCSLEs) exactly quantized on a finite interval by *classical* Jacobi polynomials. The subset was constrained by the requirement that the appropriate Liouville transformation results in the Schrödinger equation on the line. To stress that the given RCSLE has three regular singular points (including infinity), we refer to the resultant Liouville potential as 'regular Gauss-Reference' (*r*-GRef) potential, in contrast with the 'confluent Gauss-Reference' (*c*-GRef) potentials analyzed in Part II.

The common remarkable feature of the selected nodeless solutions co-existent with the discrete energy spectrum is that they can be used as seed functions for multi-step 'canonical Liouville-Darboux transformations' (CLDTs) to convert the (appearing in the resultant Schrödinger equation) into its isospectral rational SUSY partners conditionally exactly quantized by the so-called 'Jacobi-Seed' ($\mathcal{J}$S) Heine polynomials.

## Introduction

As recently shown by the author [1] almost-everywhere holomorphic (AEH) solutions of the rational canonical Sturm-Liouville equations (RCSLEs) exactly quantized by classical Jacobi, classical generalized Laguerre, and Romanovski-Routh [2-4] polynomials can be used as seed functions for constructing multi-step SUSY ladders of rational Liouville potentials (RLPs) conditionally exactly quantized by polynomials. We refer to the RLPs associated with these three RCSLEs as *r*-, *c*-, and *i*-Gauss-reference (GRef) potentials to indicate that we deal with Gauss-type equations. Similarly AEH solutions of these equations are referred to as 'Gauss-seed' (GS) or, to be more precise, Jacobi-seed ($\mathscr{J}$S), Laguerre-seed ($\mathscr{L}$S), or Routh-seed[x)] ($\Re$S) depending on which (generally non-orthogonal) polynomials are used to form the given set of GS solutions. (So far [4] we were able construct irregular-at-infinity $\Re$S solutions only in the particular case of the *i*-GRef potential constructed by Milson [7] using the symmetric change of variable for the Liouville transformation. It remains uncertain whether such solutions exist for the general RLP inherent in the generic RCSLE with three regular singular points $-i$, $+i$, and $\infty$.)

Here and in the following paper referred below as Part II we restrict our analysis solely to nodeless regular AEH solutions of the RCSLEs associated with the *r*- and *c*-GRef potentials [8] accordingly. The common remarkable feature of these solutions is that they all lie below the ground energy level. Since the multi-step Darboux transformations (DTs) using regular solutions as seed functions engender a sequence of isospectral SUSY partners and the appropriate factorization functions (FFs) themselves are regular solutions of the transformed Schrödinger equation the latter must lie below the ground energy level of the deformed potential and therefore are necessarily nodeless. It has been proven in [1] that the canonical Liouville-Darboux transformation (CLDT ) using any nodeless AEH as its FF converts the GRef potential into its rational SUSY partner quantized by the so-called 'GS Heine polynomials'.

---

[x)] See [4] for the precise definition of this two-parameter (generally non-orthogonal) set of polynomials introduced by Routh [5] in the end of 19th century but only recently brought back to life (though in a historically inaccurate context) in Ismail's monograph [6].

In this paper we only discuss ∮S solutions for RCSLEs with three regular singular points. The RCSLE associated with the *c*-GRef Liouville potential will be studied in a separate publication referred to below as Part II.

The paper is organized as follows. In Section 2 we write RCSLEs of our interest in Bose's form [9] and outline two alternative schemes (the so-called double-step algorithms) for computing characteristic exponents (ChExps) of all possible ∮S solutions at *real* energies. The first step requires computation of real roots of the appropriate quartic polynomial in one of the signed exponent differences (signed ExpDiffs). The second signed ExpDiff can be then determined using a fraction with a generally non-zero denominator. For the generic *r*-GRef potential denominator vanishes along the curves where one of the quartic polynomials has a double root so two ∮S solutions formed by Jacobi polynomials of the same order m co-exist at the same nonzero energy.

In Section 3 we derive algebraic formulas for signed ExpDiffs in the limiting case when the zero-energy ExpDiff $\lambda_o$ at the origin vanishes giving rise to the asymptotically-leveled (AL) potential curves. The most precious advantage of this simplification is that the resultant quartic polynomials can be analytically decomposed into the products of two quadratic polynomials which makes it possible to explicitly identify all regular ∮S solutions below the ground energy level. Examination of these quadratic polynomials revealed that the $m^{th}$ eigenfunction **c**m is always accompanied by ∮S solutions of three distinct types: two ∮S solutions **a**m and **b**m regular at 0 and 1, respectively and the ∮S solution **d**m irregular at both ends *provided* that the tangent polynomial (TP) used to construct the change of variable for the Liouville transformation has positive discriminant,

In Section 4 we use this ramification as a starting point for its extension to the region (referred to as Area $A_m$) where the *r*-GRef potential in question has at least m bound energy levels, again assuming that the TP has two distinct real roots. A similar analysis for the TP with a double root (DRtTP) has been already performed by us in [10] using the Williams-Levai potential [11] as the starting point.

As $\lambda_o$ increases at a fixed value of the zero-energy ExpDiff $\mu_o$ at ∞ the upper bound energy state disappears along one of the 'zero-factorization-energy' (ZFE) separatrix (see Appendix A for details). After the m$^{th}$ bound energy state disappears no analytical prediction can be generally made for types of ∮S solutions formed by Jacobi polynomials of the given order m, except the region near the point where two ZFE separatrices meet at $\lambda_o = 0$. It is shown in Appendix B that a pair of nodeless regular ∮S solutions necessarily disappears at least for some *r*-GRef potential curves and therefore such curves serve as natural boundaries for the isospectral SUSY partners generated using these solutions as seed functions.

The second precise case allowing the analytical decomposition of the quartic polynomial is presented by the points where one of the quartic polynomials of our interest has the double root of $-2m-1$ so two other polynomials roots can be easily computed. Since the latter points form the cutoff curve for the appropriate double step algorithm we found it useful to outline main features of these curves in Appendix C while postponing a more detailed analysis of the appropriate closed-form expressions for a separate paper referred to below as Part III.

The analysis presented in Sections 2-4 is focused solely on the generic *r*-GRef potential generated using second-order TP. The common remarkable feature of the linear TP (LTP) *r*-GRef potentials examined in Section 5 is that the quartic polynomial used to obtain ChExps of ∮S solutions can be analytically decomposed into product of quadratic polynomials, by analogy with AL potential curves. As a result one can directly compute signed ExpDiffs for all possible nodeless regular ∮S solutions.

The survey of regular nodeless ∮S solutions is finally completed in Section 6 by a brief analysis of the 'shape-invariant' limit of the *r*-GRef potential on the line: the Rosen-Morse (RM) potential [12] obtained by making the linear coefficient of the LTP tend to 0. The appropriate sets of all the nodeless regular ∮S solutions have been cautiously studied by Quesne [13] so we simply correlate our general approach with her more specific results for this very special case.

The paper is concluded by a brief outline of future developments followed by three aforementioned appendices.

## 2. Two alternative double-step algebraic algorithms for computing characteristic exponents of ∮S solutions

The purpose of this section is to formulate the general algorithm for computing ChExps of all possible ∮S solutions of the RCSLE

$$\left\{\frac{d^2}{dz^2}+I[z;\varepsilon\,|\,{}_1\boldsymbol{G}^{K\Im}]\right\}\,\Phi[z;\varepsilon\,|\,{}_1\boldsymbol{G}^{K\Im}]=0 \tag{2.1}$$

exactly solvable by hypergeometric functions [8] on the interval $0 \le z \le 1$. (An extension of this algorithm to *c*-GRef potentials on the line will be presented in Part II.) In following Milson [7] we refer to the energy-dependent polynomial fraction (PFr)

$$I[z;\varepsilon\,|\,{}_1\boldsymbol{G}^{K\Im}] \equiv {}_1I^{o}[z;{}_1Q^{o}]+{}_1\wp[z;{}_1T_K]\,\varepsilon \tag{2.2}$$

as the Bose invariant (BI) to emphasize Bose's impact [9] on our original studies [8] in this field. Superscripts K and $\Im$ used to identify the given PFr beam (PFrB) ${}_1\boldsymbol{G}^{K\Im0}$ indicate that the rational density function

$${}_1\wp[z;{}_1T_K]=\frac{{}_1T_K[z]}{4z^2(1-z)^2} \tag{2.3}$$

is generated using a $K^{th}$-order tangent polynomial (TP) with $\Im$ roots. The last symbol 0 in superscript K$\Im$0 implies that the TP in question remains positive at the origin is omitted in this paper since we solely focus on the rational potentials on the line. The substitution z(x) defined via the first-order differential equation

$$z'(x)={}_1\wp^{-1/2}[z(x);{}_1T_K] \tag{2.4}$$

(with prime marking the derivative with respect to x) thus converts RCSLE (2.1) into the 1D Schrödinger equation on the real axis: $-\infty< x <+\infty$. (The TP ${}_1T_K[z]$ is always chosen to be positive at the second singular point z = 1, with the Darboux/Pöschl-Teller (D/PT) potential [14, 15] as the only exception.)

The RLP obtained by means of the transformation z(x) has the form [9]

$$V[z(x)\,|\,{}_1\boldsymbol{G}^{K\Im}] = -[z'(x)]^2\,{}_1I^o[z(x);{}_1Q^o] - \tfrac{1}{2}\{z,x\}, \tag{2.5}$$

where {z, x} stands for so-called [16] ‘Schwarzian derivative’

$$\{z,x\} \equiv (z''/z')' - \tfrac{1}{2}(z''/z')^2. \tag{2.6}$$

We refer to the parameters ${}_1Q^o$ as ‘ray identifiers’ (RIs), with the term ‘ray’ used an alternative name for BI (2.2) evaluated at some fixed values of all the parameters excluding the energy ε which will be chosen to vary from $-\infty$ to 0.

The reference PFr (RefPFr) ${}_1I^o[z;{}_1Q^o]$ in the right-hand side of (2.2) has the form [8]

$$I^o[z;{}_1Q^o] \equiv -\frac{h_{o;0}}{4z^2(1-z)} - \frac{h_{o;1}}{4z(1-z)^2} + \frac{f_o}{4z(1-z)}. \tag{2.7}$$

Note that the last term in the right-hand side of (2.5) dominates at large z so $h_{o;0}$ , $h_{o;1}$, and $f_o$ are nothing but minus residues of the second-order poles at 0, 1, and ∞.

By parametrizing the TP in question as

$${}_1T_K[z] = {}_1a_2 z(z-1) + {}_1c_0(1-z) + {}_1c_1 z \qquad ({}_1a_2 \neq 0 \text{ for } K=2) \tag{2.8}$$

the author [8] obtained the following expression:

$$\{z,x\} = -\frac{2}{{}_1T_K[z]} - \left\{ {}_1a_2 - \frac{{}_1a_2 + ({}_1c_1 - {}_1c_0)(2z-1)]}{z(1-z)} - \tfrac{5}{4}\frac{\Delta_T}{{}_1T_K[z]} \right\} \frac{2z^2(1-z)^2}{{}_1T_K^2[z]}, \tag{2.9}$$

for the Schwarzian derivative in the right-hand side of (2.5), where $\Delta_T$ denotes the TP discriminant. One can easily verify that the ChExps of Fuschian solutions near the singular point z =1 satisfy the indicial equation

$$\rho_1(\varepsilon \mid {}_1\mathcal{G}^{K\Im}, \pm))[\rho_1(\varepsilon \mid {}_1\mathcal{G}^{K\Im}, \pm) - 1] - \tfrac{1}{4}(h_{o;1} - {}_1c_1\,\varepsilon) \tag{2.10}$$

which has real roots

$$\rho_1(\varepsilon \mid {}_1\mathcal{G}^{K\Im}, \pm) = \tfrac{1}{2}[1 \pm \sqrt{h_{o;1} - {}_1c_1\,\varepsilon} \tag{2.11}$$

at energies

$${}_1c_1\,\varepsilon \leq h_{o;1} + 1. \tag{2.12}$$

Making the transformation $z \to 1 - z$ if necessary, one can also choose:

$$h_{o;0} \geq h_{o;1} \tag{2.13}$$

and thereby setting $h_{o;1}$ to $-1$ assures that indicial equation (2.3) for the singular point 1 has two different real roots at any negative value of the energy parameter $\varepsilon$. In this paper we only consider the fragment of the *r*-GRef potential $V[z \mid {}_1\mathcal{G}^{K\Im 0}]$ restricted by the constraint

$$f_o \geq -1 \tag{2.14}$$

so each of the indicial equations for two other regular singular points z = 0 and ∞ have two different real roots for any negative value of the energy parameter $\varepsilon$. As far as conditions (2.13), and (2.14) hold the zero energy ExpDiffs for these two singular points

$$\lambda_o \equiv \sqrt{h_{o;0} + 1} \tag{2.15}$$

and

$$\mu_o \equiv \sqrt{f_o + 1}\,, \tag{2.15*}$$

are necessarily real and therefore can be used as two alternative (nonnegative) RIs. Setting

$$\lambda(\varepsilon; \nu) = \sqrt{\nu^2 - \varepsilon}, \tag{2.16}$$

one can verify that energy-dependent ExpDiffs $\lambda({}_1c_0\varepsilon;\lambda_o)$ and $\lambda({}_1a_2\varepsilon;\mu_o)$ for the singular points 0 and ∞ of the given RCSLE remain real at any negative value of the energy parameter $\varepsilon$.

In following our original studies [17] on the Darboux transformations of the generic centrifugal-barrier potential (years before the birth of the SUSY quantum mechanics [18, 19]), we label the four possible types of $\mathcal{J}$S solutions as

$$\mathbf{t}=\mathbf{a} \text{ for } {}_1\lambda_{0;\mathbf{t}m} > 0,\ {}_1\lambda_{1;\mathbf{t}m} < 0, \tag{2.17a}$$

$$\mathbf{t}=\mathbf{b} \text{ for } {}_1\lambda_{0;\mathbf{t}m} < 0,\ {}_1\lambda_{1;\mathbf{t}m} > 0, \tag{2.17b}$$

$$\mathbf{t}=\mathbf{c} \text{ for } {}_1\lambda_{0;\mathbf{t}m} > 0,\ {}_1\lambda_{1;\mathbf{t}m} > 0, \tag{2.17c}$$

$$\mathbf{t}=\mathbf{d} \text{ for } {}_1\lambda_{0;\mathbf{t}m} < 0,\ {}_1\lambda_{1;\mathbf{t}m} < 0 \tag{2.17d}$$

($T_3$, $T_4$, $T_1$, and $T_2$ in Sukumar's terms [20]).

Let

$$\phi_{\mathbf{t}m}[z\,|\,{}_1\boldsymbol{G}^{K\Im 0}_{\downarrow\mathbf{t}m}] = z^{\frac{1}{2}({}_1\lambda_{0;\mathbf{t}m}+1)}\,|1-z|^{\frac{1}{2}({}_1\lambda_{1;\mathbf{t}m}+1)} \times P_m^{({}_1\lambda_{1;\mathbf{t}m},\,{}_1\lambda_{0;\mathbf{t}m})}(2z-1) \tag{2.18}$$

$$= \frac{\Gamma({}_1\lambda_{0;\mathbf{t}m} + {}_1\lambda_{1;\mathbf{t}m} + 2n + 1)}{\Gamma({}_1\lambda_{0;\mathbf{t}m} + {}_1\lambda_{1;\mathbf{t}m} + n + 1)}\, z^{\rho_{0;\mathbf{t}m}}\,|1-z|^{\rho_{1;\mathbf{t}m}}\,\Pi_m[z;\bar{z}_{\mathbf{t}m}] \tag{2.18*}$$

be an $\mathcal{J}$S solution of RCSLE (2.1), where the monomial product

$$\Pi_n[z;\bar{e}] \equiv \prod_{k=1}^{n}(z-e_k) \tag{2.19}$$

is nothing but a generic substitute for scaled hypergeometric polynomials in z (or alternatively scaled Jacobi polynomials in 2z–1) keeping in mind that the latter may have only simple zeros at any point other than 0 and 1 [1]. (Note that there is a misprint in the second Jacobi index of the Jacobi polynomial in the right-hand side of (21) in the English translation of [8].)

An analysis of asymptotics of $\mathcal{J}$S solution (2.18) at large z shows that the signed ExpDiffs

$$ {}_1\lambda_{r;\dagger m} \equiv 2\,{}_1\rho_{r;\dagger m} - 1 \qquad (2.20^{\dagger})$$

$$ = \sigma_{r;\dagger}\sqrt{h_{o;r} + 1 - {}_1c_r\,{}_1\varepsilon_{\dagger m}} \quad (r = 0 \text{ or } 1) \qquad (2.20)$$

must satisfy the following necessary condition:

$$ {}_1\lambda_{0;\dagger m} + {}_1\lambda_{1;\dagger m} + 2m + 1 = \sigma_{\infty;\dagger m}\sqrt{\mu_o^2 - {}_1a_2\,{}_1\varepsilon_{\dagger m}} \qquad (2.21)$$

$$ \equiv \mu_{\dagger m}, \qquad (2.21^{\dagger})$$

where $\mu_{\dagger m}$ is nothing but the signed ExpDiff at $\infty$. One can easily extend the arguments presented in [1] for eigenfunctions to confirm that this is also the sufficient condition for the hypergeometric series to be truncated into a polynomial.

Taking into account that the Schwarzian derivative (see [21] for a convenient list of related formulas) is invariant with respect to scaling of the variable z, one can easily verify that

$$ \lim_{z \to r} \{z, x\} = -2 / {}_1c_r \qquad (r = 0 \text{ or } 1) \qquad (2.22)$$

so condition ($2.12^{\dagger}$) is equivalent to the requirement that the *r*-GRef potential vanishes as $x \to +\infty$, whereas inequality (2.12) assures that the potential has a reflection barrier at large negative x, excluding the limiting case of the *r*-AL potential curves

$$ h_0^o = h_1^o = -1 \quad (\lambda_o = 0). \qquad (2.23)$$

Substituting

$$ {}_1\varepsilon_{\dagger m} = -\,{}_1c_1\,{}_1\lambda_{1;\dagger m}^2 \qquad (2.24)$$

into (2.20) for $r = 0$ one finds

$$ {}_1\lambda^2_{0;\dagger m} = \lambda^2_o + {}_1c_0\,{}_1\lambda^2_{1;\dagger m} \tag{2.25}$$

so radical equation (2.21) can be represented as the polynomial equation

$$ \Lambda^{(m)}_2({}_1\lambda_{1;\dagger m};\lambda_o,\mu_o;{}_1d) = 2\,{}_1\lambda_{0;\dagger m}({}_1\lambda_{1;\dagger m} + 2m + 1) \tag{2.26}$$

linear in ${}_1\lambda_{0;\dagger m}$ and quadratic in ${}_1\lambda_{1;\dagger m}$, where the quadratic polynomial in ${}_1\lambda_{1;\dagger m}$ is defined via the identity

$$ \Lambda^{(m)}_2(\lambda;\lambda_o,\mu_o;{}_1d) \equiv (1+{}_1d)\lambda^2 - \left(\lambda + 2m + 1\right)^2 + \mu^2_o - \lambda^2_o. \tag{2.27}$$

As demonstrated below the parameter

$$ {}_1d \equiv {}_1a_2 - {}_1c_0 - 1 \tag{2.27'}$$

associated with the alternative parametrization of the second-order TP:

$$ T_2[z] = {}_1d\,z(z-1) + {}_1c_0(1-z)^2 + {}_1c_1z^2 \tag{2.28}$$

naturally appears in the formulas for threshold points of the generic double-step algorithm. It must be smaller than $-2\sqrt{{}_1c_0}$ for TP (2.28) to have positive discriminant $\Delta_T$. (The region $-2\sqrt{{}_1c_0} < {}_1d < +2\sqrt{{}_1c_0}$ where the TP has a pair of complex conjugated roots has some very distinctive features which will be discussed in a separate publication. The border case of the TP with a double root (DRt) with ${}_1d = -2\sqrt{{}_1c_0}$ has been examined in detail in [10].)

Equation (2.26) needs to be solved together with polynomial equation (2.25) quadratic in both ${}_1\lambda_{0;\dagger m}$ and ${}_1\lambda_{1;\dagger m}$. Taking square of both sides of (2.26) and substituting (2.25) into the right-hand side of the resultant expression, one finds that the signed ExpDiff ${}_1\lambda_{1;\dagger m}$ at the upper end z = 1 must coincide with a real root of the quartic polynomial:

$$G_4^{1;m}[\lambda \mid {}_1\boldsymbol{G}^{K\Im}] \equiv G_4^{1;m}[\lambda; \lambda_o, \mu_o; {}_1T_2] \tag{2.29a}$$

$$= [{}_1d\lambda^2 - 2(2m+1)\lambda + \mu_o^2 - \lambda_o^2 - (2m+1)^2]^2 - 4(\lambda + 2m + 1)^2({}_1c_0\,\lambda^2 + \lambda_o^2).$$

It directly follows from (2.29a) that the quartic equation of our interest is reduced to the two quadratic equations in the limiting case of the radial [22-24] GRef potentials (${}_1c_0 = 0$) as well as for AL *r*-GRef potential curves $V[z \mid {}_1\breve{\boldsymbol{G}}^{K\Im}]$ obtained by setting $\lambda_0$ to 0 so

$$V[z(-\infty) \mid {}_1\breve{\boldsymbol{G}}^{K\Im}] = V[z(+\infty) \mid {}_1\breve{\boldsymbol{G}}^{K\Im}]$$

(see next Section for details). As explained in Section 4 the AL potential curves $V[z \mid {}_1\breve{\boldsymbol{G}}^{K\Im}]$ play a crucial role in our analysis of $\oint$S solutions for a broad family of *r*-GRef potentials on the line. Namely, we will be able to make some quantitative predictions concerning $\oint$S solutions $\dagger$m coexistent with the m$^{th}$ bound energy level provided that the TP used to construct the *r*-GRef potential in question has nonnegative discriminant ${}_1\Delta_T$.

Quartic polynomial (2.29a) can be also represented in the following alternative form:

$$G_4^{1;m}[\lambda; \lambda_o, \mu_o; {}_1T_2] = [({}_1d+1)\lambda^2 + \mu_o^2 - \lambda_o^2 + (\lambda + 2m + 1)^2]^2 - 4(\lambda + 2m + 1)^2({}_1a_2\lambda^2 + \mu_o^2), \tag{2.29b}$$

which can be analytically decomposed into the product of second-order polynomials at ${}_1a_2 = 0$. Some remarkable features of PFrBs generated by means of the LTP will be explored Section 5.

One can also convert the set of radical equations (2.25) and (2.26) into the quartic equation with respect to $\mid {}_1\varepsilon_{\dagger m} \mid$:

$$(G_{4;4}^{1;m} \mid {}_1\varepsilon_{\dagger m} \mid^2 + G_{4;2}^{1;m} \mid {}_1\varepsilon_{\dagger m} \mid + G_{4;0}^{1;m})^2 = \mid {}_1\varepsilon_{\dagger m} \mid (G_{4;3}^{1;m} \mid {}_1\varepsilon_{\dagger m} \mid + G_{4;1}^{1;m})^2. \tag{2.29$\dagger$}$$

In a slightly different form the latter quartic equation has been already suggested by Grosche [26] as the quantization condition for bound energy levels ($\dagger$ = **c**) in the generic *r*-GRef potential.

It should be however stressed that use of Grosche's quartic polynomial to determine energies of bound states the discrete energy spectrum is complicated by the fact that quartic polynomial ($2.29^{\dagger}$) has generally 4 positive roots and selection of the one associated with a bound energy level is by no means a trivial problem.

Note that the leading coefficient of the quartic polynomial has the same sign as the TP discriminant and in particular vanishes if the TP has a double root [10].

For each real root ${}_1\lambda_{1;\mathbf{t}m}$ of quartic polynomial (2.29a) the signed ExpDiff ${}_1\lambda_{0;\mathbf{t}m}$ at the origin is unambiguously determined by the fraction formula

$$ {}_1\lambda_{0;\mathbf{t}m} = \frac{\mu_o^2 - \lambda_o^2 - (2m+1)^2 + {}_1d\,{}_1\lambda_{1;\mathbf{t}m}^2 - 2(2m+1)\,{}_1\lambda_{1;\mathbf{t}m}}{2({}_1\lambda_{1;\mathbf{t}m} + 2m + 1)}, \tag{2.30} $$

unless the denominator of the fraction in the right-hand side of (2.30) vanishes

$$ {}_1\lambda_{1;\mathbf{t}m} = -2m - 1. \tag{2.31} $$

The latter condition holds iff two m-dependent parameters

$$ \mu_m \equiv \sqrt{\mu_o^2 + {}_1a_2(2m+1)^2} \tag{2.32a} $$

and

$$ \lambda_m \equiv \sqrt{\lambda_o^2 + {}_1c_0\,(2m+1)^2} \tag{2.32b} $$

coincide. It then directly follows from the structure of quartic polynomial (2.29a) that the latter has the double root $-2m-1$ at each point of the curve

$$ \mu_o = {}_{\mathbf{ad}}\mu_{o;m}(\lambda_o; {}_1d) \equiv \sqrt{\lambda_o^2 - (1 + {}_1d)(2m+1)^2} \tag{2.33} $$

in the $\lambda_o\mu_o$ plane. Double index **ad** in the right-hand side of (2.33) indicates that a pair of the $\mathcal{S}$S solutions co-existent at the same energy have types **a** and **d** at any point of the curve (see Appendix C for more details).] We thus infer that the number of $\mathcal{S}$S solutions **t**m for the given order m of the seed polynomial $\Pi_m[z;\bar{z}_{\dagger m}]$ may not exceed 4 despite the mentioned ambiguity along **ad**-DRt hyperbola (2.33).

The cutoff represented by the **ad**-DRt hyperbola does not allow one to analytically continue the regular $\mathfrak{J}$S solution of type **a** from one side of the curve to another using fraction formula (2.30). One can however use holomorphic representation (2.18) for this solution by using real roots of the partner quartic polynomial

$$G_4^{0;m}[\lambda \,|\, {}_1\boldsymbol{G}^{K\Im 0}] \equiv G_4^{0;m}[\lambda; \lambda_o, \mu_o; {}_1T_2] \tag{2.34}$$

$$= {}_1c_0[(\lambda+2m+1)^2 - \mu_o^2 + (1-{}_1a_2)(\lambda^2-\lambda_o^2)/{}_1c_0]^2 - 4(\lambda+2m+1)^2(\lambda^2-\lambda_o^2)$$

to determine signed ExpDiffs (2.30). For each real root ${}_1\lambda_{0;\dagger m}$ (including the two with the same absolute value along the **ad**-DRt hyperbola) the signed ExpDiff ${}_1\lambda_{1;\dagger m}$ is then unambiguously determined by the fraction formula

$${}_1\lambda_{1;\dagger m} = \frac{\mu_o^2 + ({}_1a_2-1)({}_1\lambda_{0;\dagger m}^2 - \lambda_o^2)/{}_1c_0 - ({}_1\lambda_{0;\dagger m} + 2m+1)^2}{2({}_1\lambda_{0;\dagger m} + 2m+1)} \tag{2.35}$$

unless the denominator of the fraction in the right-hand side of (2.35) vanishes

$${}_1\lambda_{0;\dagger m} = -2m-1. \tag{2.36}$$

It then directly follows from the structure of quartic polynomial (2.34) that the latter has the double root −2m−1 at each point of the DRt curve

$$\mu_o = {}_{\mathbf{bd}}\mu_{o;m}(\lambda_o; {}_1T_2) \equiv \sqrt{(1-{}_1a_2)[(2m+1)^2 - \lambda_o^2]/{}_1c_0} \tag{2.37}$$

in the $\mu_o\lambda_o$ plane.

Since each of quartic polynomials (2.29a) and (2.34) has the known DRt along curves (2.33) and (2.37) one can analytically decompose each polynomial into the product of quadratic polynomials at any point of the given curve. A more detailed analysis of the signed ExpDiffs ${}_1\lambda_{1;\mathbf{t}\mathrm{m}}$ and ${}_1\lambda_{0;\mathbf{t}\mathrm{m}}$ along these curves will be presented in Part III.

Along the threshold curves

$$\mu_{\mathrm{o}} = \mu_{\Diamond;\mathrm{m},\mathrm{m}_*}(\lambda_{\mathrm{o}};{}_1T_2) \tag{2.38}$$

such that

$$\lim_{\mu_{\mathrm{o}} \to \mu_{\Diamond;\mathrm{m},\mathrm{m}_*}(\lambda_{\mathrm{o}};{}_1T_2)} {}_1\lambda_{r;\mathbf{t}\mathrm{m}} = -\mathrm{m}_* \quad (r = 0 \text{ or } 1) \tag{2.38*}$$

for any positive integer $\mathrm{m}_* < \mathrm{m}$ the $\mathcal{J}$S solution $\mathbf{t}\mathrm{m}$ turns into the solution $\mathbf{t}_\Diamond\,\mathrm{m}_\Diamond$ of a different type $\mathbf{t}_\Diamond$, namely,

$$\mathbf{t}_\Diamond = \begin{cases} \mathbf{c} \text{ if } \mathbf{t} = \mathbf{a} \text{ or } \mathbf{b}, \\ \mathbf{a} \text{ if } \mathbf{t} = \mathbf{d} \text{ and } r = 0, \\ \mathbf{b} \text{ if } \mathbf{t} = \mathbf{d} \text{ and } r = 1, \end{cases} \tag{2.38$_\Diamond$}$$

because, as a direct consequence of (22.3.3) in [25], the Jacobi polynomial $P_{\mathrm{m}}^{(-\mathrm{m}_*,\nu)}(\eta)$ has zero of order $\mathrm{m}_*$ at $\eta = 1$ so $\mathrm{m}_\Diamond = \mathrm{m} - \mathrm{m}_*$. Along the threshold curve (and only there) the $\mathcal{J}$S solution $\mathbf{t}\mathrm{m}$ turns into the $\mathcal{J}$S solution $\mathbf{t}_\Diamond\mathrm{m}_\Diamond$ with the ChExp

$${}_1\rho_{r;\mathbf{t}_\Diamond\mathrm{m}_\Diamond} = \tfrac{1}{2}(1+\mathrm{m}) = 1 - {}_1\rho_{r;\mathbf{t}\mathrm{m}}\Big|_{\mu_{\mathrm{o}} \to \mu_{\Diamond;\mathrm{m},\mathrm{m}_*}(\lambda_{\mathrm{o}};{}_1T_2)} \tag{2.38$^\dagger$}$$

at the end-point z = $r$.

The necessary and sufficient condition for a regular $\mathcal{J}$S solution not to have zeros inside the quantization interval is equivalent to the so-called 'indexes-of-opposite-sign' (IOS) rule [24]:

*Any Jacobi polynomial with indexes of opposite sign does not have zeros between* –1 *and* +1 *iff its order is smaller than absolute value of the negative index.*

(Note that Quesne [13] splits this rule into the two depending which index happened to be negative. However, since the indexes can be always interchanged by the reflection of the polynomial argument so we really deal with a single rule.)

By solving the equation

$$G_4^{1;m}[-m;\lambda_o,\mu_o;{}_1T_2]=0 \tag{2.39$_1$}$$

for ℘S solutions $\breve{\mathbf{t}}_m$ of type **a** or

$$G_4^{0;m}[-m;\lambda_o,\mu_o;{}_1T_2]=0 \tag{2.39$_0$}$$

for a ℘S solution $\breve{\mathbf{t}}_m$ of type **b** under the constraint

$${}_1\lambda_{0;\breve{\mathbf{t}}_m}=\lambda_{\Diamond\mathbf{a};m}(\lambda_o;{}_1c_0)\equiv\sqrt{{}_1c_0\,m^2+\lambda_o^2}>0 \tag{2.39a}$$

or correspondingly

$${}_1\lambda_{1;\breve{\mathbf{t}}_m}=\lambda_{\Diamond\mathbf{b};m}(\lambda_o;{}_1c_0)\equiv\sqrt{(m^2-\lambda_o^2)/{}_1c_0}>0, \tag{2.39b}$$

one finds that there are two threshold curves

$$\mu_o=\mu_{\Diamond m;\mathbf{a}}(\lambda_o;{}_1T_2)\equiv\sqrt{-{}_1dm^2+2m+1+\lambda_o^2+2(m+1)\sqrt{{}_1c_0\,m^2+\lambda_o^2}} \tag{2.39a$^\dagger$}$$

and

$$\mu_o=\mu_{\Diamond m;\mathbf{b}}(\lambda_o;{}_1T_2)\equiv\sqrt{[\sqrt{(m^2-\lambda_o^2)/{}_1c_0}+m+1]^2-{}_1a_2(m^2-\lambda_o^2)/{}_1c_0} \tag{2.39b$^\dagger$}$$

where one of the regular ℘S solution $\breve{\mathbf{t}}_m$ turns into the ground-state eigenfunction ($m_*=m$, $m_\Diamond=0$).

In the vicinity of the threshold curves the roots of our interest can be approximated as

$$2({}_1\lambda_{1;\breve{t}_m}+m)\approx\frac{\mu_o^2-\mu^2_{\Diamond\mathbf{a};m}(\lambda_o;{}_1T_2)}{({}_1d+2)m+1+(\lambda_o^2-{}_1c_0 m)/\lambda_{\Diamond\mathbf{a};m}(\lambda_o;{}_1c_0)} \qquad (2.39a^*)$$

and

$$2({}_1\lambda_{0;\breve{t}_m}+m)/{}_1c_0$$

$$\approx\frac{\mu_o^2-\mu^2_{\Diamond m;\mathbf{b}}(\lambda_o;{}_1T_2)}{({}_1d+2\,{}_1c_0)m+{}_1c_0-(m+1)m/\lambda_{\Diamond\mathbf{b};m}(\lambda_o;{}_1c_0)+\lambda_{\Diamond\mathbf{b};m}(\lambda_o;{}_1c_0)}, \qquad (2.39b^*)$$

accordingly. Depending on sign of the denominator of the fraction in the right-hand side of (2.39a*) or (2.39b*) the regular ℱS solution $\breve{t}_m$ in question is nodeless either above or below the corresponding threshold curve. On the contrary, it acquires at least one node on other side of the curve and as a result the potential constructed using this solution as the FF is no more quantized by polynomials beyond the region curved by the threshold curve.

An analysis of the free term,

$${}_1G^{1;m}_{4;0}(\lambda_o,\mu_o)=[\mu_o^2-(\lambda_o-2m-1)^2]\times[\mu_o^2-(\lambda_o+2m+1)^2], \qquad (2.40)$$

of quartic polynomial (2.29a) shows that it vanishes along the straight-lines

$$\mu_o=\lambda_o+2m+1, \qquad (2.41a)$$

$$\mu_o+\lambda_o=2m+1, \qquad (2.41b)$$

$$\mu_o+2m+1=\lambda_o. \qquad (2.41c)$$

referred to as the zero-factorization-energy (ZFE) separatrices since one of ℱS solutions has zero energy at any point of each straight-line.

As illustrated by Fig. 2.1 for m = 2 the ZFE) separatrices split the $\lambda_o\ \mu_o$ plane into four major Areas such that each ZFE separatrix (2.41a), (2.41b), or (2.41c) serves as the border line between this central region (Area $D_m$) and Areas $A_m$, $B_m$, or $C_m$, respectively, namely,

Area $A_m$: $\lambda_o < \mu_o - 2m - 1$; (2.42A)

Area $B_m$: $\mu_o + \lambda_o < 2m + 1$; (2.42B)

Area $C_m$: $\lambda_o > \mu_o + 2m + 1$; (2.42C)

Area $D_m$: otherwise. (2.42D)

As proven in Appendix A the eigenfunction associated with the highest bound energy level is the first one to change its type so there are two ℘S of type **a** on the D-side of straight line (2.42A), with $m = {}_1n_o - 1$. We refer to straight lines (2.42A) as the zero-factorization-energy (ZFE) **c**/**a**′-separatrices where prime indicates that we deal with ∮S solutions from the supplementary sequence which does not start from the basic solution **a**0. The analysis of regular ∮S solutions outside Area $A_m$ is more cumbersome. Let us point only to one general presumption concerning ∮S solutions of type **b** in Area $D_m$, namely, any ∮S solution of type **b** (if exists) is nodeless in the region of Area $D_m$ below the straight line $\mu_o = m+1$ in the $\lambda_0\mu_o$ plane. The cited presumption directly follows from the IOS rule based on the inequality

$$|\,{}_1\lambda_{0;\mathbf{b}m}\,| > \lambda_o > 2m+1-\mu_o > m \tag{2.43}$$

which holds at any point outside Area $B_m$ for $\mu_0 < m+1$. In Section 5 we independently confirm this conclusion for the LTP both for the primary ∮S solution **b**m and for a pair of secondary ∮S solutions of type **b** (**b**'m and **b**"m) which exist in Area $D_m$ if ${}_1c_0 > 1$.

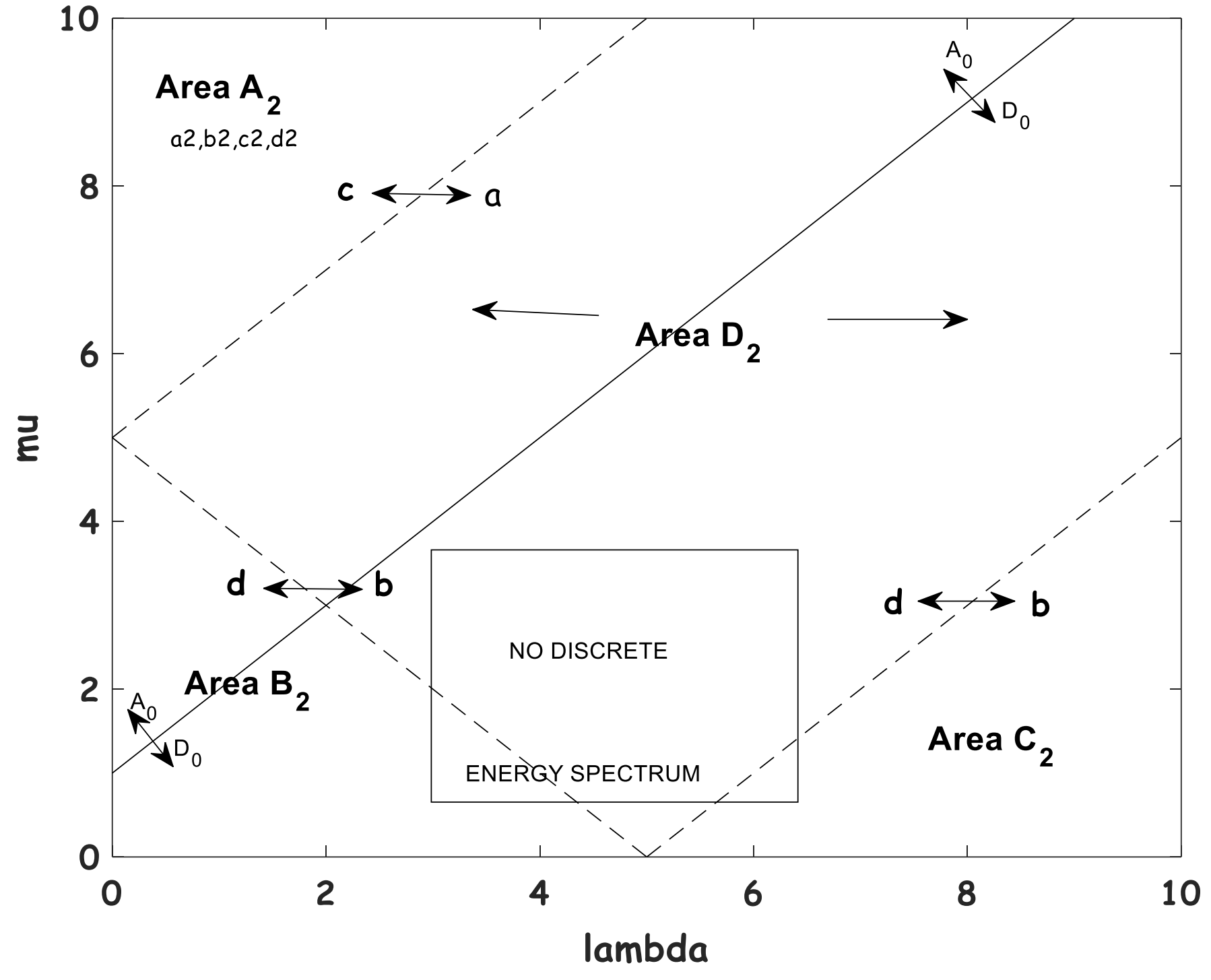

**Figure 2.1 Four major Areas carved by ZFE separatrices (dashed straight-lines) in the plane $\lambda_o\mu_o$ for the TP with positive discriminant**

## 3. Algebraic formulas for energies of ∮S solutions in the limiting case of the AL potential curves

The AL potential curves in question are obtained from the generic *r*-GRef potential $V[z\,|\,{}_1\boldsymbol{G}^{220}]$ by setting the ray identifier (RI) $\lambda_0$ to 0, i.e.,

$$V[z\,|\,{}_1\breve{\boldsymbol{G}}^{220}] = -(z')^2\, {}_1\breve{I}^o[z;\mu_o] + \tfrac{1}{2}\{z,x\}, \tag{3.1}$$

where

$${}_1\breve{I}^o[z;\mu_o] = \frac{1}{4z^2} + \frac{1}{4(1-z)^2} + \frac{\mu_o^2+1}{4z(1-z)} \tag{3.2}$$

and an explicit expression of the Schwarzian derivative in terms of the TP coefficients is given by general formula (2.3).

In the limiting case of the AL potential curves ($\lambda_o$ = 0) quadratic formula (2.22) turns into the linear relations

$$ {}_1\breve{\lambda}_{0;\breve{t}_\pm m} = \pm\sqrt{{}_1c_0}\,{}_1\breve{\lambda}_{1;\breve{t}_\pm m}, \qquad (3.3^{\pm}) $$

where

$$ \breve{\mathbf{t}}_+ = \mathbf{c} \text{ or } \mathbf{d} \qquad (3.4^{+}) $$

and

$$ \breve{\mathbf{t}}_- = \mathbf{a} \text{ or } \mathbf{b} \qquad (3.4^{-}) $$

Setting $\lambda_o$ to 0 in the right-and side of (2.29a) allows one to analytically decompose this quartic polynomial into the product of two quadratic polynomials

$$ {}^{\pm}\breve{G}_2^{(m)}[\lambda;\mu_o;{}_1T_2] = {}^{\pm}_1\breve{g}_2({}_1T_2)\lambda^2 + {}^{\pm}_1\breve{g}_1^{(m)}({}_1c_0)\lambda + {}_1\breve{g}_0^{(m)}(\mu_o) \qquad (3.5^{\pm}) $$

so the signed ExpDiffs ${}_1\breve{\lambda}_{1;\breve{t}_\pm m}$ and ${}_1\breve{\lambda}_{0;\breve{t}_\pm m}$ coincide with roots of the quadratic equations:

$$ {}^{\pm}_1\breve{g}_2({}_1d,{}_1c_0)\,{}_1\breve{\lambda}^2_{1;\breve{t}_\pm m} + {}^{\pm}_1\breve{g}_1^{(m)}({}_1c_0)\,{}_1\breve{\lambda}_{1;\breve{t}_\pm m} + {}_1\breve{g}_0^{(m)}(\mu_o) = 0, \qquad (3.6^{\pm}_1) $$

and

$$ {}^{\pm}_1\breve{g}_2({}_1d/{}_1c_0,{}_1c_0^{-1})\,{}_1\breve{\lambda}^2_{0;\breve{t}_\pm m} + {}^{\pm}_1\breve{g}_1^{(m)}({}_1c_0^{-1})\,{}_1\breve{\lambda}_{0;\breve{t}_\pm m} + {}_1\breve{g}_0^{(m)}(\mu_o) = 0, \qquad (3.6^{\pm}_0) $$

where

$$ {}^{\pm}_1\breve{g}_2({}_1d,{}_1c_0) = \pm 2\sqrt{{}_1c_0} - {}_1d = (1 \pm \sqrt{{}_1c_0})^2 - {}_1a_2, \qquad (3.7^{\pm}_2) $$

$$ {}^{\pm}_1\breve{g}_1^{(m)}({}_1c_0) = 2(\pm\sqrt{{}_1c_0} + 1)(2m+1), \qquad (3.7^{\pm}_1) $$

and

$$ {}_1\breve{g}_0^{(m)}(\mu_o) = g_m(\mu_o) \equiv (2m+1)^2 - \mu_o^2. \qquad (3.7^{\pm}_0) $$

In this paper we only consider TPs with positive discriminant ${}_1\Delta_T$ (sending the reader to [10] for a survey of nodeless $\mathfrak{J}$S solutions in the limiting case of the TP with a double root). This implies that leading coefficients (3.$7^{\pm}_2$) of quadratic equations (3.$6^{\pm}_1$) are positive so each equation has roots of opposite sign for $\mu_o > 2m+1$. We [1] refer to a sequence of $\mathfrak{J}$S solutions of the given type **t** as 'primary' if it starts from the basic solution **t**0. We thus conclude that the CRSLE with the Bose invariant $I^o[z;\varepsilon\,|\,{}_1\breve{\boldsymbol{G}}^{220}]$ has exactly four primary sequences **t**m (m = 0, 1,...) of $\mathfrak{J}$S solutions of four distinct types **t**=**a**, **b**, **c**, and **d**. In particular the positive root of quadratic polynomial (3.$6^{+}_1$),

$$ {}_1\breve{\lambda}_{1;\mathbf{c}m} = \frac{\sqrt{{}_1\breve{\Delta}^{+}_{m}(\mu_o;{}_1T_2)} - 2(\sqrt{{}_1c_0}+1)(2m+1)}{2\,{}^{+}_{1}\breve{g}_2({}_1T_2)} > 0, \tag{3.8c} $$

identifies the m[th] eigenfunction whereas its negative counter-part,

$$ {}_1\breve{\lambda}_{1;\mathbf{d}m} = -\frac{2(\sqrt{{}_1c_0}+1)(2m+1) + \sqrt{{}_1\breve{\Delta}^{+}_{m}(\mu_o;{}_1T_2)}}{2\,{}^{+}_{1}\breve{g}_2({}_1T_2)}, \tag{3.8d} $$

specifies the $\mathfrak{J}$S solution of type **d**, with

$$ {}_1\breve{\Delta}^{\pm}_{m}(\mu_o;{}_1T_2) \equiv 4[{}_1a_2(2m+1)^2 + {}^{\pm}_{1}\breve{g}_2({}_1T_2)\mu_o^2] \tag{3.9$^{\pm}$} $$

standing for discriminant of polynomials (3.$6^{\pm}_1$). Since the linear coefficient ${}^{+}_{1}\breve{g}^{(0)}_{1}({}_1c_0)$ is positive the signed ExpDiffs

$$ {}_1\breve{\lambda}_{1;\mathbf{c}0} = \frac{\tfrac{1}{2}\sqrt{{}_1\breve{\Delta}^{+}_{0}(\mu_o;{}_1T_2)} - 2\sqrt{{}_1c_0} - 1}{{}^{+}_{1}\breve{g}_2({}_1T_2)} > 0, \tag{3.10c} $$

and

$$\breve{\lambda}_{1;\mathbf{d}0} = -\frac{\sqrt{{}_1c_0} + 1 + \frac{1}{2}\sqrt{{}_1\breve{\Delta}_0^+(\mu_o; {}_1T_2)}}{{}_1^+\breve{g}_2({}_1T_2)} < 0 \tag{3.10d}$$

satisfy the inequality $|\breve{\lambda}_{1;\mathbf{d}0}| > {}_1\breve{\lambda}_{1;\mathbf{c}0}$ which assures that the basic (necessarily nodeless) solution $\mathbf{d}0$ lies below the ground energy level ${}_1\breve{\varepsilon}_{\mathbf{c}0} = -{}_1\breve{\lambda}^2_{1;\mathbf{c}0}$ as expected.

If ${}_1a_2 > 0$ then discriminants of both quadratic equations remain positive for any m. Therefore the RCSLE has exactly four ℘S solutions for any AL *r*-GRef potential curves generated by means of the second-order TP with a positive leading coefficient.

For $2m > \mu_o^2 - 1$ the common free term of quadratic equations (3.6$^{\pm}_{1}$) becomes positive so the quadratic equation for the signed ExpDiffs ${}_1\breve{\lambda}_{1;\breve{\mathbf{t}}_m}$ has two negative (two positive) roots if ${}_1c_0 < 1$ (${}_1c_0 > 1$). This implies that the primary sequence $\mathbf{b}$m for ${}_1c_0 < 1$ or $\mathbf{a}$m for ${}_1c_0 > 1$ contains a finite number of ℘S solutions necessarily equal to the number of bound energy states

$${}_1n_o = [\tfrac{1}{2}(\mu_o - 1)] + 1. \tag{3.11}$$

We thus come to the following algebraic formulas for the signed ExpDiffs of the regular ℘S solutions in the $\lambda_o = 0$ limit:

$${}_1\breve{\lambda}_{1;\mathbf{a}m} = -\frac{\frac{1}{2}\sqrt{{}_1\breve{\Delta}_m^-(\mu_o; {}_1T_2)} + (1 - \sqrt{{}_1c_0})(2m+1)}{{}_1^-\breve{g}_2({}_1T_2)} < 0 \quad ({}_1c_0 < 1 \text{ or } {}_1c_0 > 1,\ m < {}_1n_o), \tag{3.12a}$$

$${}_1\breve{\lambda}_{1;\mathbf{b}m} = \frac{\frac{1}{2}\sqrt{{}_1\breve{\Delta}_m^-(\mu_o; {}_1T_2)} + (\sqrt{{}_1c_0} - 1)(2m+1)}{{}_1^-\breve{g}_2({}_1T_2)} > 0 \quad ({}_1c_0 > 1 \text{ or } {}_1c_0 < 1,\ m < {}_1n_o) \tag{3.12b}$$

and

$$ {}_1\breve{\lambda}_{1;\mathbf{a}'m} = \frac{\tfrac{1}{2}\sqrt{{}_1\breve{\Delta}^{-}_{m}(\mu_o;{}_1T_2)} - (1-\sqrt{{}_1c_0})(2m+1)}{{}^{-}_{1}\breve{g}_2({}_1T_2)} < 0 \quad ({}_1c_0 < 1,\ m \ge {}_1n_o), \tag{3.12a'} $$

$$ {}_1\breve{\lambda}_{1;\mathbf{b}'m} = \frac{(\sqrt{{}_1c_0}-1)(2m+1) - \tfrac{1}{2}\sqrt{{}_1\breve{\Delta}^{-}_{m}(\mu_o;{}_1T_2)}}{{}^{-}_{1}\breve{g}_2({}_1T_2)} > 0 \quad ({}_1c_0 > 1,\ m \ge {}_1n_o). \tag{3.12b'} $$

In particular the primary sequence **a**m for ${}_1c_0 <1$ or **b**m for ${}_1c_0 >1$ is infinite if ${}_1a_2 \ge 0$.

The sym-GRef potential curves (${}_1c_0 = {}_1c_1 = 1$, ${}_1a_2 < 0$) require separate consideration because the linear coefficient of the quadratic polynomial ${}^{-}\breve{G}_2^{(m)}[\lambda \mid {}_1\breve{\boldsymbol{G}}^{K\Im 0}]$ vanishes so real roots (if any) must have opposite sign. Since use of a regular solution as a FF for a DT of a symmetric potential breaks its symmetry under reflection of its argument we postpone any discussion of nodeless $\mathcal{J}$S solutions of this type for future publications.

Two regular solutions of the same type merge at

$$ \mu_o = \breve{\mu}_{\Delta,m-}({}_1a_2 < 0, {}_1c_0) \equiv \sqrt{\frac{-{}_1a_2}{{}^{-}_{1}\breve{g}_2({}_1T_2)}}(2m+1). \tag{3.13} $$

As $\mu_o$ gets smaller absolute value of signed ExpDiff of the primary (secondary) solution of type **a** for ${}_1c_0 < 1$ or **b** for ${}_1c_0 < 1$ monotonically decreases (increases) and threfore reaches its minimum (maximum), namely,

$$ {}_1\breve{\lambda}_{1;\mathbf{a}m} = {}_1\breve{\lambda}_{1;\mathbf{a}'m} = -\frac{(1-\sqrt{{}_1c_0})(2m+1)}{{}^{-}_{1}\breve{g}_2({}_1T_2)} < 0 \quad \text{for } {}_1c_0 < 1 \ (m \ge {}_1n_o) \tag{3.13a} $$

or

$$ {}_1\breve{\lambda}_{1;\mathbf{b}m} = {}_1\breve{\lambda}_{1;\mathbf{b}'m} = \frac{(\sqrt{{}_1c_0}-1)(2m+1)}{{}^{-}_{1}\breve{g}_2({}_1T_2)} \quad \text{for } {}_1c_0 > 1 \ (m \ge {}_1n_o), \tag{3.13b} $$

before the pair of regular $\mathcal{J}$S solutions of the same type disappears for $\mu_o < \breve{\mu}_{\Delta,m-}({}_1a_2 < 0, {}_1c_0)$.

As far as the discussion is solely restricted to AL potential curves one can always restrict the coefficient ${}_1c_0$ to the finite interval (0, 1) by changing z for 1–z. It will be however shown in next Section that there are two distinct branches of the *r*-GRef potential: $0 < {}_1c_0 < 1$ and ${}_1c_0 > 1$ for nonzero values of the reflective barrier. The two intersect along the AL potential curves. Therefore (keeping in mind this extension) it seems necessary to discuss both ranges of the coefficient ${}_1c_0$: $0 < {}_1c_0 < 1$ and ${}_1c_0 > 1$ in parallel.

We can now directly prove that the IOS rule:

$$|\,{}_1\breve{\lambda}_{1;\mathbf{\dagger}m_{\mathbf{\dagger}}}\,| > m_{\mathbf{\dagger}} \text{ for } \mathbf{\dagger} = \mathbf{a} \text{ or } \mathbf{a}' \tag{3.14a}$$

and

$$|\,{}_1\breve{\lambda}_{0;\mathbf{\dagger}m_{\mathbf{\dagger}}}\,| > m_{\mathbf{\dagger}} \text{ for } \mathbf{\dagger} = \mathbf{b} \text{ or } \mathbf{b}' \tag{3.14b}$$

is precisely equivalent to the requirement,

$$|\,\breve{\lambda}_{1;\breve{\mathbf{\dagger}}_m}\,| > {}_1\breve{\lambda}_{1;\mathbf{c}0}, \tag{3.15}$$

for the regular $\wp$S solution $\breve{\mathbf{\dagger}}$_ m to lie below the ground energy level (as far as the potential has the discrete energy spectrum). To prove this assertion let us first exclude $\mu_0^2$ from the quadratic equation for ${}_1\breve{\lambda}_{1;\breve{\mathbf{\dagger}}_m}$ by subtracting the partner equation for ${}_1\breve{\lambda}_{1;\mathbf{c}0}$. This gives

$$\begin{aligned} {}_1^{-}\breve{g}_2({}_1T_2)\,|\,{}_1\breve{\lambda}_{1;\mathbf{a}m}\,|^2 - 2(1-\sqrt{{}_1c_0})(2m+1)]\,|\,{}_1\breve{\lambda}_{1;\mathbf{a}m}\,| + (2m+1)^2 \\ = {}_1^{+}\breve{g}_2({}_1T_2)\,{}_1\breve{\lambda}^2_{1;\mathbf{c}0} + 2(\sqrt{{}_1c_0}+1)\,{}_1\breve{\lambda}_{1;\mathbf{c}0} + 1 \end{aligned} \tag{3.16a}$$

and

$$\begin{aligned} {}_1^{-}\breve{g}_2({}_1T_2)\,{}_1\breve{\lambda}^2_{1;\mathbf{b}m} + 2(1-\sqrt{{}_1c_0})(2m+1)]\,{}_1\breve{\lambda}_{1;\mathbf{b}m} + (2m+1)^2 \\ = {}_1^{+}\breve{g}_2({}_1T_2)\,{}_1\breve{\lambda}^2_{1;\mathbf{c}0} + 2(\sqrt{{}_1c_0}+1)\,{}_1\breve{\lambda}_{1;\mathbf{c}0} + 1. \end{aligned} \tag{3.16b}$$

Taking into account that

$$ {}^{+}_{1}\breve{g}_2({}_1T_2) - {}^{-}_{1}\breve{g}_2({}_1T_2) = 4\sqrt{{}_1c_0} \tag{3.17} $$

we can alternatively represent (3.16a) and (3.16b) as

$$ 4(|{}_1\breve{\lambda}_{1;\mathbf{a}m}| - m)(\sqrt{{}_1c_0}\,|{}_1\breve{\lambda}_{1;\mathbf{a}m}| + m + 1) \tag{3.18a} $$

$$ = (|{}_1\breve{\lambda}_{1;\mathbf{a}m}| - {}_1\breve{\lambda}_{1;\mathbf{c}0})[{}^{+}_{1}\breve{g}_2({}_1T_2)({}_1\breve{\lambda}_{1;\mathbf{c}0} + |{}_1\breve{\lambda}_{1;\mathbf{a}m}|) + 2(\sqrt{{}_1c_0} + 1)] $$

and

$$ 4(|{}_1\breve{\lambda}_{0;\mathbf{b}m}| - m)[|{}_1\breve{\lambda}_{0;\mathbf{b}m}| + \sqrt{{}_1c_0}\,(m+1)] \tag{3.18b} $$

$$ = ({}_1\breve{\lambda}_{1;\mathbf{b}m} - {}_1\breve{\lambda}_{1;\mathbf{c}0})[{}^{+}_{1}\breve{g}_2({}_1T_2)({}_1\lambda_{1;\mathbf{b}m} + {}_1\breve{\lambda}_{1;\mathbf{c}0}) + 2(\sqrt{{}_1c_0} + 1)], $$

in agreement with the assertion that any regular $\mathcal{I}$S solution satisfying the IOS rule must lie below the ground energy spectrum as far as the AL potential in question has the discrete energy spectrum. Examination of (3.18a) and (3.18b) also confirms that both basic $\mathcal{I}$S solutions $\mathbf{a}0$ and $\mathbf{b}0$ lie below the ground energy level as expected.

Note that the derived formulas remain valid for any regular $\mathcal{I}$S solution whether it belongs to a primary or secondary sequence, i.e., one can alternatively change $\mathbf{a}$ for $\mathbf{a}'$ and $\mathbf{b}$ for $\mathbf{b}'$ in (3.16a) and (3.16b), respectively.

We thus explicitly confirmed that any regular $\mathcal{I}$S solution $\breve{\mathbf{t}}_m$ lying below the ground energy level is necessarily nodeless. This is of course the well-known feature of *non-singular* Sturm-Liouville problems quantized on finite intervals [27]. In [1] we have presented a sketchy proof of this assertion for the 1D Schrödinger equation on the line assuming that the equation has a limit-point singularity at infinity. However we were unable to find an accurate corroboration for the reverse statement, namely, *that nodeless regular solutions may not exist above the ground energy level*. Below (mainly to simplify the discussion) we will silently assume that this is true at least for the RCSLEs of our interest though we will never rely on this conjecture in our arguments.

$$\bar{{}_1\breve{g}}_2({}_1d/{}_1c_0,{}_1c_0^{-1})({}_1\breve{\lambda}_{0;\mathbf{b}m}+m)^2+[\bar{{}_1\breve{g}}_1^{(m)}({}_1c_0^{-1})-2m\bar{{}_1\breve{g}}_2({}_1d/{}_1c_0,{}_1c_0^{-1})]({}_1\breve{\lambda}_{0;\mathbf{b}m}+m)$$

$$-\bar{{}_1\breve{g}}_2({}_1d/{}_1c_0,{}_1c_0^{-1})m^2-m[\bar{{}_1\breve{g}}_1^{(m)}({}_1c_0^{-1})-2m\bar{{}_1\breve{g}}_2({}_1d/{}_1c_0,{}_1c_0^{-1})]+{}_1\breve{g}_0^{(m)}(\mu_o)=0,$$

$$\bar{{}_1\breve{g}}_2({}_1d/{}_1c_0,{}_1c_0^{-1})({}_1\breve{\lambda}_{0;\mathbf{b}m}+m)^2+[\bar{{}_1\breve{g}}_1^{(m)}({}_1c_0^{-1})-2m\bar{{}_1\breve{g}}_2({}_1d/{}_1c_0,{}_1c_0^{-1})]({}_1\breve{\lambda}_{0;\mathbf{b}m}+m)$$

$$-m[\bar{{}_1\breve{g}}_1^{(m)}({}_1c_0^{-1})-m\bar{{}_1\breve{g}}_2({}_1d/{}_1c_0,{}_1c_0^{-1})]+{}_1\breve{g}_0^{(m)}(\mu_o)=0,$$

$$\bar{{}_1\breve{g}}_2({}_1d,{}_1c_0)({}_1\breve{\lambda}_{1;\mathbf{a}m}+m)^2+[\bar{{}_1\breve{g}}_1^{(m)}({}_1c_0)-2m\bar{{}_1\breve{g}}_2({}_1d,{}_1c_0)]({}_1\breve{\lambda}_{1;\mathbf{a}m}+m)$$

$$-m[\bar{{}_1\breve{g}}_1^{(m)}({}_1c_0)-m\bar{{}_1\breve{g}}_2({}_1d,{}_1c_0)]+{}_1\breve{g}_0^{(m)}(\mu_o)=0$$

Near the starting points of threshold curves (2.39a†) and (2.39b†),

$$\breve{\mu}_{\Diamond\mathbf{a};m}({}_1T_2)=\sqrt{(2\sqrt{{}_1c_0}-{}_1d)m^2+2(1+\sqrt{{}_1c_0})m+1} \qquad (3.19a)$$

$$\equiv\sqrt{(2m+1)^2-m[2(1-\sqrt{{}_1c_0})(m+1)+({}_1d+2)m]} \qquad (3.19a^*)$$

$$\equiv\sqrt{[(\sqrt{{}_1c_0}+1)m+1]^2-{}_1a_2m^2} \qquad (3.19a^\dagger)$$

or

$$\breve{\mu}_{\Diamond\mathbf{b};m}({}_1T_2)=\sqrt{(2/\sqrt{{}_1c_0}-{}_1d/{}_1c_0)m^2+2(1+1/\sqrt{{}_1c_0})m+1} \qquad (3.19b)$$

$$\equiv\sqrt{(2m+1)^2-m[2(1-1/\sqrt{{}_1c_0})(m+1)+({}_1d/{}_1c_0+2)m]} \qquad (3.19b^*)$$

$$\equiv\sqrt{[(1+1/\sqrt{{}_1c_0})m+1]^2-{}_1a_2m^2/{}_1c_0} \qquad (3.19b^\dagger)$$

equation (3.6$\bar{_1}$) for the root of the selected type can be approximated as

$$2({}_1\breve{\lambda}_{1;\breve{\mathbf{t}}_m}+m)\approx\frac{\mu_o^2-\breve{\mu}^2_{\Diamond m;\mathbf{a}}({}_1T_2)}{({}_1d+2)m+1-\sqrt{{}_1c_0}} \qquad (\breve{\mathbf{t}}_=\mathbf{a}\text{ or }\mathbf{a}') \qquad (3.19a^*)$$

and

$$2({}_1\breve{\lambda}_{0;\breve{\mathbf{t}}_m}+m)\approx\frac{\mu_o^2-\breve{\mu}^2_{\Diamond m;\mathbf{b}}({}_1T_2)}{({}_1d/{}_1c_0+2)m+1-1/\sqrt{{}_1c_0}}\quad(\breve{\mathbf{t}}_=\mathbf{b}\text{ or }\mathbf{b}').\qquad(3.19b^*)$$

One can directly verify that the derived formulas are nothing but Eqs. (2.39a*) and (2.39b*) in the $\lambda_o = 0$ limit. Note that thresholds (3.19a) and (3.19b) as well as approximations (3.20a*) and (3.20b*) are interrelated via the same transformation of the TP coefficients,

$$_1d \to {}_1d/{}_1c_0,\ {}_1c_0 \to {}_1c_0^{-1}, \qquad (3.21)$$

as the leading coefficients of quadratic polynomials $(3.6^{\pm}_1)$ and $(3.6^{\pm}_0)$. Again, this is the direct consequence of the above observation that one (without loss of the generality can restrict the analysis of AL potential curves only to one range of the TP free term: $_1c_0 < 1$ or $_1c_0 > 1$.

However two branches will react differently as $\lambda_o$ increases toward its positive values.

Assuming that the TP has positive discriminant, namely, that

$$_1d + 2 < {}_1d + 2\sqrt{{}_1c_0} < 0 \text{ for } {}_1c_0 > 1 \qquad (3.22a)$$

and

$$_1d + 2\,{}_1c_0 < {}_1d + 2\sqrt{{}_1c_0} < 0 \text{ for } {}_1c_0 < 1, \qquad (3.22b)$$

denominator of the fraction in the right-hand sides of (3.19a*) or (3.19b*) is necessarily positive when only one regular ℘S solution of the given type exists at least in the $\lambda_o = 0$ limit, i.e., if $_1c_0 > 1$ for $\breve{\mathbf{t}}_ = \mathbf{a}$ or $_1c_0 < 1$ for $\breve{\mathbf{t}}_ = \mathbf{b}$, respectively. We thus conclude that the primary regular ℘S solution **a**m if $_1c_0 > 1$ or **b**m if $_1c_0 < 1$ is nodeless for any value of $\mu_o$ above point (3.19a) or (3.19b) accordingly.

Evaluating discriminant $(3.8^-)$ at points (3.19a) and (3.19b) one finds

$$_1\breve{\Delta}^-_m\big(\mu_{\Diamond\mathbf{a};m}({}_1T_2);{}_1T_2\big) = 4[({}_1d+2)m+1-\sqrt{{}_1c_0}\,]^2 \qquad (3.23a)$$

and

$$ {}_1\breve{\Delta}_{\bar{m}}(\breve{\mu}_{\lozenge \mathbf{b};m}({}_1T_2);{}_1T_2) = 4{}_1c_0[({}_1d/{}_1c_0+2)m+1-1/\sqrt{{}_1c_0}]^2 \qquad (3.23b) $$

so denominator of the fraction in the right-hand sides of (3.19a*) or (3.19b*) vanishes iff two regular ∮S solutions of the same type (**a** for ${}_1c_0 < 1$ or **b** for ${}_1c_0 > 1$) merge at the threshold point. One can also verify that

$$ \breve{\mu}^2_{\lozenge m;\mathbf{a}}({}_1T_2) - \breve{\mu}^2_{\Delta,m-}({}_1a_2 < 0, {}_1c_0) = -\frac{{}_1\breve{\Delta}_{\bar{m}}(\mu_{\lozenge \mathbf{a};m}({}_1T_2);{}_1T_2)}{4({}_1d + 2\sqrt{{}_1c_0})} \geq 0 \qquad (3.23a^*) $$

and

$$ \breve{\mu}^2_{\lozenge m;\mathbf{b}}({}_1T_2) - \breve{\mu}^2_{\Delta,m-}({}_1a_2 < 0, {}_1c_0) = -\frac{{}_1\breve{\Delta}_{\bar{m}}(\mu_{\lozenge \mathbf{b};m}({}_1T_2);{}_1T_2)}{4({}_1d + 2\sqrt{{}_1c_0})} \geq 0 \qquad (3.23b^*) $$

which implies that the denominators retain their sign as far as discriminants (3.23a) and (3.23b) remain finite.

As mentioned above absolute value of signed ExpDiff of the primary (secondary) ∮S solution of type **a** for ${}_1c_0 < 1$ reaches its minimum (maximum) at the point where both regular ∮S solutions merge. The primary ∮S solution **a**m is thus nodeless for any value of $\mu_o$ iff the right side of (3.13a) is smaller than –m, i.e., iff

$$ -2-(1-\sqrt{{}_1c_0})/m < {}_1d < -2\sqrt{{}_1c_0} \text{ for } {}_1c_0 < 1 \ (m \geq {}_1n_o) \qquad (3.24a) $$

The secondary ∮S solution **a**'m thus becomes nodeless for $\mu_o < \breve{\mu}_{\lozenge \mathbf{a};m}({}_1T_2)$. Therefore two nodeless solutions of type **a** co-exist (at least at small values of $\lambda_o$) for

$$ \breve{\mu}_{\Delta,m-}({}_1a_2 < 0, {}_1c_0) < \mu_o < \breve{\mu}_{\lozenge \mathbf{a};m}({}_1T_2) \qquad (3.24a') $$

insofar as PFrBs under consideration are restricted by constraint (3.24a). Note that denominator of the fraction in the right-hand sides of (3.19a*), with $\breve{\mathbf{t}}_- = \mathbf{a}'$, is positive in this case so the fraction is negative below the threshold as anticipated. On other hand, if

$$ {}_1d < -2 - (1-\sqrt{{}_1c_0})/m < -2\sqrt{{}_1c_0} \quad \text{for } {}_1c_0 < 1 \ (m \geq {}_1n_o). \tag{3.24a†} $$

then the primary ∮S solution **a**m is nodeless only above threshold (3.19a) whereas the secondary ∮S solution **a**'m necessarily has at least one node inside the quantization interval for any value of $\mu_o$ (at least if $\lambda_0$ is chosen to be small enough). Under the latter constraint denominator of the fraction in the right-hand sides of (3.19a*), with $\breve{\mathbf{t}}_- = \mathbf{a}$, becomes negative so the fraction is positive below the threshold in this case.

Similarly, signed ExpDiff of the primary (secondary) ∮S solution of type **b** for ${}_1c_0 < 1$ reaches its minimum (maximum) at the point where both regular ∮S solutions merge. The primary ∮S solution **b**m is thus nodeless for any value of $\mu_o$ iff the right side of (3.13b) is larger than $m/\sqrt{{}_1c_0}$ i.e., iff

$$ -2-(1-1/\sqrt{{}_1c_0})/m < {}_1d/{}_1c_0 < -2/\sqrt{{}_1c_0} \quad \text{for } {}_1c_0 > 1 \ (m \geq {}_1n_o). \tag{3.24b} $$

The secondary ∮S solution **b**'m is thus nodeless for $\mu_o < \breve{\mu}_{\Diamond\mathbf{b};m}({}_1T_2)$. Therefore two nodeless solutions of type **b** co-exist (at least at small values of $\lambda_0$) for

$$ \breve{\mu}_{\Delta,m-}({}_1a_2 < 0, {}_1c_0) < \mu_o < \breve{\mu}_{\Diamond\mathbf{b};m}({}_1T_2) \tag{3.24b'} $$

insofar as PFrBs under consideration are restricted by constraint (3.24b). Note that denominator of the fraction in the right-hand sides of (3.19b*), with $\breve{\mathbf{t}}_- = \mathbf{b}'$, is positive in this case so the fraction is negative below the threshold as anticipated. On other hand, if

$$ {}_1d/{}_1c_0 < -2-(1-1/\sqrt{{}_1c_0})/m < -2/\sqrt{{}_1c_0} \quad \text{for } {}_1c_0 > 1 \ (m \geq {}_1n_o) \tag{3.24b†} $$

then the primary ∮S solution **b**m is nodeless only above threshold (3.19b) whereas the secondary ∮S solution **b**'m necessarily has at least one node inside the quantization interval for any value of $\mu_o$ (at least in the $\lambda_0 = 0$ limit). Under the latter constraint denominator of the fraction in the right-hand sides of (3.19a*), with $\breve{\mathbf{t}}_- = \mathbf{b}$, becomes negative so the fraction is positive below the threshold as expected.

We thus conclude that one and only one of two co-existent regular ℑS solutions of the same type is nodeless near the $\lambda_0 = 0$ edge of Area $A_m$, regardless of the value of $\mu_o$ (excluding the point where the nodeless ℑS solution turns into the ground-energy eigenfunction.

As for ℑS solutions of type **d** we were unable to make any practically significant use of the Klein formula [28, 29] to select any nodeless solutions from either primary or secondary sequence:

$$ {}_1\breve{\lambda}_{1;\mathbf{d}m} = -\frac{\sqrt{{}_1\breve{\Delta}^+_m(\mu_o;{}_1T_2)} + {}^+_1\breve{g}^{(m)}_1({}_1c_0)}{2\,{}^+_1\breve{g}_2({}_1T_2)} \quad (m = 0,1,...) \tag{3.25}$$

or

$$ {}_1\breve{\lambda}_{1;\mathbf{d}'m'} = \frac{\sqrt{{}_1\breve{\Delta}^+_{m'}(\mu_o;{}_1T_2)} - {}^+_1\breve{g}^{(m')}_1({}_1c_0)}{2\,{}^+_1\breve{g}_2({}_1T_2)} \quad (m' \geq {}_1n_o\,). \tag{3.25'}$$

For this reason we will simply skip examination of solutions of this type in further discussions.

## 4. Manifolds of nodeless regular ℑS solutions for the generic *r*-GRef potential on the line

In previous Section we explicitly took advantage of the fact that quartic polynomial (2.29a) can be analytically decomposed into the product of quadratic polynomials in the AL limit ($\lambda_o = 0$). It has been demonstrated that the RCSLE with the *r*-GRef Bose invariant generated using a TP with positive discriminant ${}_1\Delta_T > 0$ has the quartet of ℑS solutions $\mathbf{t}m$ of four distinct types ($\mathbf{t}$ = **a**, **b**, **c**, **d**) for $2m < \mu_o - 1$ in the limit $\lambda_o \to 0$, excluding the *sym*-GRef potential (${}_1c_0 = 1$). As proven below this assertion also holds for positive values of the RI $\lambda_o$ as far as the *r*-GRef potential in question has at least m bound energy levels.

An extension of the results of previous Section to the GRef potentials with nonzero reflective barriers is essentially based on the observation that no two real roots ${}_1\lambda_{1;\mathbf{t}_1 m}$ and ${}_1\lambda_{1;\mathbf{t}_2 m}$ of the quartic polynomial $G_4^{1;m}[\lambda\,|\,{}_1\boldsymbol{G}^{220}]$ for two distinct types $\mathbf{t}_1$ and $\mathbf{t}_2$ may merge with each other giving rise to pair of complex-conjugated roots. This implies that a pair of ℑS solutions of two different types may not simply disappear. As far as there are four ℑS solutions **t**m of distinct types **t** = **a**, **b**, **c**, and **d** the type change is only possible along the straight-line

$$\mu_o = 2m + 1 + \lambda_o \tag{4.1}$$

where one of the roots of quartic polynomial (2.29a) vanishes. (Remember that the ExpDiffs at the singular points 0 and 1 obey the inequality $|\,{}_1\lambda_{0;\mathbf{t}m}\,| > |\,{}_1\lambda_{1;\mathbf{t}m}\,|$ for $\lambda_o > 0$ since ${}_1c_0 > 0$ for any *r*-GRef potential on the line.) The only possible exclusion from the stated rule is the 'threshold' segments (2.38*) where the Jacobi polynomials in question have zeros at one of the end-points of the quantization interval.

In particular, since there is no threshold segments for the quartets formed by basic solutions (m=0) the ground energy state **c**0 in the generic *r*-GRef potential $V[z\,|\,{}_1\boldsymbol{G}^{220}_{\downarrow\mathbf{c}0}]$ in Area $A_0$ ($\lambda_o < \mu_o - 1$) is always accompanied by three basis ℑS solutions **a**0, **b**0, and **d**0 nodeless by definition. In particular this implies that the Cooper-Ginocchio- Khare (CGK) potential [30] $V[z\,|\,{}^1_1\boldsymbol{G}^{220}_{\mathbf{c}0}]$ is always accompanied by three other exactly-quantized-by-polynomials (P-EQ) 'siblings' $V[z\,|\,{}^1_1\boldsymbol{G}^{220}_{\mathbf{t}0}]$ or, to be more precise, by two isospectral potentials $V[z\,|\,{}^1_1\boldsymbol{G}^{220}_{\mathbf{a}0}]$ and $V[z\,|\,{}^1_1\boldsymbol{G}^{220}_{\mathbf{b}0}]$ as well as by another SUSY partner $V[z\,|\,{}_1\boldsymbol{G}^{220}_{\mathbf{d}0}]$, with an extra bound energy state inserted below the energy level ${}_1\varepsilon_{\mathbf{c}0}$.

Let us study more carefully the $A_m$-segments of curves (2.39a†) and (2.39b†) where a regular ℑS solution nodeless on one side of the curve (**a**m and **b**m, accordingly) turns into the ground-state eigenfunction. We refer to these segments as **a**m- and **b**m-thresholds. In Figs. 4.1 and 4.2 below the fragments of these thresholds lying in Area $A_m$ are marked by diamonds; however they can be extended beyond Area $A_m$ (see Section 5 for example) iff the appropriate regular ℐS solution exists there.

Keeping in mind that each threshold starts at $\lambda_o = 0$ it is convenient to represent curves (2.39a†) and (2.39b†) in the alternative form

$$\mu_{\Diamond\mathbf{a};m}(\lambda_o;{}_1T_2) = \sqrt{\breve{\mu}^2_{\Diamond\mathbf{a};m}({}_1T_2) + \lambda_o^2 + 2[\sqrt{\lambda_o^2 + {}_1c_0 m^2} - \sqrt{{}_1c_0}\,m](m+1)} \qquad (4.2a)$$

and

$$\mu_{\Diamond\mathbf{b};m}(\lambda_o;{}_1T_2)$$
$$= \sqrt{\breve{\mu}^2_{\Diamond\mathbf{b};m}({}_1T_2) + ({}_1a_2 - 1)\lambda_o^2/{}_1c_0 + 2(m+1)\sqrt{(m^2 - \lambda_o^2)/{}_1c_0} - 2m(m+1)/\sqrt{{}_1c_0}}, \qquad (4.2b)$$

with the starting points

$$\breve{\mu}_{\Diamond\mathbf{a};m}({}_1T_2) = \sqrt{(2m+1)^2 - m[2(1-\sqrt{{}_1c_0})(m+1) + ({}_1d + 2)m]} \qquad (4.2\breve{a})$$

and

$$\breve{\mu}_{\Diamond\mathbf{b};m}({}_1T_2) = \sqrt{(2m+1)^2 - m[2(1-1/\sqrt{{}_1c_0})(m+1) + ({}_1d/{}_1c_0 + 2)m]}. \qquad (4.2\breve{b})$$

It has been proven in previous section that the regular ℐS solution **a**m or **b**m lying above the starting point of the appropriate threshold curve is necessarily nodeless and therefore this must be true for any point of Area $A_m$ above the threshold curve in question.

$$\mu_o^2 - \breve{\mu}^2_{\Diamond\mathbf{a};m}({}_1T_2) \approx -2({}_1\breve{\lambda}_{1;\mathbf{a}m} + m)[-({}_1d + 2)m + \sqrt{{}_1c_0} - 1]$$

$$\lambda_{\Diamond\mathbf{a};m}(\lambda_o;{}_1c_0) \equiv \sqrt{\lambda_o^2 + {}_1c_0 m^2}$$

$$\lambda_{\Diamond\mathbf{b};m}(\lambda_o;{}_1c_0) \equiv \sqrt{(m^2 - \lambda_o^2)/{}_1c_0}$$

$$[({}_1d+2)m+1]\lambda_{\Diamond\mathbf{a};m}(\lambda_o;{}_1c_0)+\lambda_o^2-{}_1c_0m=0$$

$$[({}_1d+2)m+1]\lambda_{\Diamond\mathbf{a};m}(\lambda_o;{}_1c_0)={}_1c_0m-\lambda_o^2$$

$$\lambda_{\Diamond\mathbf{a};m}(\lambda_o;{}_1c_0)=-\frac{\lambda_o^2-{}_1c_0m}{({}_1d+2)m+1}>0$$

As far as the given real root of quartic polynomial (2.29a) or (2.34) retains its sign in the vicinity of threshold curve (2.39a†) or (2.39b†), respectively, the regular ℑS solution in question must have the same type on both sides of the curve. In particular the mentioned requirement is always fulfilled for any real root of quartic polynomial (2.34) assuming that the potential has a nonzero reflective barrier. It also holds for the ℑS solution **a**m inside Area $A_m$ as well as inside three other Areas $B_m$, $C_m$, or $D_m$ if threshold curve (2.39a†) crosses them.

Examination of function (4.2a) shows that it monotonically grows as $\lambda_o$ increases. Taking into account that

$$\mu_{\Diamond\mathbf{a};m}(\lambda_o;{}_1T_2)\frac{d\mu_{\Diamond\mathbf{a};m}}{d\lambda_o}-\lambda_o-2m-1=\frac{\lambda_o(m+1)}{\sqrt{\lambda_o^2+{}_1c_0m^2}}-2m-1<-m \tag{4.3}$$

we infer that any curve (4.2a) starting in Area $B_m$ must remain below the borderline $A_m|D_m$. In particular this implies that this curve never crosses Area $A_m$ if condition (3.20a) holds.

$$2(1-\sqrt{{}_1c_0})(m+1)+({}_1d+2)m>0$$

$$-2(1-\sqrt{{}_1c_0})(1+1/m)<{}_1d+2<2(1-\sqrt{{}_1c_0})$$

$$-(1-\sqrt{{}_1c_0})/m<{}_1d+2<2(1-\sqrt{{}_1c_0})\text{ for }{}_1c_0<1\ \ (m\geq{}_1n_o) \tag{3.24a}$$

$$2(1-1/\sqrt{{}_1c_0})(m+1)+({}_1d/{}_1c_0+2)m>0$$

$$-2(1-1/\sqrt{{}_1c_0})(1+1/m)<{}_1d/{}_1c_0+2<2(1-1/\sqrt{{}_1c_0})$$

$$-(1-1/\sqrt{{}_1c_0})/m < {}_1d/{}_1c_0 + 2 < 2(1-1/\sqrt{{}_1c_0}) \text{ for } {}_1c_0 > 1 \quad (m \geq {}_1n_o). \tag{3.24b}$$

Therefore the regular $\mathfrak{J}$S solution $\mathbf{a}m$ is nodeless for $\mu_o > \lambda_o + 2m + 1$ if

$$-2-(1-\sqrt{{}_1c_0})/m < \tfrac{1}{2}\,{}_1d < -\sqrt{{}_1c_0}. \tag{4.4}$$

In particular this implies that any of n+1 $\mathfrak{J}$S solutions $\mathbf{a}m$ (m=0, …, n) for the subset of PFrBs selected by the constraint

$$-4 \leq {}_1d < -2\sqrt{{}_1c_0} \quad ({}_1c_0 < 1) \tag{4.4*}$$

is nodeless in Area $A_n$ ($\mu_o > \lambda_o + 2n + 1$) where the appropriate *r*-GRef potential has at least n bound energy levels. (Remember that $A_m \subset A_{m-1}$ for any m.) Therefore any combination of these n+1 solutions can be used as seed functions for constructing a multi-step SUSY partner conditionally exactly quantized by GS Heine polynomials [1]. Subplots a) of Figs. 4.1 and 4.2 illustrate this important case using the LTP (${}_1a_2 = 0$) as an example. We mark the type of the given regular $\mathcal{J}$S solution $^*\mathbf{a}$m or $^*\mathbf{b}$m by asterix to indicate that it is nodeless in the selected region. Similarly circle used as superscript explicitly emphasizes that the regular $\mathcal{J}$S solution $^{o}\mathbf{a}$m or $^{o}\mathbf{b}$m lie above the ground energy level at any point of Area $A_m$ under the given threshold.

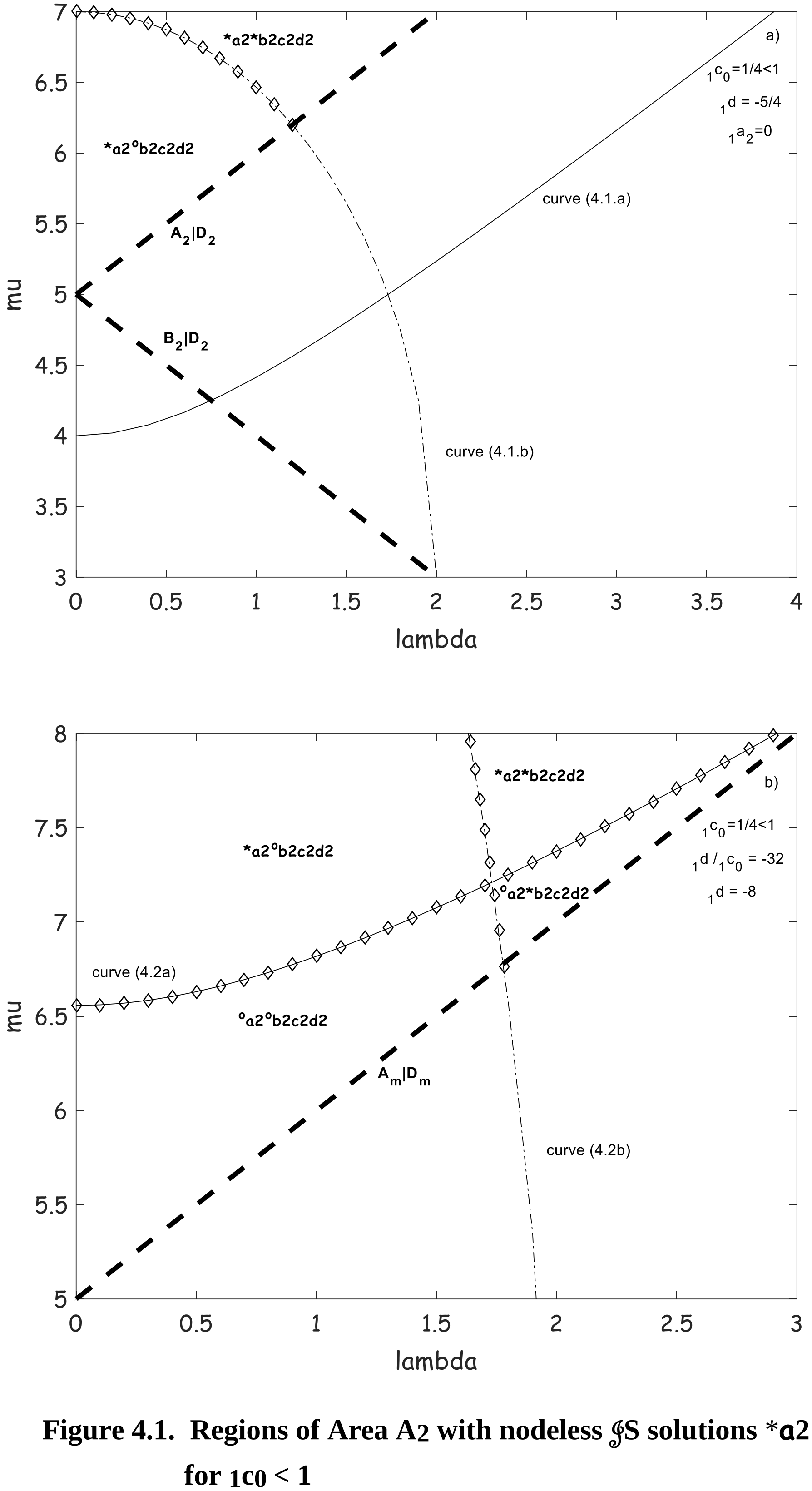

**Figure 4.1. Regions of Area A2 with nodeless ∮S solutions *a2 and *b2 for $_1c_0 < 1$**

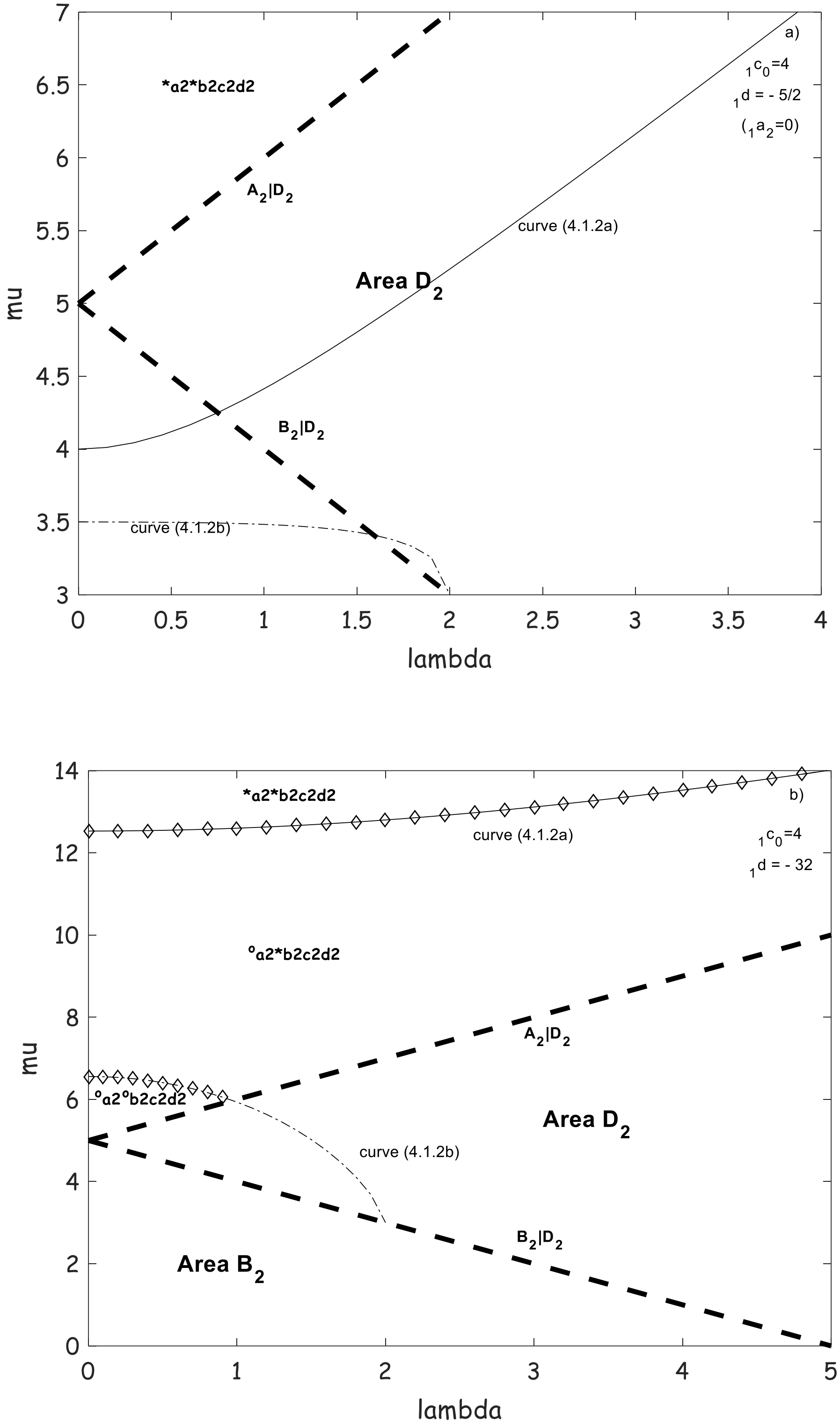

**Figure 4.2. Regions of Area $A_2$ with nodeless regular $\mathcal{J}$S solutions for $_1c_0 > 1$**

If ${}_1a_2 \le 1$ (${}_1d \le -{}_1c_0$) curve (4.2b) drawn in the $\lambda_o\mu_o$ plane monotonically goes down until it ends at the point $\lambda_o = m$, $\mu_o = m+1$ lying on the border line $B_m|D_m$.

Subtracting squares of (3.15a) and (3.15b) we find that

$$\breve{\mu}^2_{\Diamond\mathbf{b};m}({}_1d, {}_1c_0) - \breve{\mu}^2_{\Diamond\mathbf{a};m}({}_1d, {}_1c_0) = 2(1-{}_1c_0)m[(2-{}_1d/\sqrt{{}_1c_0})m+1]/\sqrt{{}_1c_0} \qquad (4.5)$$

Therefore curve (4.2b) always starts above (below) curve (4.2a) for ${}_1c_0 < 1$ (${}_1c_0 > 1$). In particular this implies this implies that two curves never cross if ${}_1c_0 > 1$ since they go in the opposite directions as $\lambda_o$ increases.

Subplots b) of Figs. 4.1 and 4.2 depict illustrative cases when both curves start in Area $A_m$. Note that subplot b) of Fig. 4.1 depict the most complicated case when the thresholds intersect in Area $A_m$ so the area is split into four distinct regions so each of the regular $\mathcal{J}$S solutions has at least one node within the quantization interval at any point of the region lying below both thresholds.

An analysis of the function

$$\begin{aligned}\mu^2_{\Diamond\mathbf{a};m}(\lambda_o; {}_1T_2) - (\lambda_o + 2m+1)^2 = \breve{\mu}^2_{\Diamond\mathbf{a};m}({}_1T_2) - (2m+1)^2 - 2\sqrt{{}_1c_0}\,m(m+1) \\ + 2[(m+1)\sqrt{\lambda_o^2 + {}_1c_0 m^2} - (2m+1)\lambda_o]\end{aligned} \qquad (4.6)$$

shows that curve (4.2a) always lies in Area $D_m$ at large $\lambda_o$ therefore it necessarily crosses the borderline $A_m|D_m$ if it starts in Area $A_m$. However these crossing points for the examples depicted in subplots b) of Figs. 4.1 and 4.2 lie outside the selected range of the RIs $\lambda_o$ and $\mu_o$.

As the final result of this subsection let us prove that the set of algebraic equations (2.21) and (2.26) have at least two real roots at sufficiently large m. Indeed, expressing the cited equations in terms of the new m-dependent sought-for variables

$$ {}_1\tau_{r;\mathbf{t}\mathrm{m}} \equiv \frac{{}_1\lambda_{r;\mathbf{t}\mathrm{m}}}{2\mathrm{m}+1} \qquad (r = 0, 1) \tag{4.7} $$

and making m tend to $\infty$ in both equations

$$ ({}_1\tau_{0;\mathbf{t}\mathrm{m}} + {}_1\tau_{1;\mathbf{t}\mathrm{m}} + 1)^2 = \frac{\mu_o^2}{2\mathrm{m}+1} + {}_1a_2\, {}_1\tau^2_{1;\mathbf{t}\mathrm{m}} \tag{4.8a} $$

and

$$ {}_1\tau^2_{0;\mathbf{t}\mathrm{m}} = \frac{\lambda_o^2}{(2\mathrm{m}+1)^2} + {}_1c_0\, {}_1\tau^2_{1;\mathbf{t}\mathrm{m}} \tag{4.8b} $$

we come to two quadratic equations

$$ {}^{\pm}_{1}\breve{g}_2({}_1T_2)\,{}_1\breve{\tau}^2_{1;\breve{\mathbf{t}}_{\pm}\infty}({}_1T_2) + 2(\pm\sqrt{{}_1c_0}+1)\,{}_1\breve{\tau}_{1;\breve{\mathbf{t}}_{\pm}\infty}({}_1T_2) + 4 = 0 \tag{4.9$^{\pm}$} $$

accompanied by the following linear relations

$$ {}_1\breve{\tau}_{0;\breve{\mathbf{t}}_{\pm}\infty}({}_1T_2) = \pm\,{}_1\breve{\tau}_{1;\breve{\mathbf{t}}_{\pm}\infty}({}_1T_2) \tag{4.$\tilde{9}^{\pm}$} $$

between the new m-independent sought-for parameters

$$ {}_1\breve{\tau}_{r;\mathbf{t}\infty}({}_1T_2) \equiv \lim_{\mathrm{m}\to\infty} {}_1\tau_{r;\mathbf{t}\mathrm{m}} \quad (r = 0, 1), \tag{4.10} $$

where the types $\breve{\mathbf{t}}_{\pm}$ are defined via (3.4$^+$) and (3.4$^-$). The quadratic equation for $\breve{\tau}_{1;\breve{\mathbf{t}}_{+}\infty}({}_1T_2)$ has two negative roots and therefore determine two ℘S solutions irregular at both ends of the quantization intervals, regardless of the value of the coefficient ${}_1c_0$. On other hand, the quadratic equation for $\breve{\tau}_{1;\mathbf{t}_\infty}({}_1T_2)$ has two negative or positive roots for ${}_1c_0 < 1$ or ${}_1c_0 > 1$, accordingly. We label these pairs of the roots in such a way that the appropriate signed ExpDiffs turn into (3.18a) and (3.24a′) or into (3.18b) and (3.24b′) for ${}_1c_0 < 1$ or ${}_1c_0 > 1$, respectively, in the limiting case $\lambda_o = 0$, i.e., we choose

$$ {}_1\breve{\tau}_{1;\mathbf{a}\infty}({}_1T_2) = -\frac{\sqrt{{}_1a_2} + 1 - \sqrt{{}_1c_0}}{{}_{\bar{1}}\breve{g}_2({}_1T_2)} < 0 \tag{4.11a} $$

and

$$ {}_1\breve{\tau}_{1;\mathbf{a}'\infty}({}_1T_2) = \frac{\sqrt{{}_1a_2} - 1 + \sqrt{{}_1c_0}}{{}_{\bar{1}}\breve{g}_2({}_1T_2)} < 0 \tag{4.11a′} $$

if ${}_1c_0 < 1$ or

$$ {}_1\breve{\tau}_{1;\mathbf{b}\infty}({}_1T_2) = \frac{\sqrt{{}_1a_2} + \sqrt{{}_1c_0} - 1}{{}_{\bar{1}}\breve{g}_2({}_1T_2)} > 0 \tag{4.11b} $$

and

$$ {}_1\breve{\tau}_{1;\mathbf{b}'\infty}({}_1T_2) = \frac{\sqrt{{}_1c_0} - 1 - \sqrt{{}_1a_2}}{{}_{\bar{1}}\breve{g}_2({}_1T_2)} > 0 \tag{4.11b′} $$

if ${}_1c_0 > 1$.

Note that the infinite secondary sequence $\mathbf{a}'\mathrm{m}_{\mathbf{a}'}$ for ${}_1c_0 < 1$ starts from

$$ \mathrm{m}_{\mathbf{a}'} = [\tfrac{1}{2}(\mu_o - \lambda_o - 1)] + 1 \tag{4.12} $$

such that $|\,{}_1\tau_{1;\mathbf{a}'\mathrm{m}_{\mathbf{a}'}}| \approx 0 < |\,{}_1\tau_{1;\mathbf{a}\mathrm{m}_{\mathbf{a}'}}|$. Comparing (4.11a) and (4.11a′) one can verify that the latter inequality also holds for sufficiently large $\mathrm{m}_{\mathbf{a}'}$. As far as there are just two regular $\mathcal{S}$S solutions $\mathbf{t}$m of type $\mathbf{a}$ we will refer to the one with a smaller ChExp as ‘primary’ to account for ambiguity which arises if a pair of negative roots ${}_1\lambda_{1;\mathbf{a}\tilde{m}}$ and ${}_1\lambda_{1;\mathbf{a}'\tilde{m}}$ of quartic polynomial (2.29a) turn into a pair of complex conjugated roots within a certain range of $\tilde{m}$.

Taking into account that absolute values of the signed ExpDiffs ${}_1\lambda_{1;\mathbf{t}m}$ for solutions from infinite secondary sequences $\mathbf{t} = \mathbf{a}'$ for ${}_1c_0 < 1$ and $\mathbf{t} = \mathbf{b}'$ for ${}_1c_0 > 1$ increase proportionally to m for sufficiently large m we conclude that there are two infinite subsets of nodeless ℑS solutions regular at the singular points z =0 and z =1 for ${}_1c_0 < 1$ and ${}_1c_0 > 1$, respectively.

## 5. Algebraic formulas for energies of ℑS solutions in the LTP limit

In the LTP limit (${}_1a_2 = 0$) quartic polynomial (2.29b) can be analytically decomposed into the product of two quadratic polynomials

$$G_4^{1;m}[\lambda\,|\,{}_1\boldsymbol{G}^{110}] = {}^{+}G_2^{1;m}[\lambda\,|\,{}_1\boldsymbol{G}^{110}] \times {}^{-}G_2^{1;m}[\lambda\,|\,{}_1\boldsymbol{G}^{110}] \qquad (5.1_1)$$

where

$${}_1^{\pm}G_2^{1;m}[\lambda\,|\,{}_1\boldsymbol{G}^{110}] = ({}_1c_0 - 1)\lambda^2 - 2(2m+1\pm\mu_o)\lambda + \lambda_o^2 - (2m+1\pm\mu_o)^2 \qquad (5.1_1^{\pm})$$

and

$${}_1^{\pm}G_2^{0;m}[\lambda\,|\,{}_1\boldsymbol{G}^{110}] = (1/{}_1c_0 - 1)\lambda^2 - 2(2m+1\pm\mu_o)\lambda - \lambda_o^2/{}_1c_0 - (2m+1\pm\mu_o)^2. \qquad (5.1_0^{\pm})$$

Taking into account that discriminant of the quadratic polynomials ($5.1_1^{\pm}$) and ($5.1_0^{\pm}$) differ only by the scale ${}_1c_0^2$:

$$\Delta_{\pm}^{1;m}(\mu_o,\lambda_o;{}_1c_0) = {}_1c_0^2\Delta_{\pm}^{0;m}(\mu_o,\lambda_o;{}_1c_0) = 4[{}_1c_0(2m+1\pm\mu_o)^2 + (1-{}_1c_0)\lambda_o^2] \qquad (5.2^{\pm})$$

one can verify that the roots of these polynomials

$${}_1\lambda_{1;\tilde{\mathbf{t}}_{\pm;j}m} = -[\pm\mu_o + 2m + 1 + \tfrac{1}{2}(-1)^j\sqrt{\Delta_{\pm}^{1;m}(\lambda_o,\mu_o;{}_1c_0)}\,]/(1-{}_1c_0) \qquad (5.3_1^{\pm})$$

and

$${}_1\lambda_{0;\tilde{\mathbf{t}}_{\pm,j}m} = {}_1c_0[\pm\mu_o + 2m + 1 + \tfrac{1}{2}(-1)^j\sqrt{\Delta_{\pm}^{0;m}(\lambda_o,\mu_o;{}_1c_0)}\,]/(1-{}_1c_0) \qquad (5.3_0^{\pm})$$

identifying the ∮S solutions $\tilde{\mathbf{t}}_{\pm,j}m$ are related via the linear formulas

$$_1\lambda_{0;\tilde{\mathbf{t}}_{\pm,j}m} + {}_1\lambda_{1;\tilde{\mathbf{t}}_{\pm,j}m} \pm \mu_o + 2m + 1 = 0 \qquad (5.\,3^{\pm})$$

as required by (2.21), with $_1a_2 = 0$.

Let us start identification of the types $\tilde{\mathbf{t}}_{\pm,j}$ from a simpler case of the LTP with a negative root

$$z_{T;1} = \frac{_1c_0}{_1c_0 - 1} < 0 \qquad (5.\,4)$$

($_1c_0 < 1$). Since both discriminants ($5.2^+$) and ($5.2^-$) are positive in this case the quartic polynomials (2.29a) and (2.34) have four real roots at any point of the quadrant $\mu_o, \lambda_o \geq 0$ and as a result the quartet of ∮S solutions $\mathbf{t}_j m$ of definite types $\mathbf{t}_j$ (j=1,2,3,4) exists inside each Area carved by the border lines $A_m|D_m$, $B_m|D_m$, and $C_m|D_m$, Since any of ∮S solution $\mathbf{t}_j m$ can change its type only at the border lines the general arguments presented in Appendix A make it possible to predict the solution type without the solving the appropriate quartic equations. Namely, the eigenfunction **c**m in the quartet **abcd** turns into the regular ∮S solution **a**'m below the border line $A_m|D_m$. The latter solution retains its type in all three Areas $B_m$, $C_m$, and $D_m$, At the border lines $B_m|D_m$ and $C_m|D_m$ the regular ∮S solution **b**m turns into the irregular ∮S solution **d**'m so the ∮S solutions $\mathbf{t}_j m$ form the quartet **aa'dd'** in both Areas $B_m$ and $C_m$

Keeping in mind that free terms of quadratic polynomials ($5.1_0^{\pm}$) are both negative (while the common leading coefficient is positive) signs of the sums in brackets in the right-hand sides of ($5.3_0^{\pm}$) are determined by sign of the second summand in each sum, i.e., $\tilde{\mathbf{t}}_{\pm,1} =$ **b** or **d** and $\tilde{\mathbf{t}}_{\pm,2} =$ **a** or **c**. In particular, taking into account $_1\lambda_{1;\tilde{\mathbf{t}}_{+,2}m} < 0$ we infer that $\tilde{\mathbf{t}}_{+,2} =$ **a** at any point of the quadrant $\mu_o, \lambda_o \geq 0$. Since both roots of quadratic polynomial ($5.1_1^+$) in Area $A_m$ are negative the second ∮S solution $\tilde{\mathbf{t}}_{+,1}$m must be of type **d** there.

Since the roots of quadratic polynomial ($5.1_1^-$) have the same sign outside Area $C_m$ and in particular are both positive in Area $A_m$, i.e., $\tilde{\mathbf{t}}_{-,1} = \mathbf{b}$ and $\tilde{\mathbf{t}}_{-,2} = \mathbf{c}$ there. We thus confirmed existence of the **abcd** quartet of ∮S solutions at any point of Area $A_m$ if ${}_1c_0 < 1$.

As expected the eigenfunction **c**m turns into the secondary regular ∮S solution **a′**m on other side of the border line $A_m|D_m$ where the free term of polynomial ($5.1_1^-$) changes its sign so the polynomial has roots of opposite sign. The **aa′bd** quartet of ∮S solutions then retains at any point of Area $D_m$ ($\tilde{\mathbf{t}}_{+,2} = \mathbf{a}'$). The free term of polynomial ($5.1_1^-$) becomes negative again below the border line $B_m|D_m$; however since the linear coefficient changes its sign, compared with Area $A_m$, the polynomial has two negative roots so $\tilde{\mathbf{t}}_{-,1} = \mathbf{d}'$. We thus infer that the **aa′dd′** quartet already disclosed by us in the AL limit for the (generally second-order) TP with the free term ${}_1c_0 < 1$ exists at any point of Area $B_m$ at least for sufficiently small values of ${}_1a_2$.

Evaluating discriminants ($5.2^{\pm}$) at $\lambda_o = 0$ and substituting square roots of the resultant expressions,

$$\tfrac{1}{2}\sqrt{\Delta_+^{0;m}(0,\mu_o;{}_1c_0)} = \tfrac{1}{2}\sqrt{\Delta_+^{1;m}(0,\mu_o;{}_1c_0)}\,/\,{}_1c_0 = 2m+1+\mu_o \qquad (5.5^+)$$

and

$$\tfrac{1}{2}\sqrt{{}_1c_0\,\Delta_-^{0;m}(0,\mu_o;{}_1c_0)} = \tfrac{1}{2}\sqrt{\Delta_-^{1;m}(0,\mu_o;{}_1c_0)/\,{}_1c_0} \qquad (5.5)$$

$$= \begin{cases} \mu_o - 2m - 1 \text{ in Area } A_m \\ \text{or} \\ 2m + 1 - \mu_o \text{ in Area } B_m \end{cases} \qquad (5.5^-)$$

into $(5.3^{\pm}_{1})$ and $(5.3^{\pm}_{0})$ one finds

$$ {}_1\breve{\lambda}_{0;\tilde{\mathbf{t}}_{+;1}m} \equiv {}_1\breve{\lambda}_{0;\mathbf{d}m} = \sqrt{{}_1c_0}\,{}_1\breve{\lambda}_{1;\mathbf{d}m} = -\sqrt{{}_1c_0}(\mu_o + 2m + 1)/(1+\sqrt{{}_1c_0}) < 0, \qquad (5.6d) $$

$$ {}_1\breve{\lambda}_{0;\tilde{\mathbf{t}}_{+;2}m} \equiv {}_1\breve{\lambda}_{0;\mathbf{a}m} = -\sqrt{{}_1c_0}\,{}_1\breve{\lambda}_{1;\mathbf{a}m} = \sqrt{{}_1c_0}(\mu_o + 2m + 1)/(1-\sqrt{{}_1c_0}) > 0, \qquad (5.6a) $$

in both Areas $A_m$ and $B_m$,

$$ {}_1\breve{\lambda}_{0;\tilde{\mathbf{t}}_{-;1}m} \equiv {}_1\breve{\lambda}_{0;\mathbf{b}m} = -\sqrt{{}_1c_0}\,{}_1\breve{\lambda}_{1;\mathbf{b}m} = \sqrt{{}_1c_0}(2m + 1 - \mu_o)/(1-\sqrt{{}_1c_0}) < 0, \qquad (5.6b) $$

$$ {}_1\breve{\lambda}_{0;\tilde{\mathbf{t}}_{-;2}m} \equiv {}_1\breve{\lambda}_{0;\mathbf{c}m} = \sqrt{{}_1c_0}\,{}_1\breve{\lambda}_{1;\mathbf{c}m} = \sqrt{{}_1c_0}(\mu_o - 2m - 1)/(1+\sqrt{{}_1c_0}) > 0, \qquad (5.6b) $$

in Area $A_m$, and

$$ {}_1\breve{\lambda}_{0;\tilde{\mathbf{t}}_{-;1}m} \equiv {}_1\breve{\lambda}_{0;\mathbf{d}'m} = \sqrt{{}_1c_0}\,{}_1\breve{\lambda}_{1;\mathbf{d}'m} = -\sqrt{{}_1c_0}(2m + 1 - \mu_o)/(1+\sqrt{{}_1c_0}) < 0, \qquad (5.6d') $$

$$ {}_1\breve{\lambda}_{0;\tilde{\mathbf{t}}_{-;2}m} \equiv {}_1\breve{\lambda}_{0;\mathbf{a}'m} = -\sqrt{{}_1c_0}\,{}_1\breve{\lambda}_{1;\mathbf{a}'m} = \sqrt{{}_1c_0}(2m + 1 - \mu_o)/(1-\sqrt{{}_1c_0}) > 0, \qquad (5.6a') $$

in Area $B_m$. One can alternatively come to the same expressions by evaluating the leading coefficients and discriminants of quadratic polynomials $(3.5^{\pm})$ at ${}_1a_2 = 0$:

$$ {}^{\pm}_{1}\breve{g}_2({}_1a_2 = 0, {}_1c_0) = (1 \pm \sqrt{{}_1c_0})^2, \qquad (5.7) $$

and

$$ \sqrt{{}_1\breve{\Delta}^{\pm}_{m}(\mu_o; {}_1a_2 = 0, {}_1c_0)} = 2(1 \pm \sqrt{{}_1c_0})\mu_o > 0 \qquad (5.7^*) $$

and then substituting the derived formulas into (3.7c), (3.7d), (3.18a), (3.18b), (3.23a'), and (3.23b').

Since both roots of quadratic polynomial $(5.1^{+}_{1})$ are negative everywhere outside Area $C_m$ and the polynomial free term changes its sign at the border line $C_m|D_m$ the $\mathcal{J}$S solution $\tilde{\mathbf{t}}_{+,1}m$ alters its type at the border from $\mathbf{d}$ to $\mathbf{b}$, in agreement with the general arguments presented in Appendix A.

A brief summary of the cited statements correlating roots of quadratic polynomials $(5.1_1^{\pm})$ and $(5.1_0^{\pm})$ with the types of the appropriate ∮S solutions for $_1c_0 < 1$ is compiled in Table 5.1.

Table 5.1

**Classification of ∮S solutions for the LTP with a negative root**

| | Area $A_m$ | Area $B_m$ | Area $C_m$ | Area $D_m$ |
|---|---|---|---|---|
| $\tilde{\mathbf{t}}_{-,1}$ | **b** | **d**' | **b** | **b** |
| $\tilde{\mathbf{t}}_{-,2}$ | **c** | **a**' | **a**' | **a**' |
| $\tilde{\mathbf{t}}_{+,1}$ | **d** | **d** | **b**''' | **d** |
| $\tilde{\mathbf{t}}_{+,2}$ | **a** | **a** | **a** | **a** |

Note that we label the secondary ∮S solution of type **b** as **b**'''m because (as demonstrated in next Section both secondary solutions **b**'m and **b**''m existent in Areas $B_m$ and $D_m$ for $_1c_0 > 1$ disappear in the limit $_1c_0 \to 1$.

It directly follows from (5.6a) that

$$|\,{}_1\breve{\lambda}_{1;\mathbf{a}m}\,| > m + \tfrac{1}{2}(\mu_o + 1) > m \quad \text{for } {}_1c_0 < 1. \tag{5.8}$$

Since discriminant $(5.2^{\pm})$ and therefore ExpDiff $|\,{}_1\lambda_{1;\tilde{\mathbf{t}}_{+,2}m}\,|$ monotonically increase as the reflective barrier grows we infer that the ∮S solution **a**m is nodeless regardless of values of the RIs $\lambda_o$ and $\mu_o$ for any LTP with a negative root.

By expressing the free term of quadratic polynomial $(5.1_1^{-})$ in terms of the ExpDiff $_1\lambda_{1;\mathbf{c}0}$ we can represent the quadratic equation for the ExpDiff $|\,{}_1\lambda_{1;\mathbf{a}'m}\,|$ as

$$(|\,{}_1\lambda_{1;\mathbf{a}'m}| - {}_1\lambda_{1;\mathbf{c}0})[(1-{}_1c_0)(|\,{}_1\lambda_{1;\mathbf{a}'m}| + {}_1\lambda_{1;\mathbf{c}0}) - 2(\mu_o - 1)] = 4(m+1-\mu_o)(|\,{}_1\lambda_{1;\mathbf{a}'m}| - m).$$

(5.9)

Making use of $(5.3_1^-)$ for j=2 to compute the sum of the ExpDiffs $|\,{}_1\lambda_{1;\mathbf{a}'m}|$ and ${}_1\lambda_{1;\mathbf{c}0}$ and substituting

$$\Delta_-^{1;m}(\lambda_o,\mu_o;{}_1c_0) - \Delta_-^{1;0}(\lambda_o,\mu_o;{}_1c_0) = 16\,{}_1c_0 m(m+1-\mu_o) \tag{5.10}$$

into the right-hand side of the resultant expression

$$(1-{}_1c_0)(|\,{}_1\lambda_{1;\mathbf{a}'m}| + {}_1\lambda_{1;\mathbf{c}0}) = \tfrac{1}{2}\sqrt{\Delta_-^{1;m}(\mu_o,\lambda_o;{}_1c_0)} - \tfrac{1}{2}\sqrt{\Delta_-^{1;0}(\mu_o,\lambda_o;{}_1c_0)} + 2m \tag{5.11}$$

we can re-write (5.9) as follows

$$\tfrac{1}{2}\frac{|\,{}_1\lambda_{1;\tilde{\mathbf{t}}_{-;j}m}| - {}_1\lambda_{1;\mathbf{c}0}}{|\,{}_1\lambda_{1;\tilde{\mathbf{t}}_{-;j}m}| - m} = \frac{4\,{}_1c_0 m}{\sqrt{\Delta_-^{1;0}(\lambda_o,\mu_o;{}_1c_0)} + \sqrt{\Delta_-^{1;m}(\lambda_o,\mu_o;{}_1c_0)}} + 1 > 0, \tag{5.12}$$

with $j = 2$ and $\tilde{\mathbf{t}}_{-;2} = \mathbf{a}'$,

which confirms that the secondary regular ℑS solution **a**'m lies below the ground energy level iff $|\,{}_1\lambda_{1;\mathbf{a}'m}| > m$.

$$\pm{}^{1;m}_{1\;2}G[\lambda\,|\,{}_1\boldsymbol{G}^{110}] = ({}_1c_0 - 1)\lambda^2 - 2(2m+1\pm\mu_o)\lambda + \lambda_o^2 - (2m+1\pm\mu_o)^2$$

for $\mu_o < \lambda_o + 1$

Prove that any regular ℑS solution is nodeless if the LTP r-GRef potential does not have the discrete energy spectrum. Does not support bound energy levels.

The signed ExpDiff

$${}_1\lambda_{1;\tilde{\mathbf{t}}_{-;2}m} = {}_1\lambda_{1;\mathbf{a}'m} = [\mu_o - 2m - 1 - \tfrac{1}{2}\sqrt{\Delta_-^{1;m}(\lambda_o,\mu_o;{}_1c_0)}\,]/(1-{}_1c_0) < 0 \tag{5.13}$$

is equal to 0 along the border line $A_m|D_m$ and then monotonically increases as $\lambda_o$ grows at a fixed value of $\mu_o$. Its absolute value becomes larger than m below the threshold curve

$$\mu_o = \mu_{\Diamond\mathbf{a};m}(\lambda_o;-_1c_0-1,{}_1c_0) \equiv \sqrt{\lambda_o^2 + {}_1c_0m^2} + m + 1 \qquad (5.14)$$

where quadratic polynomial $(5.1_{\bar{1}})$ has the root of –m. One can easily verify that the cited formula for the threshold curve directly follows from (4.2a), with

$$\breve{\mu}_{\Diamond\mathbf{a};m}(-_1c_0-1,{}_1c_0) = (1+\sqrt{{}_1c_0})m+1. \qquad (5.14^*)$$

[Since the right-hand side of (5.14*) is smaller than 2m+1 for ${}_1c_0 < 1$ the threshold lies below the border line $A_m|D_m$.] We thus conclude that the ℘S solution $\mathbf{a}'m$ is nodeless everywhere below curve (5.14) regardless of the value of $\mu_o$.

Fig. 5.1 illustrates behavior of the ℘S solution $\mathbf{a}'2$ in Areas $B_2$ and $D_2$ for ${}_1c_0 = 1/4$. We again use asterix to indicate that the regular solution in question is nodeless.

As a matter of fact, the equation

$$_{\bar{1}}G_2^{1;m}[-m\,|\,{}_1\boldsymbol{G}^{110}] = 0 \qquad (5.15)$$

is satisfied along two separate curves:

$$\mu_o = m+1 \pm \sqrt{\lambda_o^2 + {}_1c_0m^2} \qquad (5.16^{\pm})$$

such that

$$\Delta_-^{1;m}(\lambda_o,\mu_o;{}_1c_0)\Big|_{\mu_o = m+1\pm\sqrt{\lambda_o^2+{}_1c_0m^2}} = 4(\lambda_o^2 + {}_1c_0m^2) > 0. \qquad (5.17)$$

Examination of the signed ExpDiff

$$ {}_1\lambda_{1;\tilde{\mathbf{t}}_{-;j}m} = [\mu_o - 2m - 1 - (-1)^j \sqrt{\lambda_o^2 + {}_1c_0 m^2}\,]/(1 - {}_1c_0) \tag{5.18} $$

shows that the second curve corresponds to the points where the energy of the ℐS solution $\tilde{\mathbf{t}}_{-,1}m$ of type **d**, not **a**, coincides with the ground energy $\varepsilon_{\mathbf{c}0}$.

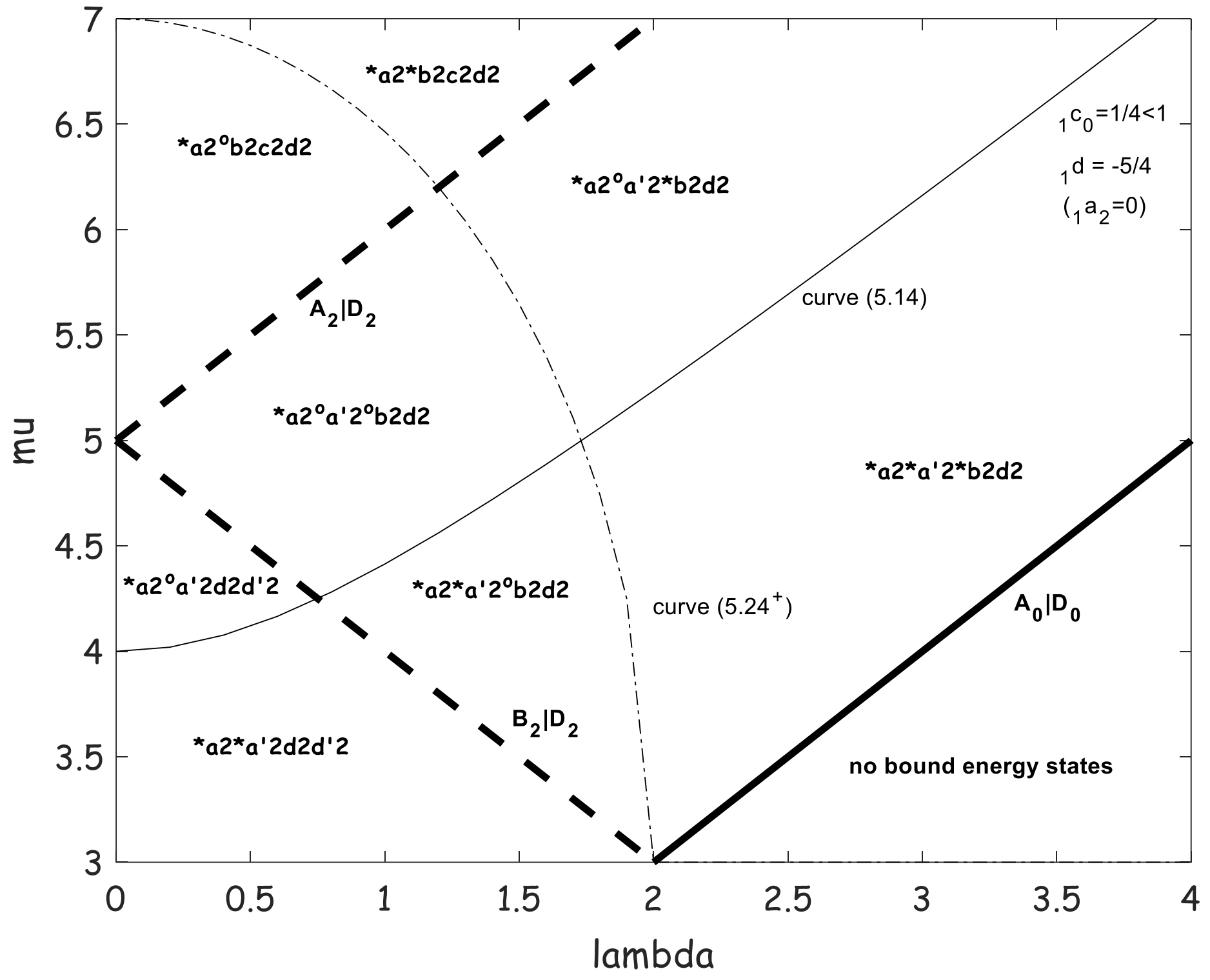

**Figure 5.1. Quartets of ℐS solutions co-existent with the discrete energy spectrum for the LTP with a negative root**

It directly follows from (5.3$^-$) for any solution $\tilde{\mathbf{t}}_{-,j}m$ of type **b** (j=1 if ${}_1c_0 < 1$) that

$$ {}_1\lambda_{1;\tilde{\mathbf{t}}_{-,j}} + m + 1 - \mu_o = |\,{}_1\lambda_{0;\tilde{\mathbf{t}}_{-,j}m}\,| - m. \tag{5.19} $$

In particular this implies that the regular ℐS solution in question is necessarily nodeless if $\mu_o \le m+1$. We thus independently confirmed that the primary ℐS solution **b**m is indeed nodeless in the region of Area $D_m$ below the straight line $\mu_o = m+1$ in the $\lambda_0\mu_0$ plane, as stated in the end of Section 2.

Examination of interrelation (5.19) between ExpDiffs ${}_1\lambda_{1;\mathbf{b}m}$ and $|{}_1\lambda_{0;\mathbf{b}m}|$ also shows that the primary $\mathcal{J}$S solution **b**m is necessarily nodeless if ${}_1\lambda_{1;\mathbf{b}m} > \mu_o - m - 1$. Representing the quadratic equation for the ExpDiff ${}_1\lambda_{1;\mathbf{b}m}$ as

$$ {}_1\lambda_{1;\mathbf{b}m} - {}_1\lambda_{1;\mathbf{c}0} = \frac{2m({}_1\lambda_{1;\mathbf{b}m} - \mu_o + m + 1)}{\mu_o - 1 - \frac{1}{2}(1 - {}_1c_0)({}_1\lambda_{1;\mathbf{b}m} + {}_1\lambda_{1;\mathbf{c}0})} \quad ({}_1c_0 < 1) \tag{5.20} $$

and using $(5.3\bar{1})$, now with j=1, to compute the sum of the ExpDiffs ${}_1\lambda_{1;\mathbf{b}m}$ and ${}_1\lambda_{1;\mathbf{c}0}$:

$$ (1 - {}_1c_0)({}_1\lambda_{1;\mathbf{b}m} + {}_1\lambda_{1;\mathbf{c}0}) = 2(\mu_o - m - 1) - \frac{8\,{}_1c_0 m(\mu_o - m - 1)}{\sqrt{\Delta_-^{1;m}(\mu_o, \lambda_o; {}_1c_0)} + \sqrt{\Delta_-^{1;0}(\mu_o, \lambda_o; {}_1c_0)}} > 0 \tag{5.21} $$

one can then directly verify that denominator of the fraction in the left-hand side is positive:

$$ 2(\mu_o - 1) - (1 - {}_1c_0)({}_1\lambda_{1;\mathbf{b}m} + {}_1\lambda_{1;\mathbf{c}0}) = 2m + \frac{8\,{}_1c_0 m(\mu_o - m - 1)}{\sqrt{\Delta_-^{1;m}(\mu_o, \lambda_o; {}_1c_0)} + \sqrt{\Delta_-^{1;0}(\mu_o, \lambda_o; {}_1c_0)}} > 0 $$

$$ {}_1c_0 < 1 \tag{5.22} $$

insofar as $\mu_o \geq m+1$. This implies that the primary $\mathcal{J}$S solution **b**m constructed using the LTP with a negative root necessarily lies below the ground energy level iff ${}_1\lambda_{1;\mathbf{b}m} > \mu_o - m - 1$, or, which is equivalent, iff ${}_1\lambda_{0;\mathbf{b}m} < -m$.

Again, the equation

$$ {}_{\bar{1}}G_2^{0;m}[-m \mid {}_1\boldsymbol{G}^{110}] = 0 \tag{5.23} $$

holds along two different curves:

$$\mu_o = m+1 \pm \sqrt{(m^2-\lambda_o^2)/{}_1c_0} \qquad (5.24^{\pm})$$

starting at the same point $\lambda_o = m$, $\mu_o = m+1$ on the borderline $B_m|D_m$. As $\lambda_o$ decrease the curve

$$\mu_{\Diamond \mathbf{b};m}(\lambda_o; -{}_1c_0 - 1, {}_1c_0) = m+1+\sqrt{(m^2-\lambda_o^2)/{}_1c_0} \qquad (5.24)$$

goes up inside Area $D_m$ whereas second curve ($5.24^-$) goes down into Area $B_m$ where the primary ∮S solution $\tilde{\mathbf{t}}_{-,1}m$ is irregular at both ends ($\tilde{\mathbf{t}}_{-,1} = \mathbf{d}'$). One can verify that general formulas (4.2b) and (4.2b*) for the threshold in question does turn into (5.24) at ${}_1a_2 = 0$.

If the LTP root $z_{T;1}$ is larger than 1 then both quadratic polynomials ($5.1_0^+$) and ($5.1_0^-$) have roots of the same sign; namely, the roots of both cases are negative unless $\mu_o > 2m+1$. In the latter case quadratic polynomial ($5.1_0^-$) has two positive roots so $\tilde{\mathbf{t}}_{-,j} = \mathbf{a}$ or $\mathbf{c}$ ($\tilde{\mathbf{t}}_{-,j} = \mathbf{b}$ or $\mathbf{d}$) for $\mu_o > 2m+1$ ($\mu_o < 2m+1$) whereas $\tilde{\mathbf{t}}_{+,j} = \mathbf{b}$ or $\mathbf{d}$ regardless of the value of $\mu_o$. Keeping in mind that both quadratic polynomials ($5.1_1^+$) and ($5.1_1^-$) has roots of opposite sign in Area $A_m$ we confirm that the $m^{th}$ eigenfunction is indeed accompanied by ∮S solutions of three distinct types or, to be more specific, $\tilde{\mathbf{t}}_{-,1} = \mathbf{a}$, $\tilde{\mathbf{t}}_{-,2} = \mathbf{c}$, $\tilde{\mathbf{t}}_{+,1} = \mathbf{d}$, $\tilde{\mathbf{t}}_{+,2} = \mathbf{b}$, with the corresponding signed ExpDiffs given by ($5.3_1^{\pm}$) and ($5.3_0^{\pm}$).

On the $D_m$-side of the $m^{th}$ **c**/**a**'-separatrix the free term of the quadratic polynomial ($5.1_1^+$) changes its sign and as a result the $m^{th}$ eigenfunction turns into the regular secondary ∮S solution **a**'m. However, contrary to the TP with a negative root, the quartet **aa**'**bd** of ∮S solutions exists only within the sector of Area $D_m$ carved by three straight lines $A_m|D_m$,

$$\mu_o = \sqrt{1-1/{}_1c_0}\,\lambda_o + 2m+1, \qquad (5.25^{+})$$

and $C_m|D_m$. [Since the linear coefficient in ($5.25^+$) is smaller than 1 the straight line necessarily ends in Area $C_m$.] Below straight line ($5.25^+$) discriminant of quadratic polynomial ($5.1_1^+$) becomes negative and the pair of ℐS solutions of the same type, **a**m and **a**'m, disappears as illustrated by Fig. 5.2 for m=2 and ${}_1c_0$=2.

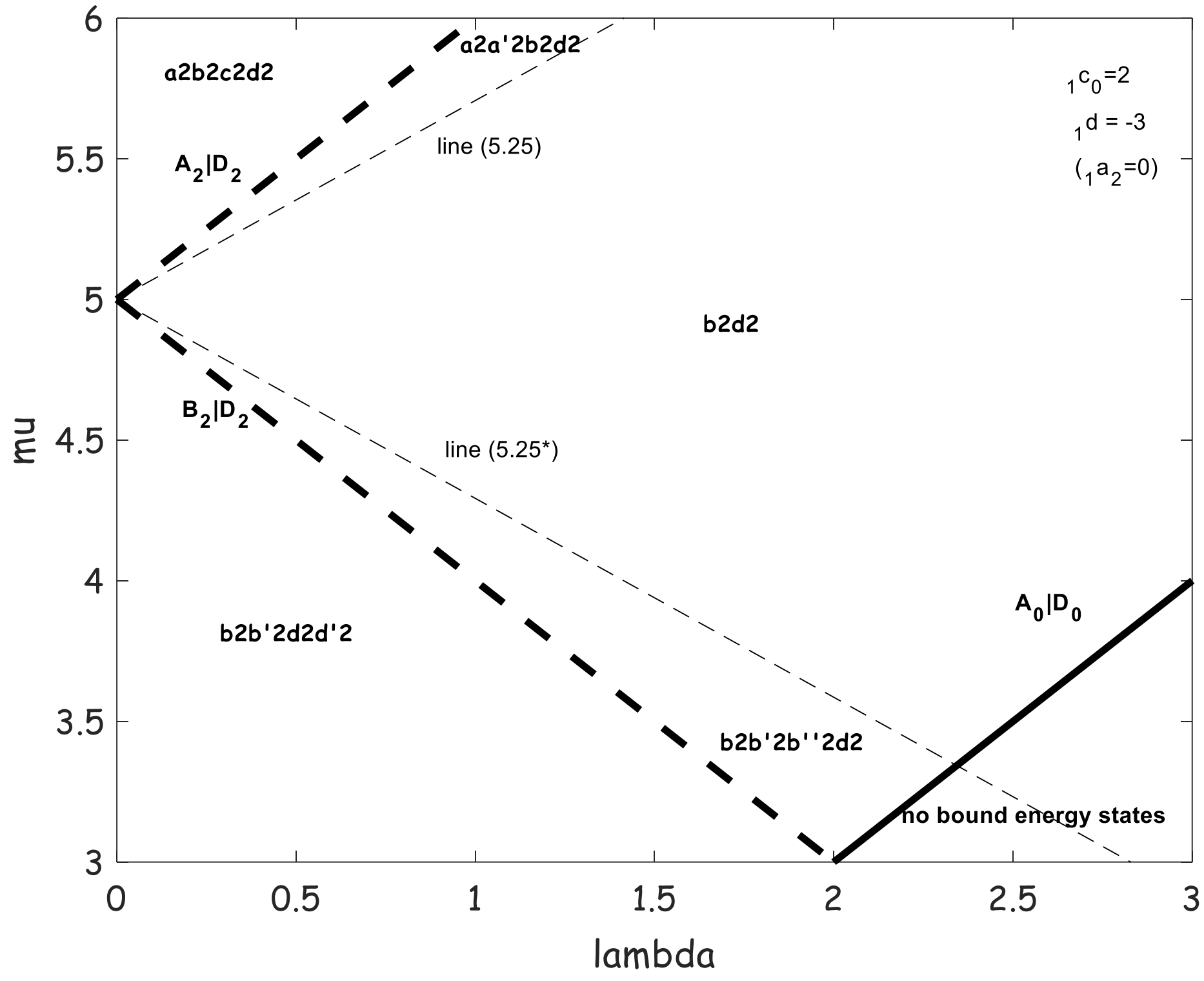

**Figure 5.2. ℐS solutions co-existent with the discrete energy spectrum for the LTP with the root $z_{T;1} > 1$**

The pair of ℐS solutions of the same type re-appears again below the straight line

$$\mu_o = -\sqrt{1 - 1/{}_1c_0}\,\lambda_o + 2m + 1; \qquad (5.25^-)$$

however, these are solutions of type **b**, not **a** because the linear coefficient of each polynomial $(5.1_1^-)$ and $(5.1_0^-)$ has different signs in the vicinity of straight lines (5.25$^+$) and (5.25$^-$). Indeed, one can easily verify that

$$ {}_1\lambda_{1;\tilde{\mathbf{t}}_{-;1}m} = {}_1\lambda_{1;\tilde{\mathbf{t}}_{-;2}m} = -\frac{\lambda_o}{\sqrt{{}_1c_0({}_1c_0+1)}} < 0, \qquad (5.26_1) $$

$$ {}_1\lambda_{0;\tilde{\mathbf{t}}_{-;1}m} = {}_1\lambda_{0;\tilde{\mathbf{t}}_{-;2}m} = \sqrt{\frac{{}_1c_0}{{}_1c_0+1}}\,\lambda_o > 0 \qquad (5.26_0) $$

along straight line (5.25$^+$) and

$$ {}_1\lambda_{1;\tilde{\mathbf{t}}_{-;1}m} = {}_1\lambda_{1;\tilde{\mathbf{t}}_{-;2}m} = \frac{\lambda_o}{\sqrt{{}_1c_0({}_1c_0+1)}} > 0, \qquad (5.26_1^*) $$

$$ {}_1\lambda_{0;\tilde{\mathbf{t}}_{-;1}m} = {}_1\lambda_{0;\tilde{\mathbf{t}}_{-;2}m} = -\sqrt{\frac{{}_1c_0}{{}_1c_0+1}}\,\lambda_o < 0 \qquad (5.26_0^*) $$

along straight line (5.25$^-$). Obviously the latter line is nothing but the LTP limit of the threshold curve where the quadratic polynomial ${}_{A|B}G_2^{1;m}[\lambda\,|\,{}_1\boldsymbol{G}^{K\Im 0}]$ in decomposition (B.1) in Appendix B has a double positive root.

As the coefficient ${}_1c_0$ increases straight lines (5.25$^+$) and (5.25$^-$) approach the border lines $A_m|D_m$ and $B_m|D_m$ accordingly so the region of Area $D_m$ with only two $\mathfrak{S}$S solutions monotonically expands. All $\mathfrak{S}$S solutions disappear below the straight line

$$ \mu_o = \sqrt{1-1/{}_1c_0}\,\lambda_o - 2m - 1 \qquad (5.27) $$

(not shown in Fig. 5.2) which lies solely in Area $C_m$ starting exactly at the point where the straight line (5.25$^-$) ends. The line fragment of straight line (5.25$^+$) ending at its intersection with line (5.27) represents the LTP limit of the threshold curve where the quadratic polynomial $_{A|B}G_2^{1;m}[\lambda \mid {}_1\boldsymbol{G}^{K\Im 0}]$ in decomposition (B.1) has a double negative root.

In Area $B_m$ both discriminants (5.2$^{\pm}$) for the LTP with a root $z_{T;1} > 1$ are always positive because

$$\lambda_o^2 < (2m+1-\mu_o)^2 < (2m+1-\mu_o)^2/(1-1/{}_1c_0) \text{ for } {}_1c_0 > 1. \tag{5.28}$$

This is the direct consequence of the fact that each of quadratic polynomials $(5.1_1^+)$ and $(5.1_1^-)$ has roots of opposite sign there. As a result the quartet **bb'dd'** of $\mathscr{J}$S solutions discussed in Section 3 in the $\lambda_o = 0$ limit retains at any non-zero value of $\lambda_o$ inside Area $B_m$.

Substituting (5.5$^+$) and (5.5$^-$) into $(5.3_1^{\pm})$ and $(5.3_0^{\pm})$ one finds

$$_1\breve{\lambda}_{0;\tilde{\mathbf{t}}_{+;1}m} \equiv {}_1\breve{\lambda}_{0;\mathbf{d}m} = \sqrt{_1c_0}\,{}_1\breve{\lambda}_{1;\mathbf{d}m} = -\sqrt{_1c_0}\,(\mu_o + 2m+1)/(1+\sqrt{_1c_0}) < 0, \tag{5.29d}$$

$$_1\breve{\lambda}_{0;\tilde{\mathbf{t}}_{+;2}m} \equiv {}_1\breve{\lambda}_{0;\mathbf{b}m} = -\sqrt{_1c_0}\,{}_1\breve{\lambda}_{1;\mathbf{b}m} = -\sqrt{_1c_0}\,(\mu_o + 2m+1)/(\sqrt{_1c_0} - 1) < 0 \tag{5.29a}$$

in both Areas $A_m$ and $B_m$,

$$_1\breve{\lambda}_{0;\tilde{\mathbf{t}}_{-;1}m} \equiv {}_1\breve{\lambda}_{0;\mathbf{a}m} = -\sqrt{_1c_0}\,{}_1\breve{\lambda}_{1;\mathbf{a}m} = \sqrt{_1c_0}\,(\mu_o - 2m-1)/(\sqrt{_1c_0} - 1) > 0, \tag{5.29b}$$

$$_1\breve{\lambda}_{0;\tilde{\mathbf{t}}_{-;2}m} \equiv {}_1\breve{\lambda}_{0;\mathbf{c}m} = \sqrt{_1c_0}\,{}_1\breve{\lambda}_{1;\mathbf{c}m} = \sqrt{_1c_0}\,(\mu_o - 2m-1)/(1+\sqrt{_1c_0}) > 0 \tag{5.29b}$$

in Area $A_m$, and

$$_1\breve{\lambda}_{0;\tilde{\mathbf{t}}_{-;1}m} \equiv {}_1\breve{\lambda}_{0;\mathbf{d}'m} = \sqrt{_1c_0}\,{}_1\breve{\lambda}_{1;\mathbf{d}'m} = -\sqrt{_1c_0}\,(2m+1-\mu_o)/(1+\sqrt{_1c_0}) < 0, \tag{5.29d'}$$

$$_1\breve{\lambda}_{0;\tilde{\mathbf{t}}_{-;2}m} \equiv {}_1\breve{\lambda}_{0;\mathbf{b}'m} = -\sqrt{_1c_0}\,{}_1\breve{\lambda}_{1;\mathbf{b}'m} = \sqrt{_1c_0}\,(2m+1-\mu_o)/(1-\sqrt{_1c_0}) < 0, \tag{5.29b'}$$

in Area $B_m$.

The free term of quadratic polynomial ($5.1_1^-$) changes its sign at the border line $B_m|D_m$ and as a result both polynomial roots are positive for $\lambda_o > 2m + 1 - \mu_o > 0$ (as far as the polynomial has positive discriminant). We thus infer that the irregular ℐS solution **d**'m turns into the third ℐS solution of type **b** above the border line $B_m|D_m$. In Fig. 5.2 we refer this regular solution as **b**"2. The analysis of signed ExpDiffs ($5.3_1^+$) and ($5.3_0^+$) shows that the ℐS solution $\tilde{\mathbf{t}}_{+,2}$m (if exists) always preserves its type **b** until it disappears below straight line ($5.25^\dagger$). Both ℐS solutions **b**'m and **b**"m disappear above straight line ($5.25^-$).

An analysis of quadratic polynomial ($5.1_1^+$) shows that its free term changes its sign at the border line $C_m|D_m$ and has two positive roots inside Area $C_m$. Since both roots of companion polynomial ($5.1_0^+$) are negative there we infer that the ℐS solution $\tilde{\mathbf{t}}_{+,1}$ alters from **d** to **b** below the border line $C_m|D_m$, as predicted by the general arguments presented in Appendix A. Note that only ℐS solutions of type **b** (if any) exist below the $C_m$-fragment of straight line ($5.25^-$).

The above observations correlating roots of quadratic polynomials ($5.1_1^\pm$) and ($5.1_0^\pm$) with the types of the appropriate ℐS solutions for ${}_1c_0 > 1$ are summarized in Table 5.2.

Table 5.2

**Classification of ∮S solutions for the LTP with a larger-than-1 root**

| | Area $A_m$ | Area $B_m$ | Area $D_m$ | Area $C_m$ above line (5.25†) |
|---|---|---|---|---|
| $\tilde{t}_{-,1}$ | c | d' | a' above line (5.25+)<br>b" below line (5.25¯) | a' above line (5.25+)<br>b" below line (5.25¯) |
| $\tilde{t}_{-,2}$ | a | b' | a above line (5.25+)<br>b' below line (5.25¯) | a above line (5.25+)<br>b' below line (5.25¯) |
| $\tilde{t}_{+,1}$ | d | d | d | b''' |
| $\tilde{t}_{+,2}$ | b | b | b | b |

For ${}_1c_0$ >1 and $\lambda_o$ = 0 the right-hand side of (5.14*) is larger than 2m+1 so threshold (5.14) always starts in Area $A_m$. Therefore the ∮S solution **a**m (being the only ∮S solution of type **a** in this region) is the one which has the energy $\varepsilon_{\mathbf{c}0}$ everywhere along the threshold fragment from the curve starting point to its crossing with the border line $A_m|D_m$. As $\lambda_o$ grows at fixed value of $\mu_o$ both discriminant (5.2¯) and therefore the ExpDiff $|\,{}_1\lambda_{1;\mathbf{a}m}\,|$ monotonically decrease. Since discriminant (5.2¯) is nonnegative along threshold curve (5.14):

$$\Delta_{-}^{1;m}(\mu_o,\lambda_o;{}_1c_0)\Big|_{\sqrt{\lambda_o^2+{}_1c_0m^2}=m+1-\mu_o} = 4[({}_1c_0+1)m+1-\mu_o]^2 \qquad (5.30)$$

the curve must lie above straight line (5.25+) just touching it at the point $\mu_o = ({}_1c_0+1)m+1$:

$$\mu_{\Diamond\mathbf{a};m}(\lambda_o;-{}_1c_0-1,{}_1c_0)-\sqrt{1-1/{}_1c_0}\,\lambda_o-2m-1=\frac{[\lambda_o/\sqrt{{}_1c_0}-(\sqrt{{}_1c_0}-1)m]^2}{\sqrt{\lambda_o^2+{}_1c_0m^2}+\sqrt{1-1/{}_1c_0}\,\lambda_o+m}\geq 0. \tag{5.31}$$

If

$$\mu_o>({}_1c_0+1)m+1 \tag{5.32}$$

then the ℐS solution **a**m remains nodeless until it disappears at straight line (5.25$^+$). Otherwise it is nodeless above threshold curve (5.14), i.e., for

$$\lambda_o<\sqrt{m^2-{}_1c_0(\mu_o-m-1)^2} \tag{5.33}$$

under constraint

$$\mu_o-m-1<{}_1c_0m. \tag{5.33*}$$

In the intermediate case $\mu_o=({}_1c_0+1)m+1$ the nodeless ℐS solution **a**m disappears as soon as the ExpDiff $|{}_1\lambda_{1;\mathbf{a}m}|$ reaches its lower bound m (see subplot a) of Fig. 5.3 for an illustrative example).

Since straight line (5.25$^+$) lies entirely in Area $A_0$ the secondary ℐS solution **a**'m exists only in the region with the discrete energy spectrum. Starting from zero value along the m$^{th}$ **c**/**a**' separatrix its energy monotonically decreases as $\lambda_o$ grows at fixed value of $\mu_o$ until the ∮S solutions of type **a**, **a**m and **a**'m, merge at straight line (5.25$^+$):

$$ {}_1\lambda_{1;\tilde{\mathbf{t}}_{-;1}m}\Big|_{\mu_o=m+1+\sqrt{\lambda_o^2+{}_1c_0m^2}} = {}_1\lambda_{1;\tilde{\mathbf{t}}_{-;2}m}\Big|_{\mu_o=m+1+\sqrt{\lambda_o^2+{}_10\,m^2}} \tag{5.34}$$
$$ = -(\mu_o - 2m - 1)/({}_1c_0 - 1) < 0 \quad ({}_1c_0 < 1) $$

and then disappear below this line. Under condition (5.32) both ∮S solution **a**m and **a**" are nodeless right above threshold (5.14).

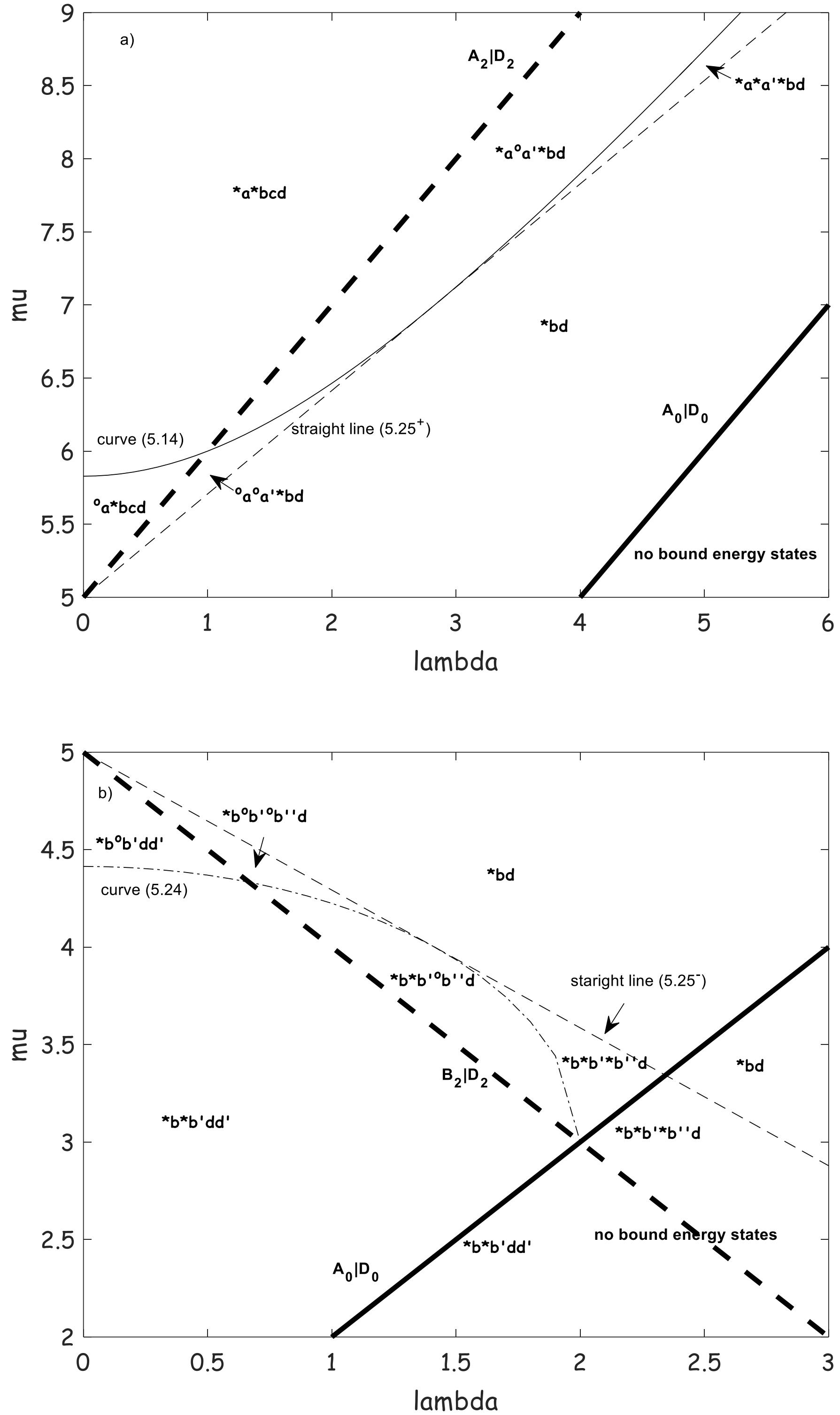

**Figure 5.3. Magnified fragments of regions with nodeless regular ∮S solutions for the same LTP as in Fig. 5.2**

As $\lambda_o$ decreases at fixed value of $\mu_o$ the signed ExpDiff ${}_1\lambda_{1;\tilde{\mathbf{t}}_{-;2}m}$ increases until it becomes equal to 0 at the border line $A_m|D_m$ so the ∮S solution $\tilde{\mathbf{t}}_{-,2}m$ turns into the eigenfunction. on other side of the border line. If condition (5.32) holds then ${}_1\lambda_{1;\mathbf{a}'m} = -m$ at the threshold and as a result the ∮S solution **a**'m must lie above the ground energy level at smaller values of $\lambda_o$. Otherwise it lies above the ground energy level regardless of the value of $\lambda_o$.

It is worth mentioning in this context that curve (5.16$^-$) lies below the horizontal straight line $\mu_o = m+1$ where no ∮S solution of type **a** exists for ${}_1c_0 > 1$ and therefore the appropriate ground-energy solution $\tilde{\mathbf{t}}_{-,1}m$ must be of type **d**. One can also verify that fraction (5.12) is positive regardless of the value of the coefficient for both j=1 and j=2 as far as the ∮S solution $\tilde{\mathbf{t}}_{-,j}m$ is of type **a**. This observation independently confirms that signed ExpDiff for any the ∮S solution of type **a** below the ground energy level must be smaller than –m.

Representing (5.3$^+$) for j=2 as

$$ |\,{}_1\lambda_{0;\mathbf{b}m}\,| - m = {}_1\lambda_{1;\mathbf{b}m} \pm \mu_o + 2m + 1 > 0 \tag{5.35}$$

we infer that the primary ∮S solution **b**m is nodeless at any point above the straight line (5.27)

where primary $\oint$S solutions **b**m and **d**m co-exist at the same energy

$$ {}_1\varepsilon_{\mathbf{b}m} = {}_1\varepsilon_{\mathbf{d}m} = -(\mu_o + 2m + 1)^2 / ({}_1c_0 - 1)^2. \tag{5.36}$$

One can directly verify that the signed ExpDiff ${}_1\lambda_{0;\mathbf{b}m}$ evaluated along straight line (5.27):

$$ {}_1\lambda_{0;\tilde{\mathbf{t}}_{+,2}m} = {}_1\lambda_{0;\mathbf{b}m} = -{}_1c_0(\mu_o + 2m + 1)/({}_1c_0 - 1) < -\mu_o - 2m - 1 \quad ({}_1c_0 > 1) \tag{5.37}$$

is smaller than –m as expected.

Since relation (5.19) holds for any $\oint$S solution of type **b** we infer that the secondary $\oint$S solution $\tilde{\mathbf{t}}_{-,j}m$ ($\tilde{\mathbf{t}}_{-,1} = \mathbf{b}''$, $\tilde{\mathbf{t}}_{-,2} = \mathbf{b}'$) is nodeless if ${}_1\lambda_{1;\tilde{\mathbf{t}}_{-,j}m} > \mu_o - m - 1.$ Representing the quadratic equation for the ExpDiff ${}_1\lambda_{1;\tilde{\mathbf{t}}_{-,j}m}$ as

$$ ({}_1\lambda_{1;\tilde{\mathbf{t}}_{-;j}m} - {}_1\lambda_{1;\mathbf{c}0})[({}_1c_0 - 1)({}_1\lambda_{1;\tilde{\mathbf{t}}_{-;j}m} + {}_1\lambda_{1;\mathbf{c}0}) + 2(\mu_o - 1)] = 4m({}_1\lambda_{1;\tilde{\mathbf{t}}_{-;j}m} - \mu_o + m + 1) \tag{5.38}$$

one can verify that the solution in question lies below the ground energy level iff ${}_1\lambda_{0;\tilde{\mathbf{t}}_{-,j}m} < -m,$ in agreement with the IOS rule.

For ${}_1c_0 > 1$ threshold curve (5.24) always lies below the horizontal straight line $\mu_o$ = 2m+1. Examination of signed ExpDiff (5.29b'), coupled with (5.36) and (5.37), shows that the secondary $\oint$S solution **b**'m lies below the ground energy level at $\lambda_o$ = 0 iff

$$ \mu_o - 1 < (1 + 1/\sqrt{{}_1c_0})\, m \tag{5.39}$$

As $\lambda_o$ grows at fixed value of $\mu_o$ discriminant (5.2¯) and therefore the ExpDiff $|\,{}_1\lambda_{0;\tilde{\mathbf{t}}_{-;2}m}\,|$, given by $(5.3\overline{\phantom{0}}_0)$ with j=2, monotonically decrease. Since discriminant (5.2¯) is nonnegative along threshold curve (5.24):

$$\Delta_{-}^{1;m}\Big|_{\sqrt{(m^2-\lambda_o^2)/{}_1c_0}=\mu_o-m-1} = 4[m-{}_1c_0(\mu_o-m-1)]^2 \ge 0 \qquad (5.40)$$

the curve must lie below straight line (5.25¯):

$$2m+1-\sqrt{1-1/{}_1c_0}\,\lambda_o-\mu_{\lozenge\mathbf{b};m}(\lambda_o;-{}_1c_0-1,{}_1c_0)$$

$$=\frac{(m\sqrt{1-1/{}_1c_0}-\lambda_o)^2}{\sqrt{(m^2-\lambda_o^2)/{}_1c_0}+m-\sqrt{1-1/{}_1c_0}\,\lambda_o}\ge 0 \qquad (5.41)$$

just touching it at $\mu_o=(1/{}_1c_0+1)m+1$ as illustrated by subplot b) of Fig. 5.3. If

$$\mu_o-m-1<m/\sqrt{{}_1c_0} \qquad (5.42^*)$$

condition (5.38) holds then the 𝒥S solution **b**'m is nodeless for

$$\lambda_o<\sqrt{m^2-{}_1c_0(\mu_o-m-1)^2}. \qquad (5.42)$$

Otherwise it always lies above the ground energy level until it merges with the 𝒥S solution **b**"m:

$${}_1\lambda_{0;\tilde{\mathbf{t}}_{-;1}m}\Big|_{\mu_o=2m+1-\sqrt{1-1/{}_1c_0}\,\lambda_o}={}_1\lambda_{0;\tilde{\mathbf{t}}_{-;2}m}\Big|_{\mu_o=2m+1-\sqrt{1-1/{}_1c_0}\,\lambda_o} \qquad (5.43)$$

$$=-{}_1c_0(2m+1-\mu_o)/({}_1c_0-1)<0$$

at the straight line (5.25¯) and then disappears on other side of the line.

Under condition (5.39) both secondary 𝒥S solution **b**'m and **b**"m are nodeless right below threshold (5.24). As $\lambda_o$ decreases at fixed value of $\mu_o$ the signed ExpDiff ${}_1\lambda_{0;\tilde{\mathbf{t}}_{-;1}m}$ increases until it becomes equal to 0 at the border line $B_m|D_m$ so the 𝒥S solution $\tilde{\mathbf{t}}_{-,1}m$ turns into the irregular solution **d**'m on other side of the border line. If condition (5.38) holds then ${}_1\lambda_{0;\mathbf{b}''m}=-m$ at the threshold and as a result the 𝒥S solution **b**"m must lie above the ground energy level at smaller values of $\lambda_o$. Otherwise it lies above the ground energy level regardless of the value of $\lambda_o$.

## 6. The CTP limit and related nodeless 'Case-I' and 'Case-II' AEH solutions for the Rosen-Morse potential

In the limiting case of the RM potential (${}_1c_0 = 1$) quadratic polynomials $(5.1_1^{\pm})$ and $(5.1_0^{\pm})$ become linear:

$$ {}_1^{\pm}G_1^{1;m}[\lambda \,|\, {}_1\boldsymbol{G}^{00}] = -2(2m+1\pm\mu_o)\lambda + \lambda_o^2 - (2m+1\pm\mu_o)^2 \qquad (6.1_1^{\pm}) $$

and

$$ {}_1^{\pm}G_1^{0;m}[\lambda \,|\, {}_1\boldsymbol{G}^{00}] = -2(2m+1\pm\mu_o)\lambda - \lambda_o^2 - (2m+1\pm\mu_o)^2 \qquad (6.1_0^{\pm}) $$

so the signed ExpDiffs for the $\wp$S solutions $\tilde{\mathbf{t}}_{\pm}m$ are given by the elementary formulas

$$ {}_1\lambda_{1;\tilde{\mathbf{t}}_{\pm}m} = -\frac{(\pm\mu_o + 2m + 1)^2 - \lambda_o^2}{2(\pm\mu_o + 2m + 1)} \qquad (6.2_1^{\pm}) $$

$$ \equiv A \pm m - \frac{B}{A \pm m}, \qquad (6.\overline{2}_1^{\pm}) $$

and

$$ {}_1\lambda_{0;\tilde{\mathbf{t}}_{\pm}m} = -\frac{\lambda_o^2 + (2m+1\pm\mu_o)^2}{2(2m+1\pm\mu_o)}. \qquad (6.2_0^{\pm}) $$

$$ \equiv A \pm m + \frac{B}{A \pm m}, \qquad (6.\overline{2}_0^{\pm}) $$

or, in Quesne's terms [13],

$$ {}_1\lambda_{1;\tilde{\mathbf{t}}_{+}m} = A + m + 1 - \frac{B}{A + m + 1}, \qquad (6.\overline{2}_1^{+}) $$

$$ {}_1\lambda_{0;\tilde{\mathbf{t}}_{+}m} = A + m + 1 + \frac{B}{A + m + 1} \qquad (6.\overline{2}_0^{+}) $$

and

$$ {}_1\lambda_{1;\tilde{\mathbf{t}}_m} = A - m - \frac{B}{A-m}, \qquad (6.\bar{2}_1^-) $$

$$ {}_1\lambda_{0;\tilde{\mathbf{t}}_m} = A - m + \frac{B}{A-m} \qquad (6.\bar{2}_0^-) $$

where

$$ A = \tfrac{1}{2}(\mu_o - 1) > 0, \quad B = \tfrac{1}{4}\lambda_o^2 \qquad (6.3) $$

The derived formulas the signed ExpDiffs of the $\mathcal{J}$S solution in question precisely match ChExps of solutions (2.7) and (2.8) in [13], keeping in mind that second-order differential equation (2.3) in [13] can be converted into the canonical form by the gauge transformation:

$$ \phi[\eta;\varepsilon;A,B] = (1-\eta^2)^{-\frac{1}{2}}\,\Phi[\tfrac{1}{2}(\eta+1);\varepsilon\,|\,{}_1\boldsymbol{G}^{00}], \qquad (6.4) $$

with η=2z–1 in our terms. It is worth mentioning in this connection that the factorization energies

$$ \varepsilon_{\tilde{\mathbf{t}}_\pm m} = -\frac{[(\pm\mu_o + 2m + 1)^2 - \lambda_o^2]^2}{4(\pm\mu_o + 2m + 1)^2} \qquad (6.5^\pm) $$

$$ \equiv -(A \pm m)^2 - \frac{B^2}{(A \pm m)^2} + 2B \qquad (6.\bar{5}^\pm) $$

differ from the appropriate formulas in [13] by the potential energy reference point which is equal to 2B in Quesne's representation of the RM potential in terms of the parameters A and B.

Note that signed ExpDiff $(6.\,2_0^+)$ is negative everywhere in the $\lambda_o\mu_o$ plane. Above the border line $C_m|D_m$ this is also true for its companion $(6.\,2_1^+)$. We thus conclude that the $\mathcal{J}$S solution $\tilde{\mathbf{t}}_+m$ is nothing but the primary solution of type **d**, at least inside Area $A_0$ with the discrete energy spectrum. (In Fig. 6.1 below this Area is separated from the rest of the plane by

bold solid line with label $A_0|D_0$.) Let us also point to the fact that all four signed ExpDiffs are negative in Area $B_m$ which implies that both primary and secondary ∮S solutions of type **b** existent in this Area for $_1c_0 > 1$ disappear at the limit $_1c_0 = 1$.

As already pointed to by Quesne [13], the second root $_1\lambda_{1;\tilde{t}_m}$ exhibits a more complicated behavior; namely, in the general case of an asymmetric RM potential ($\lambda_o > 0$) it changes its sign along the both border lines $A_m|D_m$ and $B_m|D_m$ as well as along the horizontal straight line $\mu_o = 2m + 1$ where it becomes infinitely large. One can directly verify that the straight lines (5.25) and (5.25*) approach this horizontal line as $_1c_0 \to 1$ whereas third straight line (5.27) disappears if $_1c_0$ set to 1.

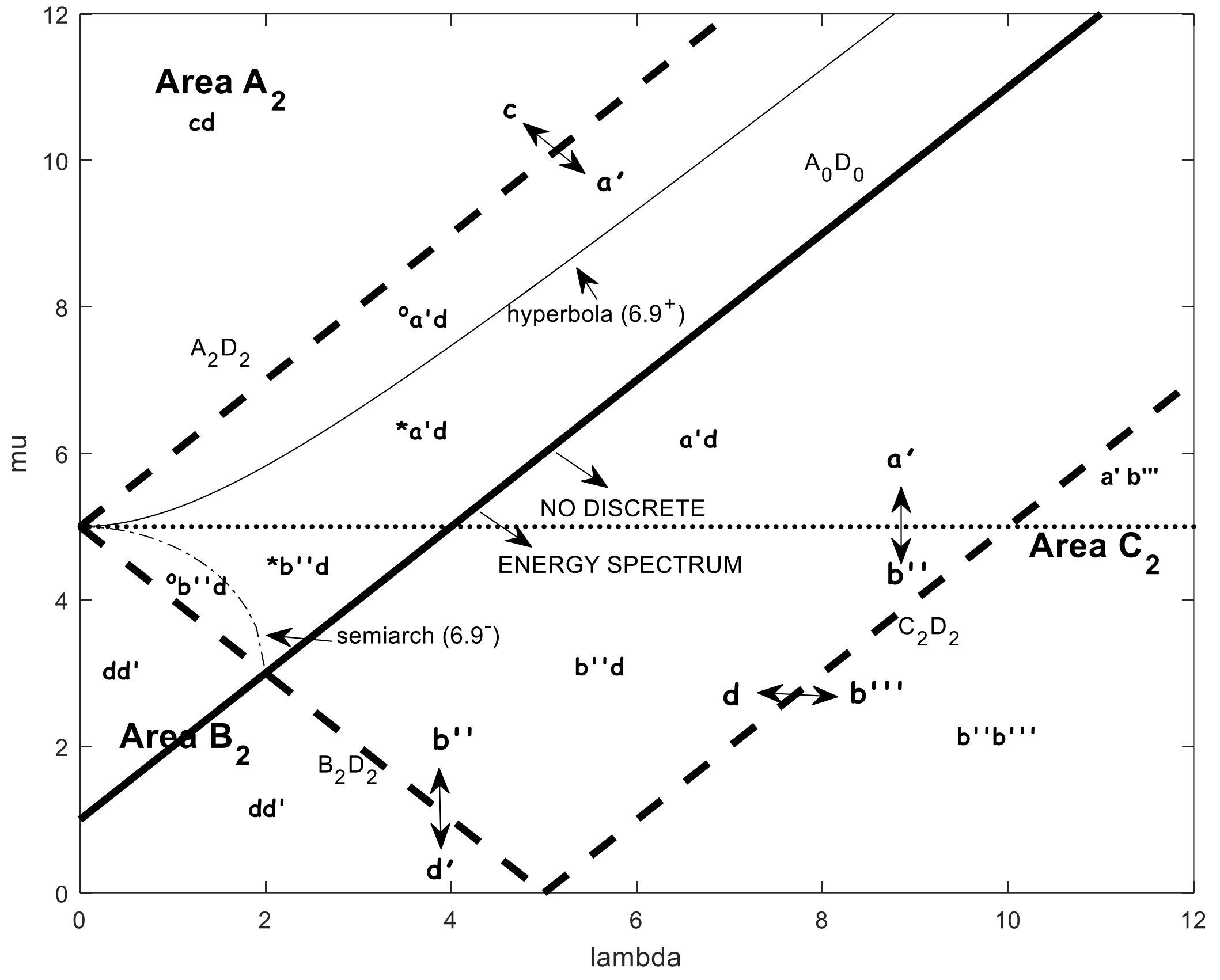

**Figure 6.1. Possible pairs of ∮S solutions in the limiting case of the RM potential**

It directly follows from $(6.2_0^-)$ that the type of the $\oint$S solution $\tilde{\mathbf{t}}_-m$ is either **a** or **c** (**b** or **d**) above (below) the horizontal straight line $\mu_o = 2m+1$ (see Fig. 6.1 for illustration). Since the numerator of the fraction in the right-hand side of $(6.2_1^-)$ is positive in Area $A_m$ we conclude that $\tilde{\mathbf{t}}_- = \mathbf{c}$ for $\mu_o > \lambda_o + 2m+1$. We thus conclude that both primary regular $\oint$S solutions of types **a** and **b** existent in both cases $_1c_0 < 1$ and $_1c_0 > 1$ disappears at the limit $_1c_0 = 1$.

Right below the border line $A_m|D_m$ the $\oint$S solution $\tilde{\mathbf{t}}_-m$ changes its type from **c** to **a** turning into the secondary regular solution **a**'m (Case I solution in [13]). It preserves its type at any point of Area $D_m$ above the horizontal straight line $\mu_o = 2m+1$. The solution type changes again below the line, this time for **b**, so the $\oint$S solution $\tilde{\mathbf{t}}_-m$ turns into another secondary regular $\oint$S solution **b**"m (Case II solution in [13]) – the only solution of type **b** which retains in Area $D_m$ when the TP coefficient $_1c_0$ approaches its limiting value of 1 from above. Finally the solution $\tilde{\mathbf{t}}_-m$ turns into the secondary irregular $\oint$S solution **d**'m below the border line $B_m|D_m$. The possibly types of the $\oint$S solutions for the asymmetric RM potential can be summarized as follows

$$\tilde{\mathbf{t}}_+ = \begin{cases} \mathbf{b}''' \text{ in Area } C_m, \\ \mathbf{d} \text{ otherwise} \end{cases} \tag{6.6$^+$}$$

and

$$\tilde{\mathbf{t}}_- = \begin{cases} \mathbf{c} \text{ in Area } A_m, \\ \mathbf{a}' \text{ in Areas } C_m \text{ and } D_m \text{ for } \mu_o > 2m+1, \\ \mathbf{b}'' \text{ in Areas } C_m \text{ and } D_m \text{ for } \mu_o < 2m+1, \\ \mathbf{d}' \text{ in Area } B_m. \end{cases} \tag{6.6$^-$}$$

One can alternatively re-write $(6.6^-)$ as

$$\tilde{\mathbf{t}}_{-} = \begin{cases} \mathbf{c} \ \text{ for } 0 \le m < A - \sqrt{B}, \\ \mathbf{a}' \ \text{ for } A - \sqrt{B} < m < A, \\ \mathbf{b}'' \ \text{ for } A < m < A + \sqrt{B}, \\ \mathbf{d}' \ \text{ for } m > A + \sqrt{B}. \end{cases} \qquad (6.\bar{6}^{-})$$

The energy of the $\mathcal{J}$S solution $\tilde{\mathbf{t}}_{-}m$ relative to the energy zero point:

$$\varepsilon_{\tilde{\mathbf{t}}_{-}m} - \varepsilon_{\mathbf{c}0} = -\frac{m(m+1-\mu_o)}{(1-\mu_o)^2(2m+1-\mu_o)^2} \times [(\mu_o - 1)^2(2m+1-\mu_o)^2 - \lambda_o^4] \qquad (6.7)$$

vanishes along the curves

$$(\mu_o - 1 - m)^2 = m^2 \pm \lambda_o^2. \qquad (6.8)$$

Here we are interested only in two of them:

$$\mu_o = m + 1 + \sqrt{m^2 \pm \lambda_o^2} \qquad (6.9^{\pm})$$

which both lie in Area $D_m$ as illustrated by Fig. 6.1 and represent the limiting cases of threshold curves (5.14) and (5.24) for ${}_1c_0 = 1$. The regular branches of the $\mathcal{J}$S solution $\tilde{\mathbf{t}}_{-}m$ are nodeless inside the sector between the two curves. In Quesne's notation

$$A = \tfrac{1}{2}m + \sqrt{\tfrac{1}{4}m^2 \pm B} \qquad (6.\bar{9}^{\pm})$$

so the secondary regular solutions $\mathbf{a}'m$ and $\mathbf{b}''m$ are nodeless within the ranges

$$0 < A - \tfrac{1}{2}m < \sqrt{\tfrac{1}{4}m^2 + B} \qquad (6.10a)$$

and

$$0 \le \sqrt{\tfrac{1}{4}m^2 - B} < A - \tfrac{1}{2}m < 0, \qquad (6.10b)$$

accordingly. One can easily associate the above inequalities with Quesne's conditions (1a) and (1b) for solution (2.7) in [13] to lie below the ground energy level

Isospectral SUSY partners of the RM potential constructed using the nodeless solutions **a**'m and **b**"m as seed functions for multi-step CLDTs will be discussed in detail in a separate publication.

## 7. Conclusions and further developments

We have presented a complete list of nodeless regular $\mathcal{J}$S solutions for the r-GRef potential on line assuming that the corresponding PFrB $_1\boldsymbol{G}^{K\Im 0}$ is constructed using either the second-order TP with two distinct real roots or its first-order reduction (referred to in our papers as 'LTP'). (Nodeless regular JS solutions for the r-GRef potential constructed using the DRtTP have been analyzed in a separate paper [10].) It has been already demonstrated in [1] that each such solution can be used as the FF for the CLDT to construct an isospectral partner potential conditionally exactly quantized by $\mathcal{J}$S Heine polynomials.

A special attention was given to the region where the *r*-GRef potential $V[z|_1\boldsymbol{G}^{K\Im 0}]$ has at least m bound energy states – the so-called 'Area $A_m$'. It has been proven that the m$^{th}$ eigenfunction **c**m is accompanied by two primary regular $\mathcal{J}$S solutions **a**m and **b**m if the TP has positive discriminant ($\Delta_T > 0$). We have explicitly selected the regions of Area $A_m$, where these solutions are nodeless and therefore can be used as seed functions to construct the ladder isospectral rational partners of the given potential *r*-GRef potential. It has been also shown that the double-step algorithm based on fraction formula (2.30) can be used at any point of Area $A_m$ to determine the signed ExpDiff $_1\lambda_{0;\dagger m}$ for each of four real roots of quartic polynomial (2.29a) if $_1c_0 < 1$ and $\Delta_T > 0$.

The LTP *r*-GRef potential which was presented in Section 5 simply as an analytically solvable illustrative example represents a very exceptional case which will require a special attention in future publications.  Namely, as already pointed to in [1] analytical representations for isospectral SUSY partners of LTP *r*-GRef potentials cannot be obtained simply by setting $_1a_2$ to 0 in the appropriate expressions for the ladders of PFrBs $^{p}_{1}G^{210}$ constructed using the second-order TP for the numerator of the fractional density function.  For this reason a family of LTP potentials conditionally exactly quantized by GS-Heun polynomials would require a special consideration in one of the forthcoming papers.  Another reason for paying special attention to the LTP  *r*-GRef potential is that  its SUSY partners generated using basis solutions †0 are quantized by 𝒥S Heun polynomials [31].  Therefore one can expect that the underlying Heun equations have some nontrivial features deserved to be investigated more scrupulously.

In [32] we will extend the above analysis to nodeless regular GS solutions of the RCSLE associated with the *c*-GRef potential exactly quantized by generalized Laguerre polynomials.  Again, as already demonstrated in [1] (and explored in more details in [33])  each regular GS solution below the ground-energy level can be used as the seed function for multi-step CLDTs to construct an isospectral partner potential conditionally exactly quantized by either 𝒥S or ℒS Heine polynomials.

Part III will focus on analytical formulas for the signed ExpDiffs along the **ad**- and **bd**-DRt curves where one of the double-step algorithms fails.  In particular we will study possible implications of this failure for the isospectral SUSY partners of the GRef potential constructed using the nodeless regular 𝒥S solutions from each DRt doublet as the FFs for the CLDT.

Both in this paper and in related studies of SUSY partners of *r*-and *c*-GRef potentials we restrict our analysis solely to nodeless *regular* GS solutions keeping in mind any multi-index FF using these solutions as seed functions for a multi-step  CLDT is necessarily nodeless.  The symmetric potential curves [34, 35], with the Ginocchio potential [36] as the commonly cited example, represent an extradionary case because they co-occur along intersection between Levai's [37] and Milson's [7] 'symmetric tangent polynomial' (sym-TP) reductions the *r*- and *i*-GRef potentials.  As a result the symmetric Milson potential is also *exactly solvable* via hypergeometric functions after the appropriate change of variable. Another important corollary directly followed from this dualism [35] is that the symmetric GRef (sym-GRef) potential is exactly quantized by both ultraspherical [29] and Masjedjamei [38] (definite-parity Romanovski-Routh) orthogonal polynomials.

In [4] we have proven general theorem stating that any irregular invariant-under-reflection solution of the 1D Schrödinger equation with a symmetric potential is necessarily nodeless if it lies below the ground energy level. The direct corollary of this finding is that the Wroskian of invariant-under-reflection seed solutions of the Schrödinger equation with an arbitrary symmetric potential does not have nodes inside the quantization interval if all the solutions lie below the ground energy level. To justify this assertion one simply needs to express the FF for the DT of the given multi-step SUSY partner of the original symmetric potential as ratio of the appropriate Crum Wroskians [39]. Since any such solution necessarily lies below the ground energy level and is itself an even function of x one can use mathematical induction to confirm that each Crum Wroskian in the given chain is nodeless as asserted.

In particular this is true for Crum Wroskians formed by any subset of symmetric Gegenbauer-seed ($\mathfrak{G}$S) solutions of type d below the ground energy level. Therefore we infer that any multi-step SUSY partner of the sym-GRef potential generated using such solutions as seed functions for the appropriate CLDT can be conditionally exactly quantized by definite-parity $\mathfrak{G}$S Heine polynomials.

It is worth mentioning in this connection that a set of irregular invariant-under-reflection AEH solutions of the Hermite equation at negative energies is formed by even Hermite functions of an imaginary argument. The fact that the latter functions do not have nodes on the real axis was explicitly exploited by Samsonov et al [40-43] to construct rational SUSY partners of the harmonic oscillator which were quantized by orthogonal polynomials of a new type – the so-called '$X_m$-Hermite' polynomials in Gomez-Ullate, Kamran, and Milson's [44] classification scheme of exceptional orthogonal polynomials. (The Samsonov-Ovcharov [40, 41] orthogonal polynomials were more recently re-discovered by Fellows and Smith [45] in a very similar context.) Based on the arguments presented above we thus conclude that Crum Wroskians of even Hermite functions of an imaginary argument *i*x may not have nodes at real values of x, in agreement with the accurate theory of multi-index exceptional Hermite polynomials developed by Gomez-Ullate, Grandati, and Milson [46] a short time ago.

Note that an extension of these ideas to shape-invariant *r*- and *c*-GRef potentials as recently suggested in [47, 48] has not been properly justified so far, with the sym-RM potential well as the only exclusion. On other hand, Quesne's conjecture [49] that Routh-seed (ℜS) solutions $\mathbf{d},2j_{\mathbf{d}}$ and $\mathbf{d}',2j_{\mathbf{d}'}$ (in our terms) can be used as FFs for constructing quantized-by-polynomials SUSY partners of the Gendenshtein potential [50] (the Scarf II potential in Cooper et al.'s classification scheme [51, 52]) has been proven [4] to be correct. In Appendix D in [1] we have briefly discussed the single-step SUSY partners of the Gendenshtein potential conditionally exactly quantized by orthogonal sets of ℜS Heine polynomials. It will be demonstrated in another forthcoming paper [53], dealing exclusively with SUSY partners of the Gendenshtein potential, that the arguments presented in [1] can be also extended to multi-step CLDTS using as seed functions some specially selected combinations of nodeless irregular solutions. for will be discussed. The common remarkable feature of the constructed rational potentials is that they are conditionally exactly quantized by *orthogonal* sets of multi-index ℜS Heine polynomials – the direct consequence of the fact [54] that the ExpDiffs for the second-order poles in the appropriate RCSLEs are all energy-independent.

## Acknowledgements

I would like to thank Géza Lévai for drawing my attention to his recent studies on rational potentials exactly quantized by Jacobi polynomials.

## Appendix A

### Disappearance of the upper energy level along the ZFE $\mathbf{c}/\mathbf{a}'$-separatrix

The common benchmark of the ZFE separatrices is that one can analytically derive some important features of $\mathfrak{J}$S solutions in the neighborhood of each separatrix without solving the given quartic equation. In particular, representing (2.30) as

$$ {}_1\lambda_{0;\dagger m} = \frac{\mu_o^2 - (\pm\lambda_o + 2m+1)^2 + {}_1d_1\lambda_{1;\dagger m}^2 + 2(2m+1)(\pm\lambda_o - {}_1\lambda_{1;\dagger m})}{2({}_1\lambda_{1;\dagger m} + 2m+1)} \qquad (A.1) $$

one finds that

$$ {}_1\lambda_{0;\dagger m}\Big|_{X_m^{\pm}|D_m} = \frac{{}_1d_1\lambda_{1;\dagger m}^2 + 2(2m+1)(\pm\lambda_o - {}_1\lambda_{1;\dagger m})}{2({}_1\lambda_{1;\dagger m} + 2m+1)} \qquad (A.2) $$

along the border lines $X_m^+ | D_m$ ($X_m^+ = A_m$) and $X_m^- | D_m$ ($X_m^- = B_m$ or $C_m$). In particular, as a direct corollary of (A.2) we conclude that

$$ {}_1\lambda_{0;\dagger_{X^{\pm}} m}\Big|_{X_m^{\pm}|D_m} = \pm\lambda_o \qquad (A.2^*) $$

for any $\mathcal{J}$S solution $\dagger_X m$ changing its type from one side of the separatrices $X_m | D_m$ ($X_m = A_m, B_m$, or $C_m$) to another. Since the signed ExpDiff ${}_1\lambda_{0;\dagger_X m}$ has a nonzero absolute value for $\lambda_o > 0$ it must have the same sign on both sides of each separatrix (except the limiting case of the AL potential curve). This implies that the only possible change of solution type from one side of the separatrix to another is either **a** $\leftrightarrow$ **c** or **b** $\leftrightarrow$ **d**.

The root ${}_1\lambda_{1;\dagger_X m}$ vanishing along the given separatrix $X_m | D_m$ can be approximated in its neighborhood $X_m \leftrightarrow D_m$ by the fraction

$$ {}_1\lambda_{1;\dagger_X m}\Big|_{X_m \leftrightarrow D_m} \approx -\frac{G_{4;0}^{1;m}(\lambda_o, \mu_o)}{G_{4;1}^{1;m}(\lambda_o, \mu_o)}, \qquad (A.3) $$

dependent only on the linear coefficient and free term of quartic polynomial (2.29a),

$$ G_{4;1}^{1;m}(\lambda_o, \mu_o) = -4(2m+1)[\mu_o^2 + \lambda_o^2 - (2m+1)^2] \qquad (A.4) $$

and

$$G^{1;m}_{4;0}(\lambda_o,\mu_o) = (\mu_o + \lambda_o + 2m + 1)(\mu_o - \lambda_o - 2m - 1) \times (\mu_o + 2m + 1 - \lambda_o)(\mu_o + \lambda_o - 2m - 1). \qquad \text{(A.4*)}$$

Evaluating the linear coefficient along each border line $X_m \mid D_m$ and comparing the derived formulas

$$G^{1;m}_{4;1}(|\,2m+1-\mu_o\,|,\mu_o) = -8(2m+1)\mu_o(\mu_o - 2m - 1) \text{ if } X_m = A_m \text{ or } B_m \qquad \text{(A.5)}$$

or

$$G^{1;m}_{4;1}(\mu_o + 2m + 1, \mu_o) = -8(2m+1)\mu_o(\mu_o + 2m + 1) < 0 \text{ if } X_m = C_m \qquad \text{(A.5C)}$$

with the free term linearized in the vicinity of the appropriate line:

$$G^{1;m}_{4;0}(\lambda_o,\mu_o)\Big|_{A_m \leftrightarrow D_m} \approx 8(2m+1)\mu_o(\mu_o - 2m - 1)(\mu_o - \lambda_o - 2m - 1), \qquad \text{(A.6A)}$$

$$G^{1;m}_{4;0}(\lambda_o,\mu_o)\Big|_{B_m \leftrightarrow D_m} \approx 8(2m+1)\mu_o(\mu_o - 2m - 1)(\lambda_o + \mu_o - 2m - 1), \qquad \text{(A.6B)}$$

and

$$G^{1;m}_{4;0}(\lambda_o,\mu_o)\Big|_{C_m \leftrightarrow D_m} \approx 8(2m+1)\mu_o(\mu_o + 2m + 1)(\lambda_o - \mu_o - 2m - 1) \qquad \text{(A.6C)}$$

gives

$${}_1\lambda_{1;\dagger_A m}\Big|_{A_m \leftrightarrow D_m} \approx \mu_o - \lambda_o - 2m - 1, \qquad \text{(A.7A)}$$

$${}_1\lambda_{1;\dagger_B m}\Big|_{B_m \leftrightarrow D_m} \approx \lambda_o + \mu_o - 2m - 1, \qquad \text{(A.7B)}$$

and

$${}_1\lambda_{1;\dagger_C m}\Big|_{C_m \leftrightarrow D_m} \approx \lambda_o - \mu_o - 2m - 1 \qquad \text{(A.7C)}$$

so the root in question is positive on the upper side of the border line $A_m \mid D_m$ or $B_m \mid D_m$ and on the C-side of the border line $C_m \mid D_m$. By combining this observation with (A.2*) we infer that the $\mathcal{J}$S solution $\dagger_X m$ changing its type from **b** to **d** along the separatrix $X^{-}_m \mid D_m$ must be of type **b** on the D-side of the border line $B_m \mid D_m$ and on the C-side of the border line $C_m \mid D_m$ (as illustrated by Fig. 2.1). We also confirm that the m^th eigenfunction turns into the regular ℑS solution of type **a** on the D-side of the separatrix $A_m|D_m$. The ground-energy eigenfunction **c**0 disappears on the D-side of the basic **c**/**a′**-ZFE separatrix $A_0|D_0$.

In theory it is possible (though quite unlikely) that normalizable ℑS solutions re-appear outside Area $A_0$ after a pair of ℑS solutions of the same type merge so quartic polynomial (2.29a) has complex roots. However all the closed-formed examples analyzed by us so far support the assumption that the discrete energy spectrum exists only in Area $A_0$.

Based on the latter assumption Area $A_2$ in Fig. 2.1 represents the region where the *r*-GRef potential generated by means of a TP with positive discriminant has at least 3 bound energy levels. The number of bound energy levels below the second-order **c**/**a′**-ZFE separatrix may not exceed 2. No discrete energy spectrum exists below the solid line representing the basic ZFE **c**/**a′**-separatrix $A_0|D_0$.

## Appendix B

### Some analytical results for ℑS solutions near the crossing point between the $A_m|D_m$ and $A_m|B_m$ border lines

Taking advantage of the fact that the ℑS solutions **d**m and $\breve{\mathbf{t}}_{-}({}_1c_0)\mathrm{m}$, where $\breve{\mathbf{t}}_{-}({}_1c_0)$ = **a** or **b** for ${}_1c_0 < 1$ or ${}_1c_0 > 1$ accordingly, are unambiguously defined $\lambda_o = 0$, $\mu_o = 2m+1$ near the point quartic polynomial (2.23a) can be decomposed as follows

$$\begin{aligned} G_4^{1;m}[\lambda \mid {}_1\boldsymbol{G}^{K\Im 0}] &\equiv G_4^{1;m}[\lambda;\lambda_o,\mu_o;T_2] \\ &= (\lambda - {}_1\lambda_{1;\mathbf{d}m})(\lambda - {}_1\lambda_{1;\breve{\mathbf{t}}_{-}({}_1c_0)m})\ {}_{A|B}G_2^{1;m}[\lambda \mid {}_1\boldsymbol{G}^{K\Im 0}]. \end{aligned} \qquad (B.1)$$

In particular, as it has been demonstrated in Section 5 decomposition (B.1) is valid for any point above straight line (5.27) for the LTP.

By expressing quartic polynomial (2.29a) in terms of a new argument y = $\lambda / \lambda_o$, introducing a temporary m-dependent parameter

$$\nu_m \equiv (\mu_o - 2m - 1) / \lambda_o, \tag{B.2}$$

and making $\lambda_o$ tend to 0 at a fixed value of $\nu_m$ one finds

$$\lim_{\lambda_o \to 0} \{G_4^{1;m}[\lambda_o y; \lambda_o, 2m+1+\lambda_o \nu_m; T_2] / \lambda_o^2\} = 4(2m+1)^2 A_2^{(1)}[y; \nu_m; {}_1c_0], \tag{B.3}$$

where

$$A_2^{(1)}[y; \nu; {}_1c_0] \equiv (1 - {}_1c_0) y^2 - 2\nu y + \nu^2 - 1 \tag{B.4}$$

so two roots of quartic polynomial (B.1) vanishing at the point $\lambda_o = 0$, $\mu_o = 2m+1$ satisfy the following limit formula:

$$\lim_{\lambda_o \to 0} \{{}_1\lambda_{1;\hat{\mathbf{t}}_\pm m} / \lambda_o\} = \frac{\nu \pm \sqrt{\Delta_A(\nu; {}_1c_0)}}{1 - {}_1c_0}. \tag{B.5$^\pm$}$$

It is worthy of note that the leading coefficient of quadratic polynomial (B.4) vanishes if ${}_1c_0 = 1$. Therefore the scheme based on decomposition (B.1) is inapplicable to sym-TP GRef potential (A.1a) introduced in Appendix A. This is a direct consequence of our observation in Section 3 that there are three (not two!) $\mathcal{J}$S solutions which change their type at the point $\lambda_o = 0$, $\mu_o = 2m+1$ in the particular case of the sym-GRef potential.

Replacing $\mu_o$ by $\nu$ in (2.30), making $\lambda_o$ tend to 0, and substituting (B.5$^\pm$) into the resultant formula then gives

$$\lim_{\lambda_o \to 0} \{{}_1\lambda_{0;\hat{\mathbf{t}}_\pm m} / \lambda_o\} = \frac{{}_1c_0 \nu \pm \sqrt{\Delta_A(\nu; {}_1c_0)}}{{}_1c_0 - 1}. \tag{B.6$^\pm$}$$

Note that quadratic polynomial (B.3) has positive discriminant

$$\Delta_A(\nu;{}_1c_0) = 4({}_1c_0\,\nu^2 + 1 - {}_1c_0) \qquad \text{(B.7)}$$

for any value of $\nu$ if ${}_1c_0 < 1$ so quartic polynomial (B.1) for ${}_1c_0 < 1$ has four real roots at small values of $\lambda_o$. To be more precise, its two roots ${}_1\lambda_{1;\hat{\mathbf{t}}_\pm m}$ vanishing at $\lambda_o = 0$, $\mu_o = 2m+1$ are both positive for $\nu > 1$ (Area $A_m$), both negative for $\nu < -1$ (Area $B_m$) and have opposite sign for $-1 < \nu < 1$ (Area $D_m$). One can verify that the right-hand sides of limit formulas (B.6$^\pm$) coincide with the roots of the quadratic polynomial

$$A_2^{(0)}[y;\nu;{}_1c_0] \equiv ({}_1c_0 - 1)y^2 - 2\,{}_1c_0\nu y + {}_1c_0\nu^2 + 1 \qquad \text{(B.8)}$$

which has the same discriminant as quadratic polynomial (B.4). Keeping in mind that quadratic polynomial (B.8) has two roots of opposite sign if its leading coefficient is negative we conclude that $\hat{\mathbf{t}}_+ = \mathbf{c}$, $\hat{\mathbf{t}}_- = \mathbf{b}$ in Area $A_m$, $\hat{\mathbf{t}}_- = \mathbf{a}'$, $\hat{\mathbf{t}}_+ = \mathbf{b}$ in Area $D_m$, and $\hat{\mathbf{t}}_+ = \mathbf{a}'$, $\hat{\mathbf{t}}_- = \mathbf{d}'$ in Area $B_m$ for ${}_1c_0 < 1$, as illustrated by subplot a) of Fig. B.1 for m=2 . In particular this implies that $\mathfrak{J}$S solutions below the separatrix $B_m|D_m$ have types **a**, **d**, **a**', and **d**' at least for small values of $\lambda_o$, in agreement with the analysis presented in Section 2 for the AL potential curves.

If ${}_1c_0 > 1$ then $\mathfrak{J}$S solutions exhibit a more complicated pattern even for small values of the reflective barrier since discriminant (B.7) becomes negative in Area $D_m$ for $-\nu_A < \nu_m < \nu_A$, where

$$\nu_A = \sqrt{1 - 1/{}_1c_0} < 1\,. \qquad \text{(B.9)}$$

As a result there are only two $\mathfrak{J}$S solutions (**b**m and **d**m) in this range of $\nu$. As a matter of fact this analysis can be extended beyond two ‘infinitesimal straight-line’ segments

$$\mu_o \approx 2m + 1 \pm \lambda_o\nu_A \quad \text{for } \lambda_o << 1 \qquad \text{(B.10}^\pm\text{)}$$

marked by dotted lines on subplot b) of Fig. B.1. Namely, discriminant (B.7) of the quadratic polynomial ${}_{A|B}G_2^{1;m}[\lambda\,|\,{}_1\boldsymbol{G}^{K\Im 0}]$ vanishes along two finite curved segments and as a result he

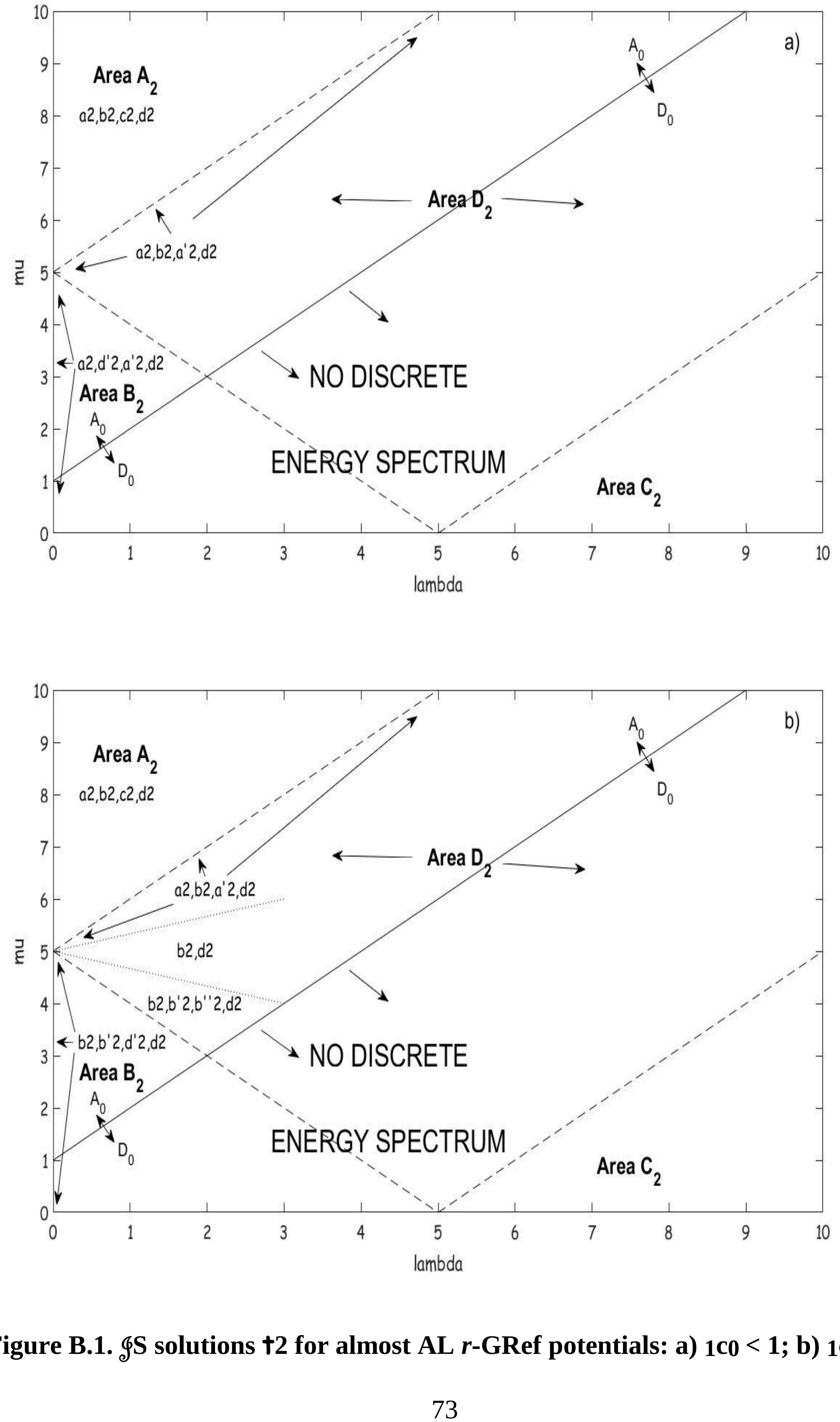

**Figure B.1. ∯S solutions †2 for almost AL *r*-GRef potentials: a) 1c0 < 1; b) 1c0 > 1**

polynomial has a double root which is negative (positive) along the upper (lower) curve for some nonzero values of the reflective barrier. This implies that two regular ℑS solutions of the same type co-exist at the same energy along each curve. The pair of the ℑS solutions of type **a** disappears below along the upper curve but then a pair of the regular ℑS solutions re-appears again below the lower curve but now these are two secondary solutions of type **b**, **b**'m and **b**"m.

As pointed to in Section 5 the sketched upper threshold curve turns at ${}_1a_2 = 0$ into the line segment cut off from straight line (5.25$^+$) by line (5.27) whereas the lower curve straightens into line (5.25$^-$). The quadratic polynomial

$${}_{A|B}G_2^{1;m}[\lambda\,|\,{}_1\boldsymbol{\mathcal{G}}^{110}] \equiv {}_{\bar{1}}G_2^{1;m}[\lambda\,|\,{}_1\boldsymbol{\mathcal{G}}^{110}] \qquad \text{(B.11)}$$

has complex roots inside the triangle formed by these three straight lines which represents the region with only two ℑS solutions, **b**m and **d**m.

Since quadratic polynomial (B.9) has two positive (negative) roots for $\nu > 0$ ($\nu < 0$) while two roots of quadratic polynomial (B.5) have opposite signs both in Area $A_m$ and in Area $B_m$ we conclude that two roots ${}_1\lambda_{1;\hat{\mathbf{t}}_{\pm}m}$ vanishing at the point $\lambda_o = 0,\ \mu_o = 2m+1$ describe a pair of ℑS solutions **a**m and **c**m in Area $A_m$ and pair of ℑS solutions **b**′m and **d**′m in Area $B_m$. As expected from the analysis presented in Section 3 for the AL potential curves ℑS solutions below the border line $B_m|D_m$ have types **b**, **d**, **b**', and **d**' at least at small $\lambda_o$.

## Appendix C

### Cutoffs of regular ℑS solutions along the **ad**- and **bd**-DRt curves

The main purpose of this appendix is to prove that the isoenergetic ℑS solutions associated with double zeros of quartic polynomials (2.29a) or (2.34) along DRt curve (2.33) or (2.37) have two distinct types: **a** and **d** or **b** and **d**, respectively.

The quartic polynomials (2.29a) and (2.34) can be analytically decomposed along **ad**-DRt

hyperbola (2.33) as

$$G_4^{1;m}[\lambda;\lambda_o,{}_{\mathbf{ad}}\mu_{o;m}(\lambda_o;{}_1c_0);{}_1T_2] = (\lambda+2m+1)^2\,{}_{\mathbf{ad}}G_2^{1;m}[\lambda;\lambda_o;{}_1T_2] \qquad (C.1_1)$$

and

$$G_4^{0;m}[\lambda;\lambda_o,{}_{\mathbf{ad}}\mu_{o;m}(\lambda_o;{}_1c_0);{}_1T_2] = [\lambda^2-\lambda_o^2-{}_1c_0(2m+1)^2]\,{}_{\mathbf{ad}}G_2^{0;m}[\lambda;\lambda_o;{}_1T_2], \qquad (C.1_0)$$

where

$${}_{\mathbf{ad}}G_2^{1;m}[\lambda;\lambda_o;{}_1T_2] = ({}_1d^2-4\,{}_1c_0)\lambda^2 - 2\,{}_1d({}_1d+2)(2m+1)\lambda + 4({}_{\mathbf{ad}}\lambda_{o;m\times}^2-\lambda_o^2) \qquad (C.2_1)$$

and

$${}_{\mathbf{ad}}G_2^{0;m}[\lambda;\lambda_o;{}_1T_2] = {}_{\mathbf{ad}}G_{4;4}^{0;m}\,|\,{}_1\boldsymbol{G}^{K\Im 0})\lambda^2 + {}_{\mathbf{ad}}G_{4;3}^{0;m}\,|\,{}_1\boldsymbol{G}^{K\Im 0})\lambda + {}_{\mathbf{ad}}G_{2;0}^{0;m}(\lambda_o;{}_1T_2). \qquad (C.2_0)$$

As a direct corollary of $(C.1_0)$ we conclude that the roots of quartic polynomial (2.34) associated with the isoenergetic $\mathfrak{J}$S solutions have opposite signs so the solutions do have distinct types **a** and **d** at each point of hyperbola (2.33) as stated above.

An analysis of the inequality

$${}_{\mathbf{ad}}\mu_{o;m}^2(\lambda_o;{}_1d) - (\lambda_o+2m+1)^2 = -2(2m+1)[\lambda_o+(1+\tfrac{1}{2}\,{}_1d)(2m+1)] \qquad (C.3)$$

shows that the hyperbola intersects the border line $A_m|D_m$ at the crossing point

$$\lambda_o = {}_{\mathbf{ad}}\lambda_{o;m\times} \equiv -\tfrac{1}{2}({}_1d+2)(2m+1) \geq 0, \qquad (C.4)$$

$$\mu_o = {}_{\mathbf{ad}}\mu_{o;m\times} \equiv -\tfrac{1}{2}\,{}_1d(2m+1) > 0 \qquad (C.4^\dagger)$$

iff ${}_1d < -2$. On the contrary, the hyperbola does not go through Area $A_m$ for the TP with non-negative discriminant ${}_1\Delta_T$ iff

$$-2 < {}_1d \leq -2\sqrt{{}_1c_0}\,. \qquad (C.5)$$

Insofar as the TP coefficient $_1d$ changes within this range for $_1c_0 < 1$ fractional relation (2.30) can be used everywhere in Area $A_m$ to compute the signed ExpDiff $_1\lambda_{0;\dagger m}$ for each of four real roots of quartic polynomial (2.29a) and as a result the appropriate double-step algorithm is applicable to any potential curve with at least m bound energy states.

Note also both leading and linear coefficients of quadratic polynomial (C.$2_0$) depend only on the TP parameters, namely,

$$G^{0;m}_{2;2} \mid {}_1\boldsymbol{G}^{K\Im 0}) \equiv G^{0;m}_{4;4} \mid {}_1\boldsymbol{G}^{K\Im 0}) = [(\sqrt{{}_1c_0} - 1)^2 - {}_1a_2] \times [(\sqrt{{}_1c_0} + 1)^2 - {}_1a_2] / {}_1c_0 \quad \text{(C.6)}$$

$$= {}_1d^2 / {}_1c_0 - 4 \quad \text{(C.6')}$$

$$= {}^{+}_{1}g_2({}_1T_2)\, {}^{-}_{1}g_2({}_1T_2) / {}_1c_0 \quad \text{(C.6'')}$$

and

$$G^{0;m}_{2;1} \mid {}_1\boldsymbol{G}^{K\Im 0}) \equiv G^{0;m}_{4;3} \mid {}_1\boldsymbol{G}^{K\Im 0}) = -4(2m+1)({}_1d+2) \quad \text{(C.7)}$$

whereas the negative free term is a function of all three parameters $\lambda_o$, $_1d$, and $_1c_0$:

$$_{ad}G^{0;m}_{2;0}(\lambda_o; {}_1T_2) = -G^{0;m}_{4;0}(\lambda_o, \mu_{o;m}(\lambda_o; {}_1d); {}_1T_2) / [\lambda_o^2 + {}_1c_0(2m+1)^2] \quad \text{(C.8)}$$

$$= -{}_1d^2\lambda_o^2 / {}_1c_0 - ({}_1d+2)^2(2m+1)^2 < 0. \quad \text{(C.8*)}$$

Inspection of discriminants of quadratic polynomials (C.$2_1$) and (C.$2_0$),

$$_{ad}\Delta_{1;m}(\lambda_o; {}_1d, {}_1c_0) = 16\, {}_1c_0 [({}_1d+2)^2(2m+1)^2 + ({}_1d^2 - 4\, {}_1c_0)\lambda_o^2 / {}_1c_0] \quad \text{(C.9}_1\text{)}$$

and

$$_{ad}\Delta_{0;m}(\lambda_o; {}_1T_2) = \tfrac{1}{4}({}_1d / {}_1c_0)^2\, {}_{ad}\Delta_{1;m}(\lambda_o; {}_1T_2), \quad \text{(C.9}_0\text{)}$$

shows that they are positive for any TP with positive discriminant ${}_1\Delta_T$ regardless of the height of the reflective barrier. Quadratic polynomial (C.2$_1$) thus has two positive roots at each point of the **ad**-DRt hyperbola inside Area $A_m$ ($0 \le \lambda_o < {}_{\mathbf{ad}}\lambda_{0;m\times}$).

Since leading coefficient (C.6) and free term (C.8*) of quadratic polynomial (C.2$_0$) have opposite signs in the case of our interest this must be also true for its two roots. Therefore the roots of the quadratic polynomials in question specify the ∮S solutions **b**m and **c**m:

$$ {}_1\lambda_{0;\mathbf{b}m}\Big|_{\lambda_m=\mu_m} = \frac{2({}_1d+2)(2m+1) - \tfrac{1}{2}\sqrt{{}_{\mathbf{ad}}\Delta_{0;m}(\lambda_o;{}_1T_2)}}{{}_1d^2/{}_1c_0 - 4} < 0, \qquad \text{(C.10b}_0\text{)} $$

$$ {}_1\lambda_{1;\mathbf{b}m}\Big|_{\lambda_m=\mu_m} = \frac{{}_1d^2({}_1d+2)(2m+1) - {}_1c_0\sqrt{{}_{\mathbf{ad}}\Delta_{0;m}(\lambda_o;{}_1T_2)}}{{}_1d({}_1d^2 - 4{}_1c_0)}, \qquad \text{(C.10b}_1\text{)} $$

and

$$ {}_1\lambda_{0;\mathbf{c}m}\Big|_{\lambda_m=\mu_m} = \frac{{}_1c_0[2({}_1d+2)(2m+1) + \tfrac{1}{2}\sqrt{{}_{\mathbf{ad}}\Delta_{0;m}(\lambda_o;{}_1T_2)}]}{{}_1d^2 - 4{}_1c_0} > 0, \qquad \text{(C.10c}_0\text{)} $$

$$ {}_1\lambda_{1;\mathbf{c}m}\Big|_{\lambda_m=\mu_m} = \frac{{}_1d^2({}_1d+2)(2m+1) + {}_1c_0\sqrt{{}_{\mathbf{ad}}\Delta_{0;m}(\lambda_o;{}_1T_2)}}{{}_1d({}_1d^2 - 4{}_1c_0)}, \qquad \text{(C.10c}_1\text{)} $$

accordingly, where (C.10b$_1$) and (C.10c$_1$) were derived from (C.10b$_0$) and (C.10c$_0$) based on the interrelation

$$ {}_1d\,{}_1\lambda_{1;\mathbf{t}m}\Big|_{\lambda_m=\mu_m} = 2\,{}_1\lambda_{0;\mathbf{t}m}\Big|_{\lambda_m=\mu_m} + ({}_1d+2)(2m+1) \qquad \text{(C.11)} $$

($\mathbf{t} \ne \mathbf{a}$ or $\mathbf{d}$).

This confirms that the **abcd**-quartet of ∮S solutions does exist at any point of the $A_m$-fragment of the **ad**-DRt hyperbola as predicted by the general arguments presented in Section 4.

Roots of quadratic polynomials (C.2$_1$) and (C.2$_0$) outside Area $A_m$ will be studied in detail in Part III.

Examination of the relation

$$\mathbf{bd}\mu^{2}_{o;m}(\lambda_o;{}_1T_2)-(\lambda_o+2m+1)^2={}_1d(\lambda_o+2m+1)(\lambda_o-\mathbf{bd}\lambda_{o;m\times})/{}_1c_0 \qquad \text{(C.12)}$$

where

$$\mathbf{bd}\lambda_{o;m\times}\equiv(1+2\,{}_1c_0/{}_1d)(2m+1), \qquad \text{(C.13)}$$

reveals that there are four distinct subsets of PFrBs generated by the TPs with nonnegative discriminants $({}_1d\le-2\sqrt{{}_1c_0})$ such that

$$0<{}_1d+{}_1c_0<{}_1d+2\,{}_1c_0\le 2\sqrt{{}_1c_0}(\sqrt{{}_1c_0}-1) \qquad \text{(C.14a)}$$

if ${}_1a_2>1$ and ${}_1c_0>1$,

$${}_1d+{}_1c_0<{}_1d+2\,{}_1c_0\le -2\sqrt{{}_1c_0}(1-\sqrt{{}_1c_0})<0 \qquad \text{(C.14b)}$$

if ${}_1a_2<1$ and ${}_1c_0<1$, and either

$${}_1d+{}_1c_0<0<{}_1d+2\,{}_1c_0\le 2\sqrt{{}_1c_0}(\sqrt{{}_1c_0}-1) \qquad \text{(C.14c)}$$

or

$${}_1d+{}_1c_0<{}_1d+2\,{}_1c_0<0 \qquad \text{(C.14c')}$$

if ${}_1a_2<1$ and ${}_1c_0>1$. The **bd**-DRt curve intersects the border line $A_m|D_m$ at the crossing point

$$\lambda_o=\mathbf{bd}\lambda_{o;m\times}\equiv(1+2\,{}_1c_0/{}_1d)(2m+1),\;\; \mu_o=\mathbf{bd}\lambda_{o;m\times}+2m+1 \qquad \text{(C.15)}$$

iff $\mathbf{bd}\lambda_{o;m\times}>0$, i.e., the cross point exists only for PFrBs (C.14b) and (C.14c'). On other hand, the right-hand side of (C.12) is negative everywhere along **bd**-DRt curve (2.37) if

$$-2\,{}_1c_0<{}_1d<-2\sqrt{{}_1c_0}, \qquad \text{(C.16)}$$

i.e., for both PFrBs (C.14a) and (C.14c). As far as the latter condition holds fractional relation (2.35) can be used everywhere in Area $A_m$ to compute the signed ExpDiff ${}_1\lambda_{1;\mathbf{a}m}$ for each of four real roots of quartic polynomial (2.34) so the double-step algorithm in question is applicable to any potential curve with at least m bound energy states.

The quartic polynomials (2.29a) and (2.34) are analytically decomposed along **bd**-DRt curve (2.37) as follows

$$G_4^{0;m}[\lambda;\lambda_o,{}_{\mathbf{bd}}\mu_{o;m}(\lambda_o;{}_1d/{}_1c_0);{}_1T_2] = (\lambda+2m+1)^2\,{}_{\mathbf{bd}}G_2^{0;m}[\lambda;\lambda_o;{}_1T_2] \tag{C.17$_0$}$$

and

$$G_4^{0;m}[\lambda;\lambda_o,{}_{\mathbf{bd}}\mu_{o;m}(\lambda_o;{}_1d/{}_1c_0);{}_1T_2] = \{\lambda^2-[(2m+1)^2-\lambda_o^2]/{}_1c_0\}\,{}_{\mathbf{bd}}G_2^{1;m}[\lambda;\lambda_o;{}_1T_2], \tag{C.17$_1$}$$

where

$$\begin{aligned}{}_{\mathbf{bd}}G_2^{0;m}[\lambda;\lambda_o;{}_1T_2] = ({}_1d^2-4{}_1c_0)\lambda^2/{}_1c_0 - 2{}_1d({}_1d+2{}_1c_0)(2m+1)\lambda/{}_1c_0 \\ +({}_1d+2{}_1c_0)^2(2m+1)^2/{}_1c_0+4\lambda_o^2\end{aligned} \tag{C.18$_0$}$$

and

$$\begin{aligned}{}_{\mathbf{bd}}G_2^{1;m}[\lambda;\lambda_o;{}_1T_2] = ({}_1d^2/{}_1c_0-4)\lambda^2 - 4(2m+1)({}_1d/{}_1c_0+2)\lambda \\ +{}_1d^2(\lambda_o^2-{}_{\mathbf{bd}}\lambda_{o;m\times}^2)/{}_1c_0^2.\end{aligned} \tag{C.18$_1$}$$

Examination of decomposition (C.17$_1$) confirms that the isoenergetic ℘S solutions do have distinct types **b** and **d** at each point of the semiarch.

For both PFrBs (C.15b) and (C.15c') **bd**-DRt curve (2.37) has a form of the semiarch which starts in Area $A_m$ at $\lambda_o = 0$, intersects the border line $A_m|D_m$ at crossing point (C.15) and ends at $\mu_o = 0$. Since the free term of quadratic polynomial (C.18$_1$) along the $A_m$-segment of the semiarch is negative the polynomial has two roots of opposite sign. Taking into account that both roots of quadratic polynomial (C.18$_0$) are positive along this segment we directly confirm that ℘S solutions **t**m form the **abcd** quartet at any point of the semiarch inside Area $A_m$.

The free term of quadratic polynomial (C.18$_1$) is negative its sign at the crossing point between the **bd**-DRt semiarch and the borderline $A_m|D_m$. Since the leading coefficient of the polynomial in question is positive for $\Delta_T > 0$ the polynomial must have two roots of opposite sign in Area $A_m$. On other hand quadratic polynomial (C.18$_0$) has two positive roots in Area $A_m$ for ${}_1d < -2c_0$ so the **abcd**-quartet of $\mathcal{J}$S solution thus exists at any point of the $A_m$-fragment of the **bd**-DRt semiarch (in agreement with the general theory).

The signed ExpDiffs ${}_1\lambda_{1;\mathbf{t}m}$ ($\mathbf{t}$ = **a** or **c**) for the $\mathcal{J}$S solutions **a**m and **c**m along the $A_m$-fragment of the **bd**-DRt semiarch are given by negative and positive zeros of quadratic polynomial (C.18$_1$):

$$ {}_1\lambda_{1;\mathbf{a}m}\Big|_{\mu_o = {}_{\mathbf{bd}}\mu_{o;m}(\lambda_o;{}_1T_2)} = \frac{2(2m+1)({}_1d + 2\,{}_1c_0) - \tfrac{1}{2}\,{}_1c_0\sqrt{{}_{\mathbf{bd}}\Delta_{1;m}(\lambda_o;{}_1T_2)}}{{}^{+}_{1}\breve{g}_2({}_1T_2)\,{}^{-}_{1}\breve{g}_2({}_1T_2)} < 0 \tag{C.19a$_1$} $$

and

$$ {}_1\lambda_{1;\mathbf{c}m}\Big|_{\mu_o = {}_{\mathbf{bd}}\mu_{o;m}(\lambda_o;{}_1T_2)} = \frac{2(2m+1)({}_1d + 2\,{}_1c_0) + \tfrac{1}{2}\,{}_1c_0\sqrt{{}_{\mathbf{bd}}\Delta_{1;m}(\lambda_o;{}_1T_2)}}{{}^{+}_{1}\breve{g}_2({}_1T_2)\,{}^{-}_{1}\breve{g}_2({}_1T_2)} > 0 \qquad (m < {}_1n_o) \tag{C.19c$_1$} $$

The supplementary relations

$$ {}_1\lambda_{0;\mathbf{a}m}\Big|_{\mu_o = {}_{\mathbf{bd}}\mu_{o;m}(\lambda_o;{}_1T_2)} = \frac{{}_1d^2(2m+1)({}_1d + 2\,{}_1c_0) - {}_1c_0^2\sqrt{{}_{\mathbf{bd}}\Delta_{1;m}(\lambda_o;{}_1T_2)}}{{}_1d\;{}^{+}_{1}g_2({}_1T_2)\,{}^{-}_{1}g_2({}_1T_2)} > 0 \tag{C.19a$_0$} $$

and

$$ {}_1\lambda_{0;\mathbf{c}m}\Big|_{\mu_o = {}_{\mathbf{bd}}\mu_{o;m}(\lambda_o;{}_1T_2)} = \frac{{}_1d^2(2m+1)({}_1d + 2\,{}_1c_0) - {}_1c_0^2\sqrt{{}_{\mathbf{bd}}\Delta_{1;m}(\lambda_o;{}_1T_2)}}{{}_1d\;{}^{+}_{1}g_2({}_1T_2)\,{}^{-}_{1}g_2({}_1T_2)} > 0 \qquad (m < {}_1n_o) \tag{C.19c$_0$} $$

are then obtained via interrelation (2.35) simplified along the **bd**-DRt semiarch as follows

$$ {}_1d\,{}_1\lambda_{0;\mathbf{\dagger}m}\Big|_{\mu_o={}_{\mathbf{bd}}\breve{\mu}_{o;m}(\lambda_o;{}_1T_2)} - 2\,{}_1c_0\,{}_1\lambda_{1;\mathbf{\dagger}m}\Big|_{\mu_o={}_{\mathbf{bd}}\breve{\mu}_{o;m}(\lambda_o;{}_1T_2)} $$

$$ =({}_1d+2\,{}_1c_0)(2m+1). \qquad \text{(C.20)} $$

Again, taking into account that discriminant of quadratic polynomial (C.$18_0$) differs from

$$ {}_{\mathbf{bd}}\Delta_{1;m}(\lambda_o;{}_1d,{}_1c_0)=4\,{}_1d^2[({}_1d/{}_1c_0+2)^2(2m+1)^2/{}_1c_0-({}_1d^2/{}_1c_0-4)\lambda_o^2/{}_1c_0^2] $$

(C.$21_1$)

only by a $\lambda_o$-independent scale:

$$ {}_{\mathbf{bd}}\Delta_{0;m}(\lambda_o;{}_1T_2)=4({}_1c_0/{}_1d)^2\,{}_{\mathbf{bd}}\Delta_{1;m}(\lambda_o;{}_1T_2) \qquad \text{(C.}21_0\text{)} $$

one can verify that (C.$20a_0$) and (C.$20c_0$) define two zeros of quadratic polynomial (C.$18_0$) as anticipated.

Examination of roots of quadratic polynomials (C.$18_0$) and (C.$18_1$) outside Area $A_m$ is postponed for Part III.